\documentclass{emulateapj}

\newcommand{\beq}{\begin{equation}}
\newcommand{\beqa}{\begin{eqnarray}}
\newcommand{\eeq}{\end{equation}}
\newcommand{\eeqa}{\end{eqnarray}}
\def\vec#1{\ensuremath{\mathchoice{\mbox{\boldmath$\displaystyle#1$}}
{\mbox{\boldmath$\textstyle#1$}}
{\mbox{\boldmath$\scriptstyle#1$}}
{\mbox{\boldmath$\scriptscriptstyle#1$}}}}

\shorttitle{Magnetospheric Accretion and Ejection of Matter in Resistive
 MHD Simulations}
\shortauthors{\v{C}emelji\'{c}, Shang \& Chiang}

\begin{document}

\title{Magnetospheric Accretion and Ejection of Matter in Resistive
Magnetohydrodynamic Simulations}
\author{M. \v{C}emelji\'{c}, H. Shang and T.-Y.
Chiang}
\affil{Academia Sinica, Institute of Astronomy and Astrophysics and
Theoretical Institute for Advanced Research in
Astrophysics, P.O. Box 23-141,
Taipei 106, Taiwan}
\email{miki@tiara.sinica.edu.tw}

\begin{abstract} 
The ejection of matter in the close vicinity of a
young stellar object is investigated, treating the accretion disk
as a gravitationally bound reservoir of matter. By solving the
resistive MHD equations in 2D axisymmetry using our version of the
Zeus-3D code with newly implemented resistivity, we study the effect
of magnetic diffusivity in the magnetospheric accretion-ejection
mechanism. Physical resistivity was included in the whole
computational domain so that reconnection is enabled by the physical
as well as the numerical resistivity. We show, for the first
time, that quasi-stationary fast ejecta of matter, which
we call {\em micro-ejections}, of small mass and angular momentum
fluxes, can be launched from a purely resistive magnetosphere. They
are produced by a combination of pressure gradient and magnetic forces,
in presence of ongoing magnetic reconnection along the boundary layer
between the star and the disk, where a current sheet is formed. Mass flux of
micro-ejection increases with increasing magnetic field strength and
stellar rotation rate, and is not dependent on the disk to corona
density ratio and amount of resistivity. 

\end{abstract}


\section{Introduction}\label{intro}
Highly collimated outflows have been observed from AGNs to
Young Stellar Objects (YSOs) and young brown dwarfs \citep{wh1,wh2}.
Accreting compact stars, like accreting white dwarfs in symbiotic binaries
\citep{sk03} and neutron stars like Cir X-1 \citep{hei07}, also show similar
outflowing phenomena. An outflow is characterized as a jet if
it is super-magnetosonic, collimated into an apparent narrow opening, and
reaches a stationary or quasi-stationary state. Such
high-velocity outflowing fluxes of matter are an integral part of stellar
evolution. Observations in multiple wavelengths are reaching closer and closer
to the objects that drive them.

Among all the systems, models of launching outflows in YSOs are closest to
scrutiny by observations due to available data from star--forming regions.
An accretion disk, through which matter accretes onto the young star with
velocities close to a free--fall, is often associated with a jet--driving
YSO \citep{e94,e06}. A correlation between the accretion rate and the
high-velocity jet power was found in many Classical T-Tauri Stars (CTTSs)
\citep{Cab90}. The ratio of mass loss in the outflow to disk accretion
rate, $\dot{M}_{\rm w}/\dot{M}_{\rm a}$, extracted from observations is
hard to constrain. It is best estimated to be approximately 0.1 through
measurements of optical forbidden lines and veiling --- see
e.g.\ \citet{hart95} and \citet{edw08}. Recently, He~I~$\lambda$10830 line
has been used as a probe into the high-velocity winds originating from
the inner region where the star interacts with the disk \citep{e03,kw07}.
Despite the potential diagnostic power of such emission lines, the actual
structure and physical conditions of outflows can be more complex.
Star-disk interaction is also investigated by means of X-rays,
which offer deepest probe into the launching region. During
flare events from the protostellar objects X-rays are emitted and
have been observed (\citet{fav05}, \citet{gia06}, \citet{aar10},
\citet{mclw11}). Production of such X-rays is possible by either
high temperature or high flow velocities, which can be
related to the stellar surface,
shocks in the stellar magnetosphere or, as we will suggest, to the
reconnection events in the star-disk magnetosphere. Such events will leave
imprint in the chemical and physical properties of the object
\citep{shuetal07}. To further interpret the observed line profiles,
and the origins of outflows from the close vicinity of CTTSs, predictions
from both theoretical and numerical models are required. 
\begin{table*}
\caption{In many resistive-MHD simulations performed up to date, assumptions
for initial conditions vary much and duration of simulations is very
different. We list some of important works to position
our work in the context. In the second column, ``fast'' stellar rotation
means that corotation radius is smaller than the disk truncation radius,
$R_{cor}<R_t$ and ``slow'' rotation means that $R_{cor}>R_t$. Disk is
rotating in all the cases. Columns about physical resistivity and
viscosity specify only if the option is at all included in the code; for
where it is really effective, reader should check the related publication.
When present, viscosity is important only in the disk.
} 
\begin{tabular}{|c||c|c|c|c|c|c|}
\hline
 & & & & & & \\
\ \  & $\Omega_\ast$ &  corona & $\kappa=\rho_d/\rho_c$ &
$T_{\mathrm max}/T_0$ (days) & resistive & viscous\\
 & & & & & & \\
\hline
 & & & & & & \\
\ \  \citet{hay96} & non-rotating & 
  non-rotating & 10$^3$ & 5 & yes & no\\
\ \  \citet{hir} & non-rotating & rotating & 10$^4$ & 16.5 & yes & no\\
 & & rotating different than disk & & & & \\
\ \  \citet{ms97} & slow & solid body rotation
& 10$^2$ & 0.3 & yes & no \\
 & & corotating with star at $R_{\rm cor}$ & & & & \\
\ \ \citet{gods99a} & fast & rotating & $10^4$ & 100 & yes & no \\
\ \  \citet{rukl02} & slow &
corotating with star & 10$^2$ & 100 & no & yes \\
 & & for R$\leq R_{\rm cor}$, else with disk & & & & \\
\ \ \citet{kuk03} & slow & not in hydrostatic balance,
& 10$^3$ & 1000 & yes & yes \\
 & & non-rotating & & & &  \\
\ \  \citet{u06} & fast & corotating with star & 10$^3$ & 2000 & yes & yes \\
 & & for R$\leq R_{\rm cor}$, else with disk & & & & \\
\ \  \citet{R09} & fast & corotating with star  & 10$^4$ & 2000 & yes & yes\\
 & and slow & for R$\leq R_{\rm cor}$, else with disk & & & & \\
\hline
 & & & & & & \\
\ \  M\v{C} et al. (present paper) & slow &
corotating with star  & 10$^4$ & 1500 & yes & no \\
 & & for R$\leq R_{\rm cor}$, else with disk & & & &  \\
\hline
\end{tabular}
\label{tabla1}
\end{table*}

Outflows driven by energy derived from accretion are particularly appealing
in the scenarios of jet launching. Many models have been proposed based on
the concept of magnetocentrifugal wind mechanisms \citep{bp82},
differing in the origins of the underlying magnetic fields and locations
of matter launching. An outflow could be a disk wind driven by magnetic
fields dragged in from the envelope or generated by the disk dynamo, or an
inner disk wind anchored to the narrow innermost region as in the X-wind
model powered by an enhanced dynamo from the star-disk interaction
\citep{shuetal94, shuetal97}, simultaneously with an accretion funnel
\citep{osshu95}. It might also be a stellar wind driven along the
open field lines from the stellar surface by thermal or magnetic pressure
\citep{mattpu05,mattpu08}, or some combination of the different possibilities.
Related to the launch of winds, magnetospheric accretion has been
described in works by \citet{koe91}, \citet{osshu95} and
\citet{kold02} in the context of a magnetosphere interacting with the
surrounding disk, sharing some similarities with the compact objects like
neutron stars \citep{ghla79a,ghla79b}. Except for the
pure disk wind models, a magnetically connected star-disk system plays an
important role in the making of the young stellar system and the evolution
of angular momentum through the generation of outflows during the main
phase of accretion.

Numerical investigations have been followed up on the time-dependent
evolution of a system where the central star is magnetically connected to 
its accretion disk and their connection to jet formation and accretion. In
one of the earliest attempts by \citet{hay96}, where simulations of only a
few rotation periods were obtained, a dipole magnetosphere corotating with
the central star threaded the accretion disk that was in Keplerian 
rotation. Magnetic field lines connecting both the disk and the star
inflate outwards due to shear, and reconnection blows out the matter along
with the field, partially opening up the originally closed dipole loops.   
Gas can outflow from those opened field lines and might form part of the  
X-emission that is often associated with flares. Reconnection as a possible
origin of X-rays from such systems has also been indicated in \citet{EdP10}.
\citet{hir} investigated a magnetized star interacting with a truncated disk
that was threaded with an initially uniform field dragged in from the outer
core, in the same direction of the magnetosphere, but separated by a
neutral current sheet  in the equatorial plane as a result of interaction
between the fields brought together. For simplicity, the star was not rotating,
but the differentially rotating disk could anyway provide enough shear
to make the field inflate outwards, followed by a reconnection event
and mass transfer onto the magnetosphere. The transferred mass diverted into
two directions: one that falls onto the star and the other that flows out
along the opened stellar field lines.

Longer simulations by \citet{gods97}
with an aligned dipole and a conducting accretion disk showed that
differential rotation of the disk can drive episodes of loop expansion.
Such expansion can drive two outflow components of gas: one hot
convergent flow along the rotation axis, and another, slower cold flow
on the disk side of the expanding loop. \citet{ms97}, on the other hand,
investigated interactions of magnetospheres with accretion disks under
three different magnetic configurations and their respective dynamical
evolution.

\citet{kuk03} solved the disk in 1D with a radiative
hydrodynamic code by \citet{kley89}, and then extrapolated
the solution to 2D as their initial condition. For the
full 2D axisymmetric MHD problem, the induction equation, Lorentz force 
and Ohmic dissipation were now included into Kley's code, with the
assumption of equal viscous and resistive dissipations. With $Pr\sim 1$
magnetic field lines would not bend towards the axis. However, because of
non-equilibrium initial conditions, they bunched close to the star. The  
main result was that, with the assumed mass accretion rate of
$\dot{M}_0=10^{-7}M_\odot$~yr$^{-1}$, for a smaller magnetic field than
1~kG the disk was not disrupted; but for a larger field of the order of
1-10 kG, an outflow could be launched from the disk. In \citet{rukl02}
and \citet{lrl05}, much better initial equilibrium has been set than in
any previous simulations, with matter continuing to inflow through the disk
because of viscosity. It was an improvement, as it was proceeding with
the viscous time-scale, because of a slow accretion of matter. Without
such initial equilibrium, the non-stationary initial conditions determine
the flow in the disk, and influence the simulation. A star and part of
the magnetosphere corotated up to the corotation radius, and the
magnetosphere corotated with the disk farther out. The simulations were
performed in the ideal MHD regime, with effective numerical resistivity
diffusing magnetic field in the radial direction. They found funnel
flows onto the central object, spinning up or down the star, depending
on the ratio of rotation rate of the star to the rotation rate of the
disk inner rim.

In \citet{u06} and \citet{R09} the effects of
physical viscosity and resistivity on the outflow were investigated. In
\citet{R09} one of studied cases is with the magnetic Prandtl number,
the ratio between the viscosity and the resistivity $\mathrm{Pr}=\nu/\eta$,
greater than one. Magnetic field lines are bent towards the magnetosphere
in the gap and magnetic energy increases, enabling the outflow. It has been
found that, in addition to the fast and light jet, there is another, new
conical wind flowing up to 30 percent of the matter from the innermost
portion of the disk. Two different cases were considered in their
simulations, one for a fast (with the setup as in \citet{u06} and a
slightly different parameters), and the other for a slowly rotating
star. In both cases, to enable the smooth start of the simulation,
initially slow rotation of the star was gradually speeded up to its
maximum value, with matter slowly inflowing from the outer boundary,
to obtain stellar magnetic field compressed towards the magnetosphere
in the gap. In the latter case, disk was not initially present in the
computational box, but was formed from the matter inflowing from the
outer boundary. Such setup was different from most other simulations in
the literature.

With non-stationary initial conditions in the disk, accretion tends
to be too fast. Most of the mentioned works involve resistive MHD models
of accretion disks. Introduction of viscosity in the disk helps to obtain
slow accretion, with a viscous time-scale. We do not include physical
viscosity in our simulations, but we set the resistivity in the whole
computational box, not only in the disk. Resistivity, which controls the
onset of magnetic reconnection, triggers the necessary change in the
magnetic field geometry needed for any launching. In our previous work, in
\citet{cf04}, we reported on the result (with the same code) when only a
disk is present, without the magnetosphere of the central object
taken into account. Propagation of the outflow through the resistive corona
(with the disk set as a boundary condition) we investigated in
\citet{FC02}. 

One important parameter to distinguish the investigated MHD regime is
the already mentioned magnetic Prandtl number. As there is no physical
viscosity included, our resistive simulations are in the regime of
$\mathrm{Pr}\lesssim 1$. Violent initial conditions now
helped to bring magnetic flux closer
towards the star, helping the launching of matter outwards. Another
parameter whose effect we study is the density ratio between the disk
and the corona. It is usually included as a free parameter of the order
of $10^2$ or $10^3$, at best $10^5$. We investigate the influence of
this ratio on the mass and angular momentum flux in the launching of
outflows. There are other possibilities in the setup, which we did not
investigate here, e.g.\ inclusion of the stellar wind, which would
probably affect the open stellar field.

Table \ref{tabla1} lists the kinematic and thermodynamic
assumptions adopted in earlier works, each of them being usually
repeated with a variety of parameters or methods, with or without
physical resistivity and viscosity included in the code. We put our work
in context of setups and assumptions of those works, as our results
differ in the mass fluxes carried in the ejected gas.

The organization of the paper is as follows. We first describe our
implementation of the boundary and initial conditions. In \S\ref{props}
we report regimes we found under a broad range of parameters. We
investigated the influence of corona to disk density ratio, strength of
magnetic field and the physical resistivity. In \S\ref{recon} we address
the role of reconnection in the launching, in \S\ref{elsase} we check a
criterion for the site of launching, and in \S\ref{trunc} we compare
position of the disk truncation radius in our simulations with some
theoretical predictions. Then we discuss investigated parameters and the
resulting ejections.

\section{Numerical Setup for the Resistive MHD System}
\begin{figure}
\includegraphics[width=8cm]{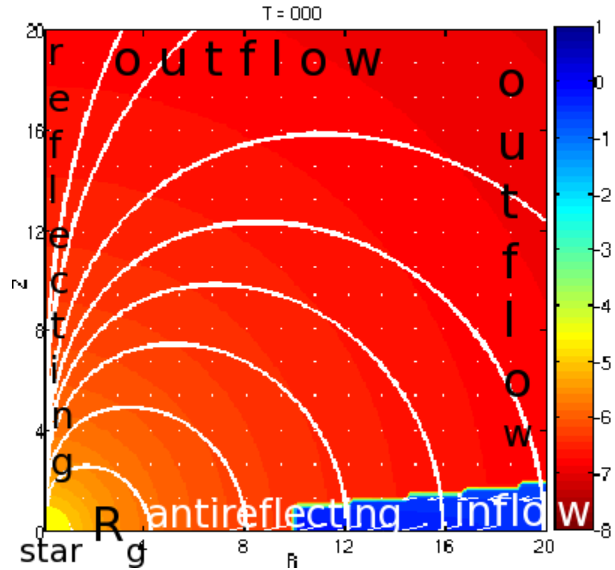}
\caption{Initial and boundary conditions in our simulations. The
initial hydrostatic density distribution in the disk corona and the disk
is plotted in logarithmic color grading. The density in the
disk is four orders of magnitude larger than in the corona. The dipole
stellar magnetic field is plotted in white solid lines, and velocity
vectors are shown in white arrows. The stellar surface is defined as a
rotating absorbing boundary layer, enclosing the origin --- see the zoom
into this region in Figure \ref{starbc}. In simulation S2, the stellar
absorbing layer extends into the small portion of the disk mid-plane
inside the disk gap, of radius $R_g$, as an outflow boundary. Along the
axis of symmetry and at the mid-plane in the disk, a reflection and
anti-reflection boundaries are imposed. An outflow boundary is imposed
on the outer boundaries of the computational domain, except for the disk
outer rim, where a small inflow into the disk is set, to mimic the
accretion flow from the portion of the disk beyond the computational box.}
\label{cbox1}
\end{figure}
\begin{figure}
\includegraphics[width=5cm]{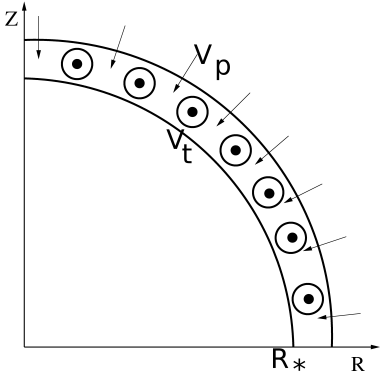}
\caption{Zoom into the setup of the stellar surface from the Figure
\ref{cbox1}. The star is set as a rotating, absorbing boundary condition
inside the computational box, enclosing the origin. Components of the
poloidal velocity $v_{\mathrm p}$ are copied from the layer immediately above
the star. The stellar rotation rate, determined by the initial toroidal
component of the velocity $v_{\mathrm t}$ at the stellar surface, is
kept constant throughout the simulation.
}
\label{starbc}
\end{figure}

We extend previous work of \citet{cf04}, who adopted a disk in the
resistive MHD regime and its halo in the ideal-MHD regime, following
\citet{CK02}. We implement an absorbing, rotating stellar surface layer
enclosing the origin, and include resistivity in the whole computational
box. It is an anomalous, turbulent resistivity parametrized by Alfv\'{e}nic
velocity as a characteristic velocity, resulting in $\eta\sim\rho^{1/3}$, as
derived in \cite{FC02}. We limited its value to few orders of magnitude
above the numerical resistivity, but leaving it large enough to facilitate
reconnection everywhere in the computational box. Our resistivity is not
limited only to the region nearby the disk, as we simulate innermost part of
the star-disk system, nearby the gap, enclosing only $0.2\times 0.2$ AU.
Hence we anticipate presence of the turbulence, and with it, resistivity,
in the whole computational box. The initial conditions of density and
magnetic field are shown in Figure \ref{cbox1}, and the setup of the
stellar surface as a boundary layer inside the computational box is shown
in Figure \ref{starbc}.

The equations of resistive MHD are solved using our version of the Zeus-3D
code\footnote{For general description and numerical methods used in Zeus
code see \citet{SN92a,SN92b}}, Zeus347 \citep{FC02}, in axisymmetry option.
They are, in the cgs system of units:
\beqa
{\frac{\partial \rho}{\partial t}} + \nabla \cdot (\rho \vec{\rm v} ) =0\\
\rho \left[ {\frac{\partial\vec{\rm v} }{\partial t}}
+ \left(\vec{\rm v} \cdot \nabla\right)\vec{\rm v} \right]
+ \nabla p +
 \rho\nabla\Phi - \frac{\vec{j} \times \vec{B}}{c} = 0 \label{mom2}\\
{\frac{\partial\vec{B} }{\partial t}}
- \nabla \times \left(\vec{\rm v} \times \vec{B}
-{\frac{4\pi}{c}} \eta\vec{j}\right)= 0 \label{farad}\\
\rho \left[ {\frac{\partial e}{\partial t}}
+ \left(\vec{\rm v} \cdot \nabla\right)e \right]
+ p(\nabla \cdot\vec{\rm v})= 0\label{ener}\\
\vec{j}=\frac{c}{4\pi}\nabla\times\vec{B} \ ,
\eeqa
where we neglected the Ohmic term in the energy equation. For a complete set
of equations, an ideal gas law is assumed. The
symbols in the equations of continuity of mass, momentum and induction
equation have their usual meaning: $\rho$ and $p$ are
the matter density and thermal pressure, $\vec{v}$ is the velocity,
$\Phi=-GM_\ast/(R^2+Z^2)^{1/2}$ is the gravitational potential of the
central object, and $\vec B$ and $\vec j$ are the magnetic field and the
electrical current, respectively. In cgs units, magnetic diffusivity is
equal to resistivity, so $\eta$ stands for the electrical resistivity.
The resistive term in the induction equation (Equation \ref{farad}) is
included in the code by subtracting $4\pi\eta\vec{j}/c$ from the
electromotive forces in the {\sc mocemfs} procedure in Zeus347 --- see
Appendix A in \citet{FC02} for tests. In the energy equation,
$e=p/(\gamma-1)$ is the internal energy per unit volume, with an adiabatic
index $\gamma=5/3$ in the initial conditions.

We solve these MHD equations in dimensionless form. The variables are
normalized to their value measured in the mid-plane of the disk, at a
fiducial radius $R_0$, which we choose to be inside the initial disk gap,
$R_0=2.86 R_\ast$, where $R_\ast$ is the stellar radius. We include radial
distances up to 20 stellar radii in our computational box. Choice of the
mass accretion rate for the disk fixes the fiducial density, from
$\dot{M}_0=R_0^2\rho_0{\rm v}_{\rm K,0}$, and from the Keplerian velocity at
$R_0$, which is ${\rm v}_{\rm K,0}=(GM_\ast/R_0)^{1/2}$, we normalize the
stellar mass and velocity unit in the code with $(GM_\ast)^{1/2}=1$. The
normalized coordinates are $R'=R/R_0$, $Z'=Z/R_0$, and $\vec{\rm
v'}=\vec{\rm v}/{\rm v}_{\rm K,0}$. The time in the code is measured in
units of the rotation rate timescale at $R_0$, $t_0= R_0/{\rm v}_{\rm K,0}$,
and the period at $R_0$ is equal to $2\pi$. The dimensionless equation of
motion can be written as: \begin{equation} \frac{\partial\vec{\rm v}'
}{\partial t'} + \left(\vec{\rm v}' \cdot \nabla'\right)\vec{\rm v}' =
\frac{\vec{j}' \times \vec{B}' }{M_{\rm A,0}^2\,\rho'} - \frac{\nabla'
p'}{\delta_0 \,\rho'} - \nabla'\Phi'\,, \label{noreqm} \end{equation} with
$\nabla'=R_0\nabla$, $t'=t/t_0$, $\rho'=\rho/\rho_0$,
$\vec{B}'=\vec{B}/\vec{B}_{\rm 0}$ and $\Phi'=-1/(R'^2+Z'^2)^{1/2}$. Primes
are omitted in equations in the rest of this paper, and quantities are
written in code units, unless otherwise specified.

We introduced the free parameters:
\beq
M_{\rm A,0}^2\equiv 4 \pi \rho_{\rm 0}{\rm v}^2_{\rm K,0} /B_{\rm 0}^2\  {\rm and}\
\delta_{\rm 0} \equiv \rho_{\rm 0} {\rm v}^2_{\rm K,0} / p_{\rm 0} \ .
\label{freep}
\eeq
The Alfv\'{e}nic Mach number, $M_{\rm A,0}$, at $R_0$ and $Z=0$, determines
the magnetic field strength. In a typical run, $M_{\rm A,0}=225$
for a dipole field of the order of 100\,G, at the surface of the star. The
kinetic to thermal energy density ratio, $\delta_{\rm 0}$, is the square of
the gas Mach number, whose fiducial value can be estimated from the
definition of the adiabatic coefficient\footnote{The sound speed is
$c_s^2=(\partial p/\partial\rho)|_S=\gamma\Re T/m$, where S
denotes the constant entropy, and $m$ the number of baryons per particle,
with inclusion of free electrons. For hot, completely ionized
hydrogen in the corona, $m=0.5$, but in a cold disk $m=1$.
$\Re$ stands for the ideal gas constant
$\Re=8.31\times 10^7$~erg~K$^{-1}$~mol$^{-1}$, from the ideal
gas law $p=\rho\Re T/m$.}. For our setup we choose $\delta_{\rm 0}=100$ or
50. The typical temperatures in the corona and the disk of the YSOs are
$10^6\,K$ and $5\times 10^3\, K$, respectively, so that at the inner edge of
the disk initially should be $c_{\mathrm s,corona}/c_{\mathrm s,disk}=20$.
Our initial disk is initially for an order of magnitude too hot, so that
this ratio is about 6 times smaller.

We use a uniform grid of $R\times Z=(90\times 90)$ cells, in the
axisymmetric cylindrical coordinates $(R,\phi,Z)$. The physical scale
corresponds to $(20\times 20)$ stellar radii. We note that our simulations
do not scale well: it was not easy to obtain results in other resolutions
or domains. We performed larger domain simulations in
$R\times Z=(40\times 40)R_\ast$ with the same resolution, and
simulations in higher resolution  with $R\times Z=(180\times 180)$ grid
cells in $(20\times 20)R_\ast$. Such simulations lasted long enough for
verification, but are more prone to numerical problems, and tend to cease
during the relaxation or soon afterwards, so that it is harder to perform
a thorough parameter study using them. This is probably because of numerical
viscosity $\nu_{\mathrm num}=v\Delta x$, which is larger with less
resolution $\Delta x$ ($v$ is the characteristic velocity), and helps the
code to go through problematic events. Inclusion of physical viscosity would
probably enable better resolution, but we focus here only on the effects of
resistivity. 

\subsection{Example of rescaling}
We give an example of rescaling for a case of an YSO with
stellar mass of $M_\ast=0.8M_\odot$ and radius $R_\ast=2R_\odot$, so that
fiducial distance is
$R_0=2.86R_\ast=5.7R_\odot=0.027$~AU. Our computational domain is then
$R\times Z\approx 0.2\times 0.2$ AU. We can rewrite the Keplerian speed at
$R_0$ in units of solar mass and radius as
\beq
{\rm v}_{\rm K,0}=\sqrt{\frac{GM_\ast}{R_{\rm 0}}}=
\sqrt{\frac{GM_\odot}{R_\odot}}\sqrt{\frac{M_\ast}{M_\odot}\big/\left(\frac{R_{\rm
0}}{R_\odot}\right)}\ , 
\eeq
giving the fiducial velocity $1.64\times 10^7$~cm~s$^{-1}=164~$km~s$^{-1}$.
The period of
Keplerian rotation at $R_0$ is $P_0=2\pi R_0/{\rm v}_{\rm K,0}=1.76$ days.
The stellar rotation rate for T-Tauri Stars is usually about $1/10$ of the
breakup rate, which we obtain from $(GM_\ast/R_\ast^3)^{1/2}=2\times
10^{-4}\mathrm{s}^{-1}=0.4$ days. This gives a typical period of rotation of
a star about 4 days. If we assume an accretion rate of
$\dot{M}_0=10^{-8}M_\odot$~yr$^{-1}$, fiducial density and pressure are
$\rho_0=2.44\times 10^{-13}$~g~cm$^{-3}$ and $p_0=109$~erg~cm$^{-3}$,
respectively. With $c_{\rm s}\sim{\rm v}_{\rm K,0}$, the reference
temperature is $T_0=m{\rm v}_{\rm K,0}^2/(\gamma\Re)\sim 10^{6}$~K.
Estimate of the initial temperature at R$_i=10R_\ast$ in the
disk gives $5.5\times 10^4$~K, and in the corona it is 5 times larger,
$2.7\times 10^6$~K. The sound speed in the disk has been estimated as
$\sim{\rm v}_{\rm K,0}$, and in the corona, in the part corotating with the
star, as $\sim 0.1R_i\Omega_\ast$. The reference value of resistivity is
$\eta_0\sim{\rm v}_{\rm K,0}R_0=10^{19}$cm$^2s^{-1}$, which is much larger
than the classical Spitzer value.

The strength of the  magnetic field we obtain from the magnetic pressure at
the mid-plane of the disk, $B_0(Z=0)=(4\pi p_0\delta_0)^{1/2}/M_{\rm A,0}$
(see Equation \ref{freep}), and the stellar dipole field at $R_0$ is
$B_0=B_\ast(R_\ast/R_0)^3$. A complete expression for the fiducial magnetic
field we can write in units of solar mass and radius as:
\beq
B_0^2=\frac{4\pi}{M_{\rm A,0}^2}
\frac{\dot{M}_{\rm a}}{M_\odot/{\mathrm yr}}\frac{M_\odot}{\mathrm yr}
\sqrt{\frac{GM_\odot}{R_\odot^5}\frac{M_\ast/M_\odot}{(R_0/R_\odot)^5}}
\label{jedna} \,.
\eeq
The factor $4\pi$ is required to obtain the Gaussian cgs value from the
implicit normalization of the magnetic field in the ZEUS code. When the
surface strength of the dipole magnetic field is combined with
Equation \ref{jedna}, it gives
\beq
B_\ast=\frac{667~{\rm Gauss}}{M_{\rm A,0}}\sqrt{\dot{M}_8}\,,
\eeq
where $\dot{M}_8$ is the disk mass accretion rate in units of
$10^{-8}M_\odot$~yr$^{-1}$. For $\dot{M}_8=100$ and
$M_{A,0}=35$, $B_\ast$ is about 200\,G.

\subsection{Boundary conditions}\label{boundc}
\begin{figure}
\includegraphics[width=7.5cm,height=7cm]{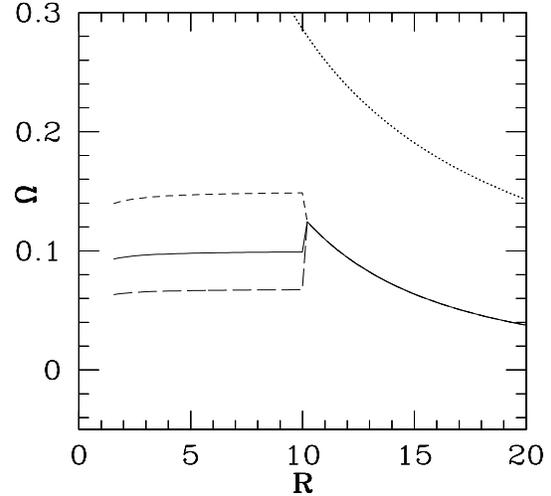}
\caption{
Initial angular velocity profiles along the disk mid-plane in our
setups. Shown are the angular velocities in simulations with the stellar
rotation rate $\Omega_\ast=$0.15, 0.1 and 0.068, in short-dashed,
solid and long-dashed line, respectively. For comparison, the
dotted line shows the Keplerian rotation profile. The kink in the
initial rotational profile is near the initial inner disk radius
$R_{\mathrm i}$.
}
\label{cbox2} 
\end{figure}
In order to mimic an absorbing stellar surface, we define the ``outflow''
boundary condition around a central object. As done by \citet{uchshib85},
we define part of the computational box surrounding the origin as a boundary
layer, for a region further above the star. We set up a rotating circular
layer of one grid cell thickness on top of the star, at a distance
$R_\ast$ from the origin, with the stellar rotation rate $\Omega_\ast$
as a free parameter -- see Figure \ref{starbc}. All the other
hydrodynamical quantities are absorbed, so that the values in this layer
are copied from the cells immediately above it. With this procedure, we
neglect the stellar wind.

The outer boundaries of our computational box are open, with the flows
extrapolated beyond the boundary. One exception is a small part at the
disk outer boundary. There, we prescribe a small mass inflow that is
consistent with the initial radial component of the velocity in the disk.
Reflection boundaries are imposed along the axis of symmetry and, in
simulation S1, at the disk mid-plane inside the disk, where the normal
component of the magnetic field is continuous, the tangential component
is reflected, and the toroidal magnetic field is anti-reflected.
Under axisymmetry for the disk mid-plane and the axis,
$B_R(R=0,Z)=B_R(R,Z=0)=0$, and with these conditions
$\nabla\cdot\vec{B}=0$ is satisfied. We use the Constrained
Transport (CT) method of \citet{ehaw88} to ensure that it is preserved
to a machine round-off precision in computations.

In simulation S2, we treat a small part of the disk mid-plane inside
the disk gap as an ``outflow'' boundary, extrapolating the flow from the
active zones into the ghost zones, effectively extending the stellar
absorbing layer into the disk gap. All the values in the ghost zones are set
equal to the values in the corresponding active zones, and the angular
velocity and toroidal component of the magnetic field are projected. This
ensures preservation of the disk gap in the simulation, even when the
magnetic field is not strong enough to truncate the disk. This means that
some other physical effects, which we do not include in our simulations,
would have to act in terminating the disk. It could be that material which
actually resides in the disk gap is not well described by our
approximations, or that influence of additional forces, as the radiation
pressure force from the central object, which we neglect
here, is large. Using such a setup, we can study if a weaker stellar dipole
can launch matter from the innermost magnetosphere. Caveat is that
the final disk truncation radius in simulation S2 is then dependent on
the boundary conditions, and is not self-consistently computed.

\subsection{Initial conditions}
We set up initial conditions for the density distribution, velocity
profiles, magnetic field and resistivity in the computational domain as
follows. Gravitationally bound disk rotates slightly sub-Keplerian. For a
self-consistent accretion disk, additional constraints such as constant
fluxes through surfaces at different radii and stability to various modes
of oscillation should be included. However, the disk stability or the
accretion process itself is not the subject of study here, and we treat the
disk only as a supply of matter into the stellar magnetosphere.

\subsubsection{Density distribution}
The initial disk density distribution is
\beqa
 \rho_{\rm d}(R,Z)=\frac{R_{\rm off}^{3/2}}{(R_{\rm off}^2+R^2)^{3/4}}
\times \nonumber \\
\left( \max \{ 10^{-6}\ ,\ \left[ 1-\frac{(\gamma-1)Z^2}{2H^2} \right] \}
\right) ^{1/(\gamma-1)}\,,
\eeqa
shown in Figure \ref{cbox1}. The density is limited by the maximum
function, and ensured to be regular by a constant offset radius
$R_{\rm off}=4$. The disk is adiabatic with an index $\gamma=5/3$, and
physically thin, with an aspect ratio of $H/R=0.1$, where $H$ is the
disk height at a given radius $R$. For the initial inner disk radius we
tried various initial positions of the initial inner disk radius
$R_{\rm i}$ in our parameter study; here it is chosen to be at half size
of the computational domain, $R_{\rm i}=10 R_\ast$. It is not a critical
parameter, as the disk will adjust its inner rim position during the
simulation, but too close positioned $R_{\rm i}$ can, especially in a case
of strong magnetic field, result in a too violent initial relaxation, which
will stop a simulation.

The corona above the star and in the disk gap corotate with the central
object, and further away, with the underlying disk. The corona is in
hydrostatic balance, with an initial coronal density\footnote{For such
setup it is essential to set a force-free initial magnetic field in the
computational box---see e.g.\ \citet{fe99,fe00}.}:
\beq
\rho_{\rm c}(R,Z) = \frac{(R^2+Z^2)^{-3/4}}{\kappa}\,,
\eeq
which is obtained from the equality of gravitational and hydrostatic
pressure. Such a corona, when rotating, is not in an equilibrium, and the
different rotation rates of the corona in the inner portion of the
computational box and above the disk is making it even further from
equilibrium. The free parameter $\kappa=\rho_{\rm d}/\rho_{\rm c}$ determines
density in the corona. In similar studies $\kappa$ is usually in the
range of $10^{2}$ to $10^{4}$. In simulations without magnetic field and S1a
we used $\kappa=10^4$, and in S1b $\kappa=10^5$ (see Table \ref{tabla2}).
We address the influence of this parameter on the launching process in the
resistive simulations, and investigate the range
in $\kappa$ from $10^{2}$ to $10^{6}$ for simulation S2.

\subsubsection{Velocity profiles}
In our simulations, the initial stellar rotation rate is a free parameter,
kept constant throughout the simulation. Since the
time scale of change in stellar rotation is much longer than duration of
simulations here, this constraint should not influence the outcome. In the
case of T-Tauri type stars, there are observational indications that stellar
rotation rate is actually constant \citep{irw07}, so that for those objects
it is a plausible assumption even for very long lasting simulations.

The corotation radius, at which matter in the disk is rotating with the
angular velocity of the stellar surface, is:
\beq
R_{\mathrm cor}=\left(\frac{GM_\ast}{\Omega_\ast^2}\right)^{1/3}\ .
\label{rcorr}
\eeq
The position of the corotation radius with respect to the disk truncation
radius $R_{\mathrm t}$ defines two regimes:
$R_{\mathrm cor}>R_{\mathrm t}$, and $R_{\mathrm cor}\le R_{\mathrm t}$.
\citet{u06} investigated the latter as a ``fast rotating'' (or
``propeller'') regime, and in \citet{R09} the former, ``slow rotating'' regime was
investigated. In this work we present results for a parameter study in a
slow rotating regime, with stellar angular velocity of 0.15, which gives the
rotation period of 11.8 days. The corresponding corotation radius is
$10.1 R_\ast$.

For the disk, we adopt the following rotation profile:
\beq
{\rm v}_\phi (R,Z)=(1-\epsilon^2)\frac{R_{\rm off}^{1/2}}{ (R_{\rm off}^2+R^2)^{1/4}}\exp
\left( -2\frac{Z^2}{H^2}\right)
\label{vphi}
\eeq
The free parameter $\epsilon$ gives the departure from the Keplerian
rotation profile, and is chosen to be 0.1 in our typical simulations.
For $\epsilon=0$ the disk would go back to the Keplerian profile. Figure
\ref{cbox2} shows the initial angular velocity profiles at the equatorial
plane of the disk in our runs. In most of the simulations shown in
Table \ref{tabla1}, the initial disk profile is sub-Keplerian in a
similar fashion, to ensure the disk equilibrium. Slightly sub- or
super-Keplerian setup, tuned with
the factor $(1\pm\epsilon^2)$ and a parameter $\epsilon$, compensates for
the pressure force, which arises when the disk is of non-negligible height. 
At $R>R_{\mathrm cor}$, the disk corona initially corotate with the disk,
with ${\rm v}_\phi$ from Equation \ref{vphi}, without the exponential part.

Both components of the initial disk poloidal velocity are given by the
requirement which has been derived for radially self-similar stationary
solution of accreting disk. The radial component scales the same way as
${\mathrm v}_\phi$:
\beqa
 {\rm v}_R(R,Z)=-m_{\rm s}\epsilon\frac{R_{\rm off}^{1/2}}{(R_{\rm off}^2+R^2)^{1/4}}\exp\left(
-2\frac{Z^2}{H^2}\right) \ ,\\ {\rm v}_Z(R,Z)={\rm v}_R(R,Z)\frac{Z}{R} \ .
\eeqa
The Z-component of the velocity is non-zero only in the disk and sets
the initial poloidal velocity in the wedge-shaped thin disk. The constant
parameter $m_{\mathrm s}<1$ is used to obtain a subsonic
inflow, and is chosen to be 0.1 in our simulations here. We also
performed runs with $m_{\mathrm s}=0.3$ and 0.6, which give larger influxes
of mass into the disk, with similar results --- for more massive disk,
simulations are more prone to instabilities and tend to cease earlier than
for lighter disk. The exponential factor in the equation effectively confines
the initial disk profile.

\subsubsection{Magnetic field}
The initial magnetic field is a pure stellar dipole, and we computed it
from the derivatives $B_{\mathrm R}=-\partial A_\phi/\partial Z$ and
$B_{\mathrm Z}=\partial({\mathrm RA}_\phi)/{\mathrm R}\partial{R}$ of a
magnetic potential:
\beq
A_\phi=\frac{\mu_\ast R}{(R^2+Z^2)^{3/2}} \ .
\label{dipopen}
\eeq
The stellar dipole magnetic moment $\mu_\ast$ we set to unity. Setup with
multipole expansion of magnetic field is feasible in
our simulations, but without stellar wind included we already neglect
effects at the surface of the star. We assume that dipole is leading term in
the disk gap.

\subsubsection{Resistivity and artificial viscosity}
\begin{figure}
\includegraphics[width=8.7cm]{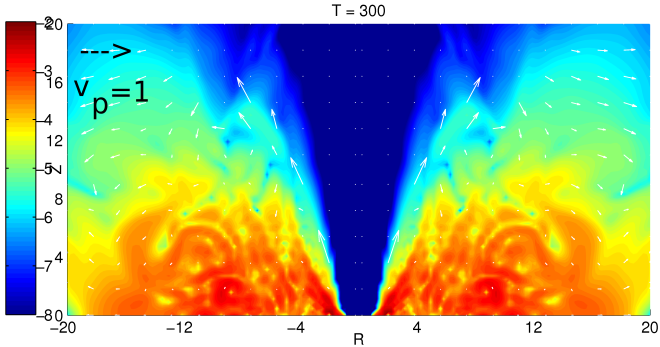}
\caption{The mass flux $\rho {\rm v}$ at T=300, in logarithmic color scale
grading in units of $\dot{M}_0$, in a simulation without magnetic field.
There is no significant
mass or angular momentum flow after the relaxation phase. Vectors show
the velocity of matter, with unit velocity in code units indicated in
black arrow. The disk reaches the stellar surface, where matter is
accreted onto the star.
}
\label{rhovb0}
\end{figure}
\begin{table}
\caption{Parameters used in our setups presented here. The initial inner
disk radius is located at R$_{\mathrm i}=10 R_\ast$. The stellar rotation
rate is $\Omega_\ast=0.15$, so that the corotation radius
$R_c=3.54R_0=10.13R_\ast$. Shown are values for $B_\ast$ in the case of
disk accretion rate $10^{-6}M_\odot$~yr$^{-1}$.} 
\begin{tabular}{|c||c|c|c|c|}
\hline
\ \  & S0 & S1a & S1b & S2 \\
\hline
\hline
\ \ $\kappa$ & $10^4$ & $10^4$ & $10^5$ & $10^4$ \\
\hline
\ \ M$_{\mathrm A,0}$ & $\infty$ & 274 & 35 & 223 \\
\hline
\ \ $\delta_0$ & 100 & 500 & 50 & 100 \\
\hline
\ \ $B_\ast$ & 0 & 24 & 190 & 30 \\
\hline
\end{tabular}
\label{tabla2}
\end{table}
\begin{figure*}
\includegraphics[width=8.5cm]{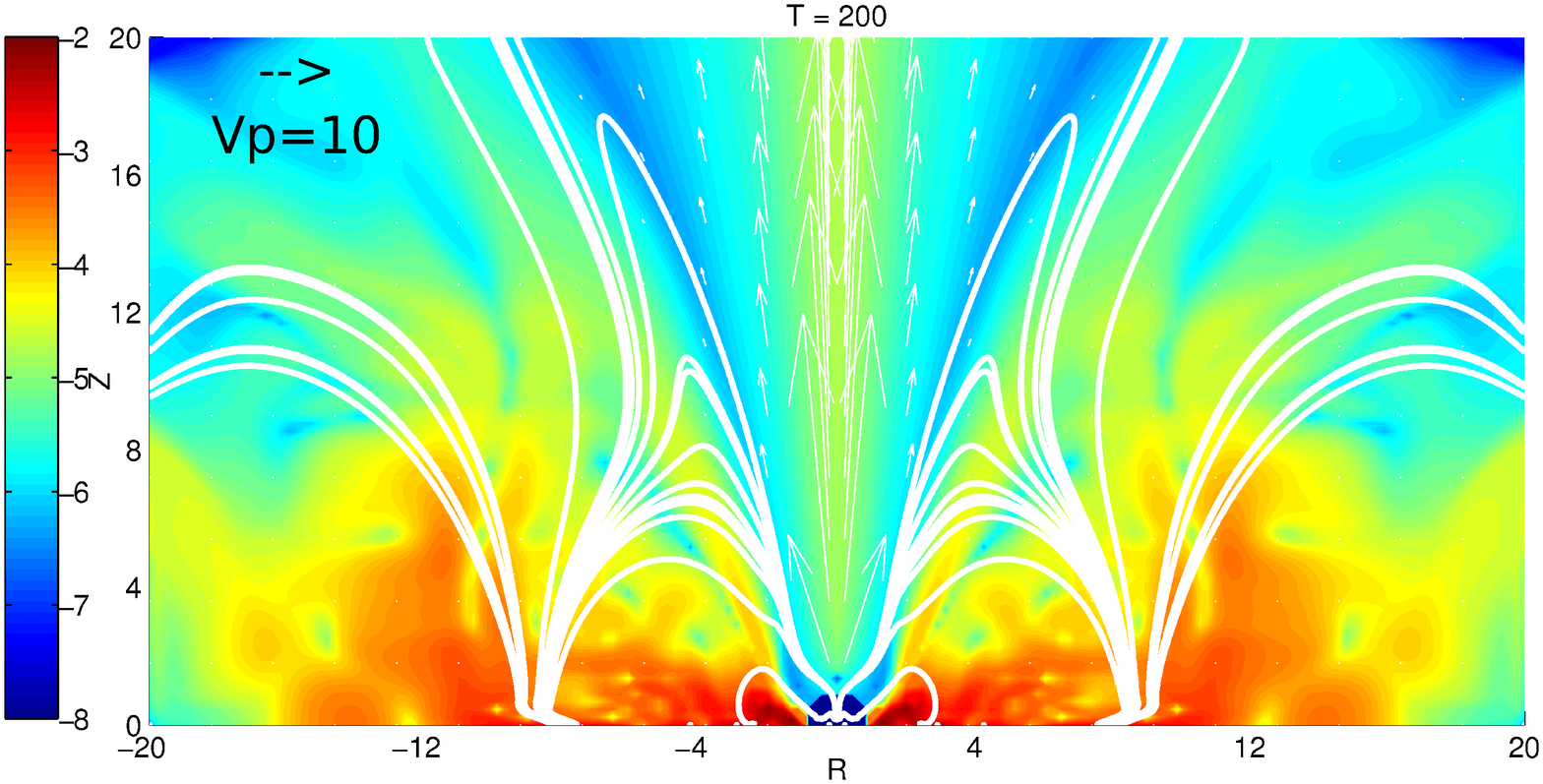}
\includegraphics[width=8.5cm]{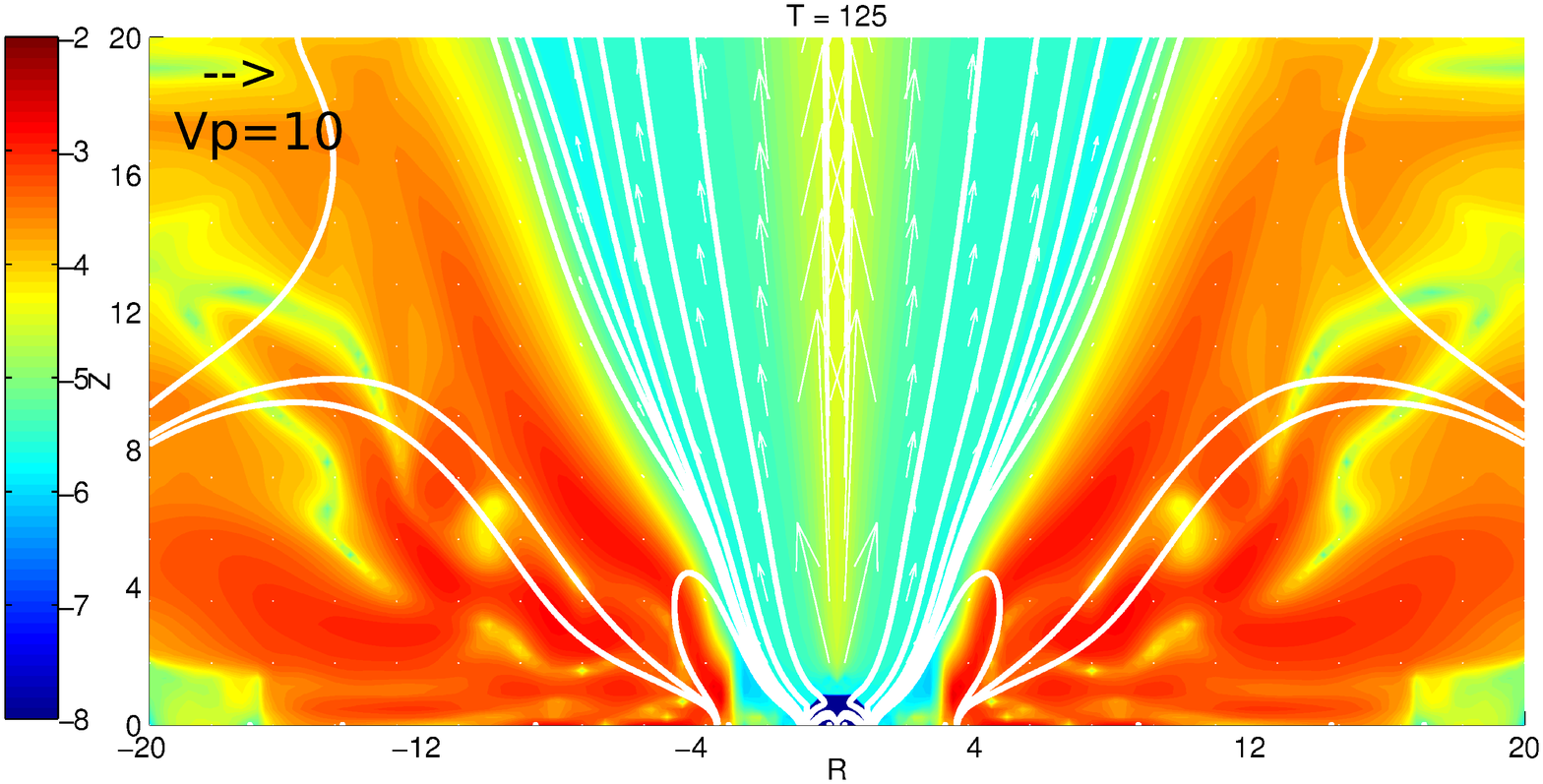}
\caption{The mass flux $\rho {\rm v}$ for the case with medium magnetic
field in the simulation S1a is shown in the {\it Left panel}, and in the
simulation S1b, with larger magnetic field, in the {\it Right panel}.
Flow inside R$<0.5$ from the axis is very fast and light, and is
artificial, of numerical origin. Mass flux is shown in a logarithmic
color grading in units of $\dot{M}_0$, vectors show the velocity of
matter, with ${\rm v}_{\rm p}=10$ in code units indicated in black
arrow. The white
solid line show the poloidal magnetic field lines. In both plots magnetic
field lines nearby the axis are removed from the shown sample, not to obscure
the underlying mass flux distribution. The disk is truncated at the radius
where its ram pressure and magnetic pressure are balanced. In the case of
simulation S1a this line is close to the stellar surface, so that in our
resolution the disk almost touches the star. In simulation S1b the gap is
well defined.
}
\label{S1abplots}
\end{figure*}
The electrical resistivity $\eta$ is defined through the electric
conductivity $\sigma$ as $\eta=c^2/4\pi\sigma$, where $c$ is the speed of
light. The ratio of the advection and diffusion terms in the induction
equation (\ref{farad}) is the magnetic Reynolds number Rm=VL/$\eta$, where V
and L stand for the characteristic speed and distance, respectively. The
characteristic speed in our problem is the Alfv\'{e}n speed
${\mathrm v_A}={\mathrm B}/(4\pi\rho)^{1/2}$, which defines the Lundquist
number\footnote{In some works, magnetic diffusivity is parametrized
through the coefficient
$\alpha_{\mathrm m}=S_{\mathrm L}^{-1}$.} S$_L$.:
\beq
S_L=\frac{{\mathrm v_{\mathrm A}} L}{\eta}\, .
\eeq

To explain the physical processes, the accretion disk requires an enhanced,
{\em anomalous} level of resistivity, which is much larger than the
classical value. The anomalous resistivity could be an effect of MHD
turbulence or ambipolar diffusion in a partially ionized
medium\footnote{For extensive discussion of physical conductivity in
partially ionized disks see e.g.\ \citet{wng}, \citet{sal07}.}. We set
the initial constant resistivity of the disk to be of the same order of
magnitude as the numerical resistivity
$\tilde{\eta}=\Delta x^2/\Delta t\sim 10^{-4}$, which gives $S_{\mathrm L}\sim
10^{4}$. We find the numerical resistivity by lowering the disk constant
resistivity in the code until it does not affect the results.

We omit the actual Ohmic term in the calculation of the MHD equations.
When the Ohmic part is included in the internal energy equation in
Zeus347, Equation \ref{ener} gains an additional Ohmic heating term $-\eta
\vec{j}^2$. The inclusion of this term is expensive computationally. However, 
the actual difference from the solution without the Ohmic term is
negligible, as the $p\nabla$v term is much larger. Similar results have been
reported in \citet{ms97} and \citet{R09}. It would take the Ohmic term many
orders of magnitude larger to produce a visible effect.

A resistive corona is essential for the magnetic reconnection to occur, and
reconnection is crucial for re-organizing the magnetic field. The
resistivity is modeled as a function of matter density, following
\citet{FC02}, so that $\eta\propto v_{\rm A}R\sim\rho^{(\gamma-1)/2}$. It
comes from taking the turbulent velocity $v_T$ as a characteristic velocity in
parametrization of resistivity, so that the resistivity is $\eta\sim v_T
L$. Turbulent beta plasma is then $\beta_T=(c_s/v_T)^2$, which gives
$v_T=(\gamma/\rho\beta_T)^{1/2}$, and the normalized
turbulent velocity is then proportional to $\rho'^{(\gamma-1)/2}$. For the
adiabatic case, $\gamma=5/3$ and $v_T\propto\rho'^{1/3}$. Inserted
back to the relation for $\eta$, this gives
\beq
\eta=\eta_0\ \rho^{1/3}\,.
\label{resis}
\eeq
To avoid unrealistically large $\eta$ and too small Ohmic timesteps
$\tau_\eta=min(L^2/\eta)$ in the densest part of the domain, we limit this
value to the order of unity, with $\eta_0=3.0$. Another restriction we
imposed on the change of resistivity in each point of the computational box
is that we do not allow it to be too steep. If in two succeeding instants
of time the ratio of the new to the old resistivity is more than order of
magnitude, we keep the old value of resistivity. It accounts for difference
in the timescales for change of density and resistivity.

Another diffusive process in our simulations is the numerical viscosity.
In finite differencing numerical scheme, it is of the same order as
numerical resistivity. We do not treat the physical viscosity, only the
von Neumann-Richtmyer artificial viscosity is included, through a
constant parameter that controls the number of zones through which shocks are
smoothed out. Such viscous term exists only in presence of shocks. For a
smooth flow, it is tiny, and for rarefactions, it is zero. The
characteristic speed for viscous effects is the sound speed $c_{\mathrm s}$.

In Table \ref{tabla2} we give parameters used for initial conditions in
each of our presented simulations.

\section{Results of simulations}\label{results}
\begin{figure}
\includegraphics[width=4.2cm,height=4.cm]{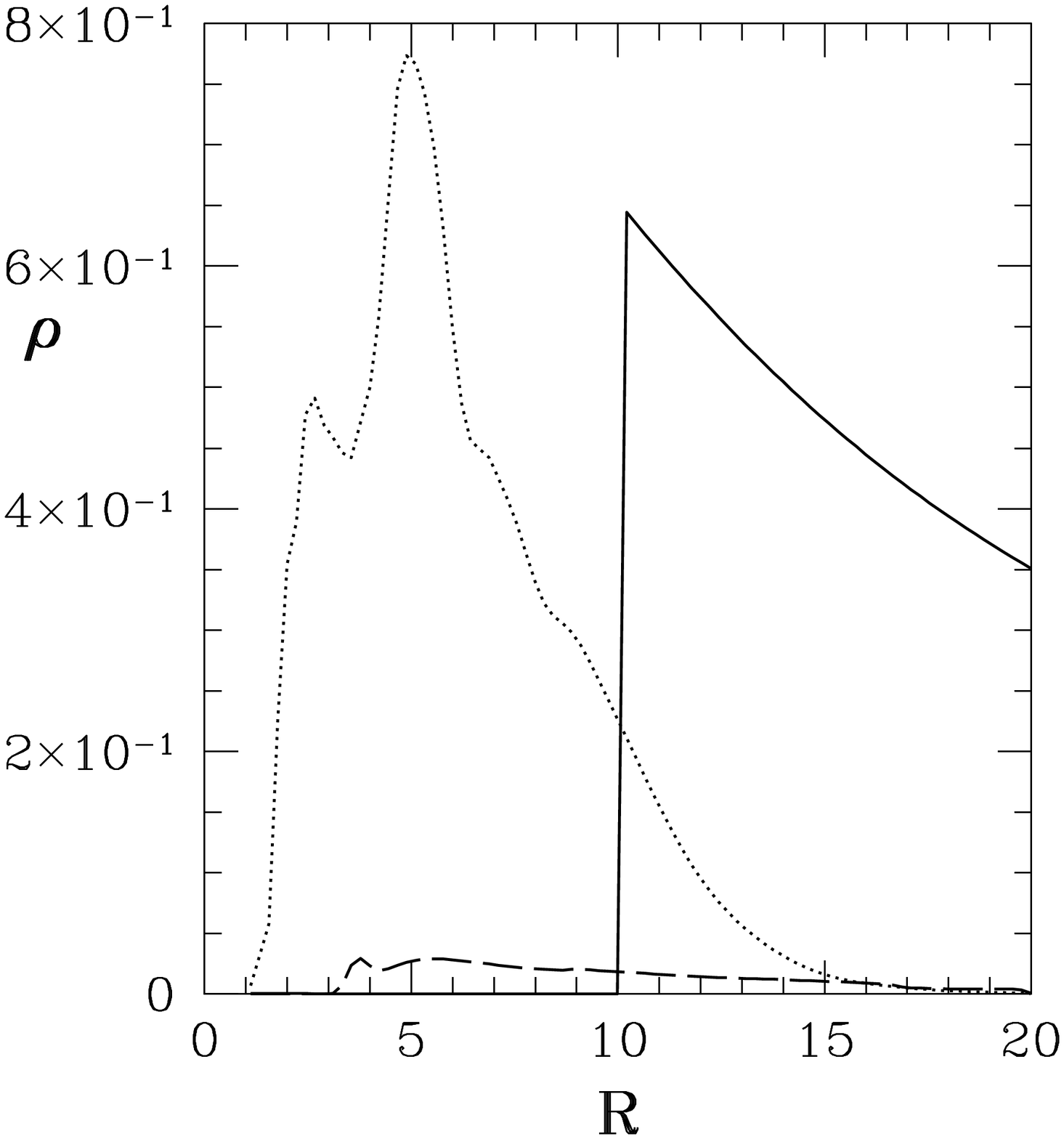}
\includegraphics[width=4.2cm,height=4.cm]{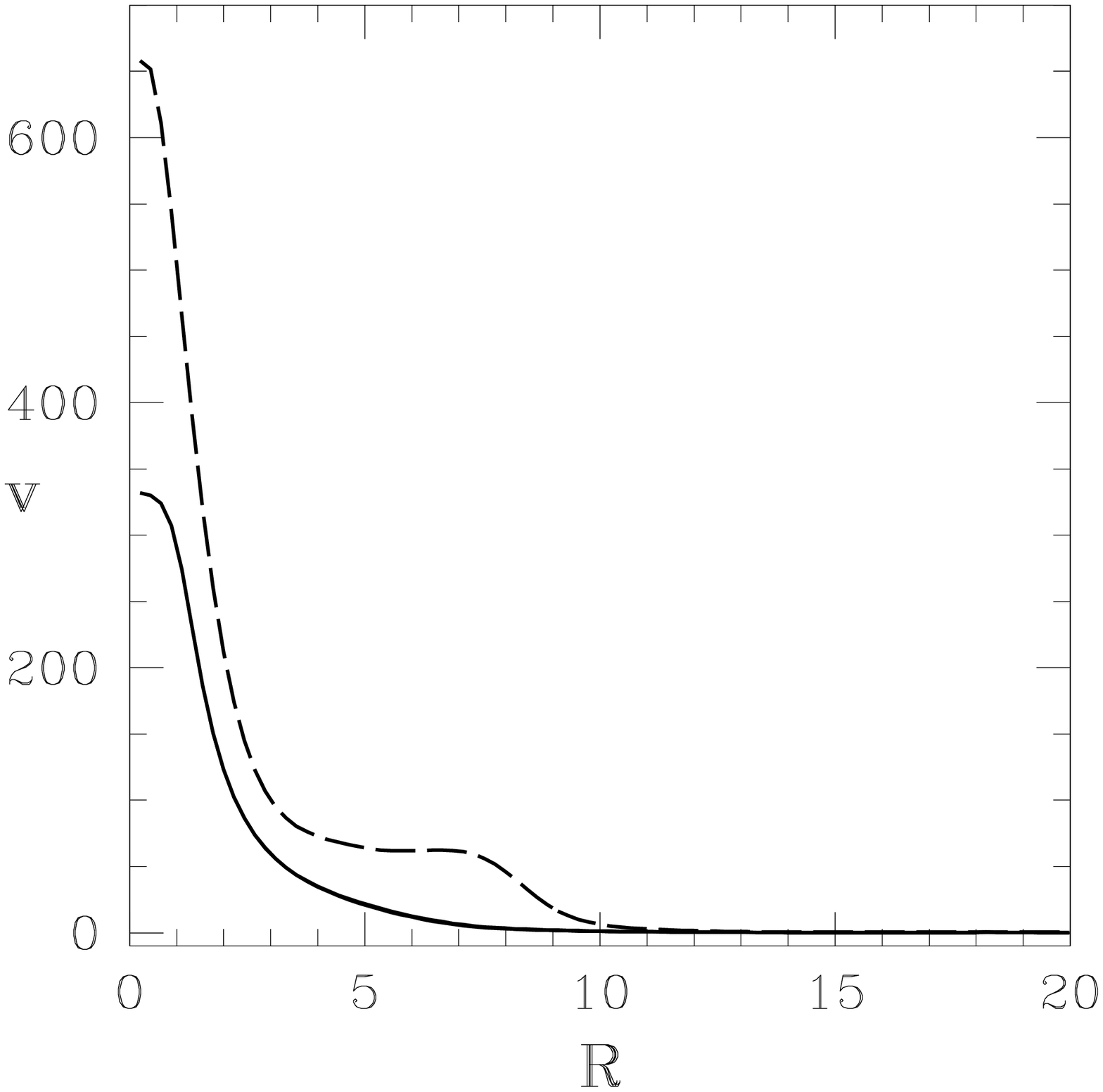}
\caption{In the {\it Left Panel} is shown the density along the disk
equatorial plane in simulations S1a and S1b at T=200 and T=125 in dotted
and dashed lines, respectively. The initial density profile is shown in
solid line. In the {\it Right panel} are shown the velocity profiles along the
outer Z-boundary in the simulations S1a and S1b at T=200 and T=125, in the
solid and dashed lines, respectively, in units of v$_{\mathrm
K,0}=164~$km~s$^-1$. In the same simulations,
Alfv\'{e}n velocity is of the order of 10 v$_{\mathrm K,0}$, escape velocity
is of the order of 0.1 v$_{\mathrm K,0}$, and sound speed is of the order of
v$_{\mathrm K,0}$.
}
\label{velS1ab}
\end{figure}
\begin{figure}
\includegraphics[width=4.2cm,height=4.cm]{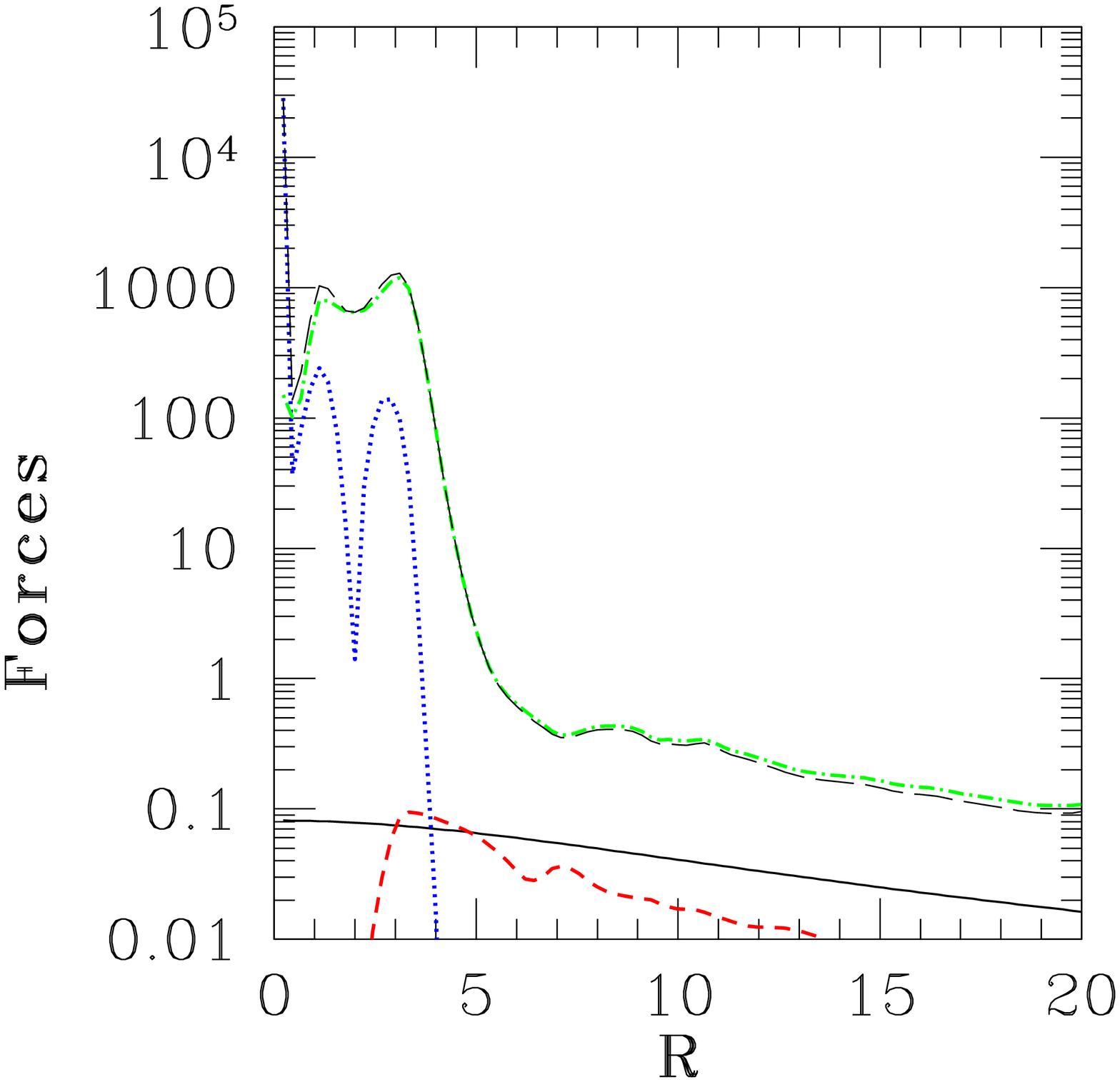}
\includegraphics[width=4.2cm,height=4.cm]{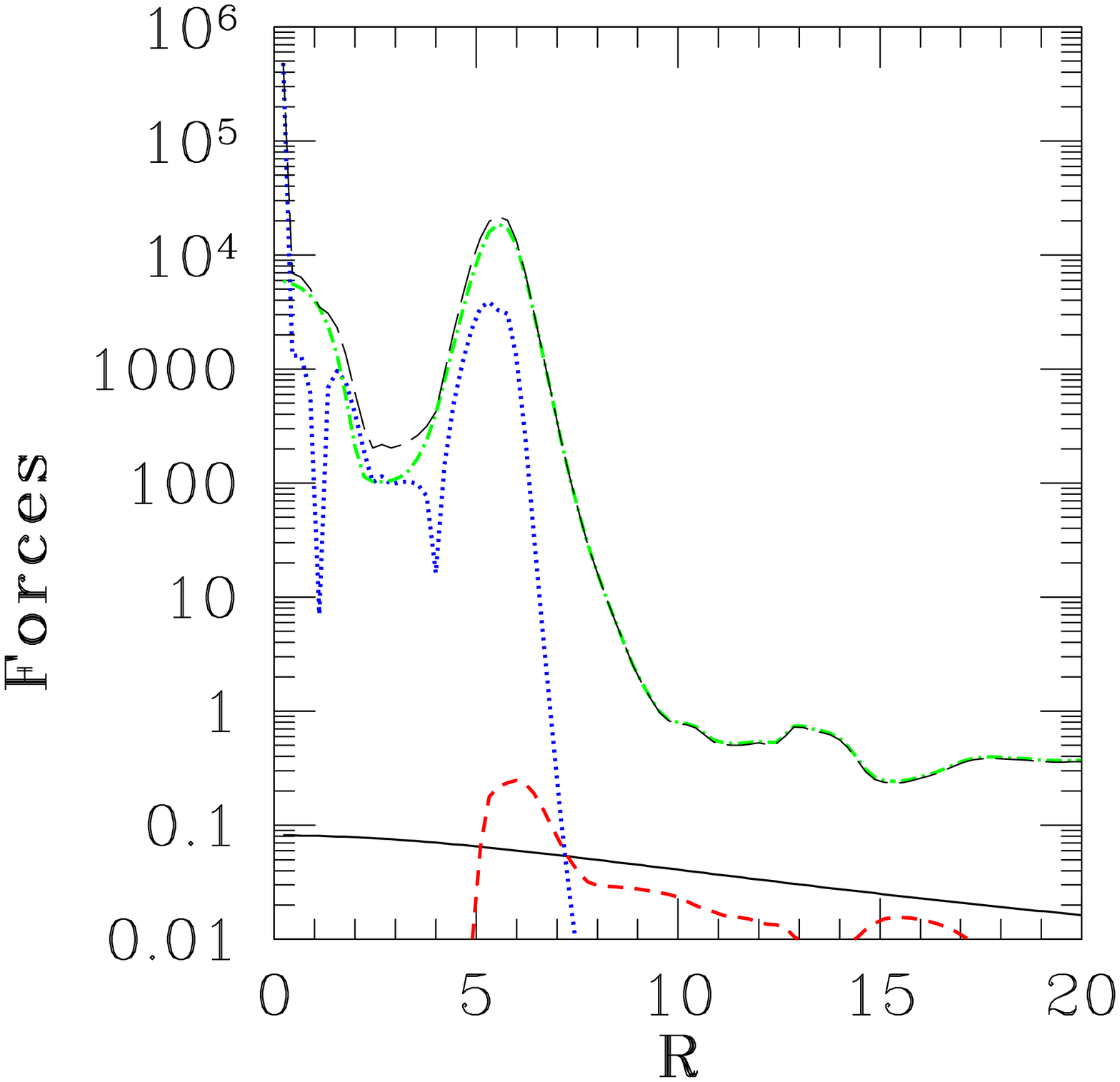}
\caption{In the {\it Left} and {\it Right} panels we show forces along
the slice parallel to the disk equator, in simulations S1a and S1b,
respectively, in a logarithmic scale. Slice is taken at half of the
computational box, at Z=10; similar result is obtained all along the
Z-axis. In blue dotted line is shown magnetic force, and in solid,
green dot-dashed, red short-dashed and black
long-dashed lines are shown absolute value of the gravitational force,
pressure gradient, centrifugal and total forces, respectively. Ejection
along the axis is artificial, produced by the steep gradient in magnetic
pressure nearby the axis. At larger radial distance,
realistic, physical ejections occur, and they are mainly launched by
pressure gradient and magnetic forces.
}
\label{forceS1abinR}
\end{figure}
\begin{figure}
\includegraphics[width=4.2cm,height=4.cm]{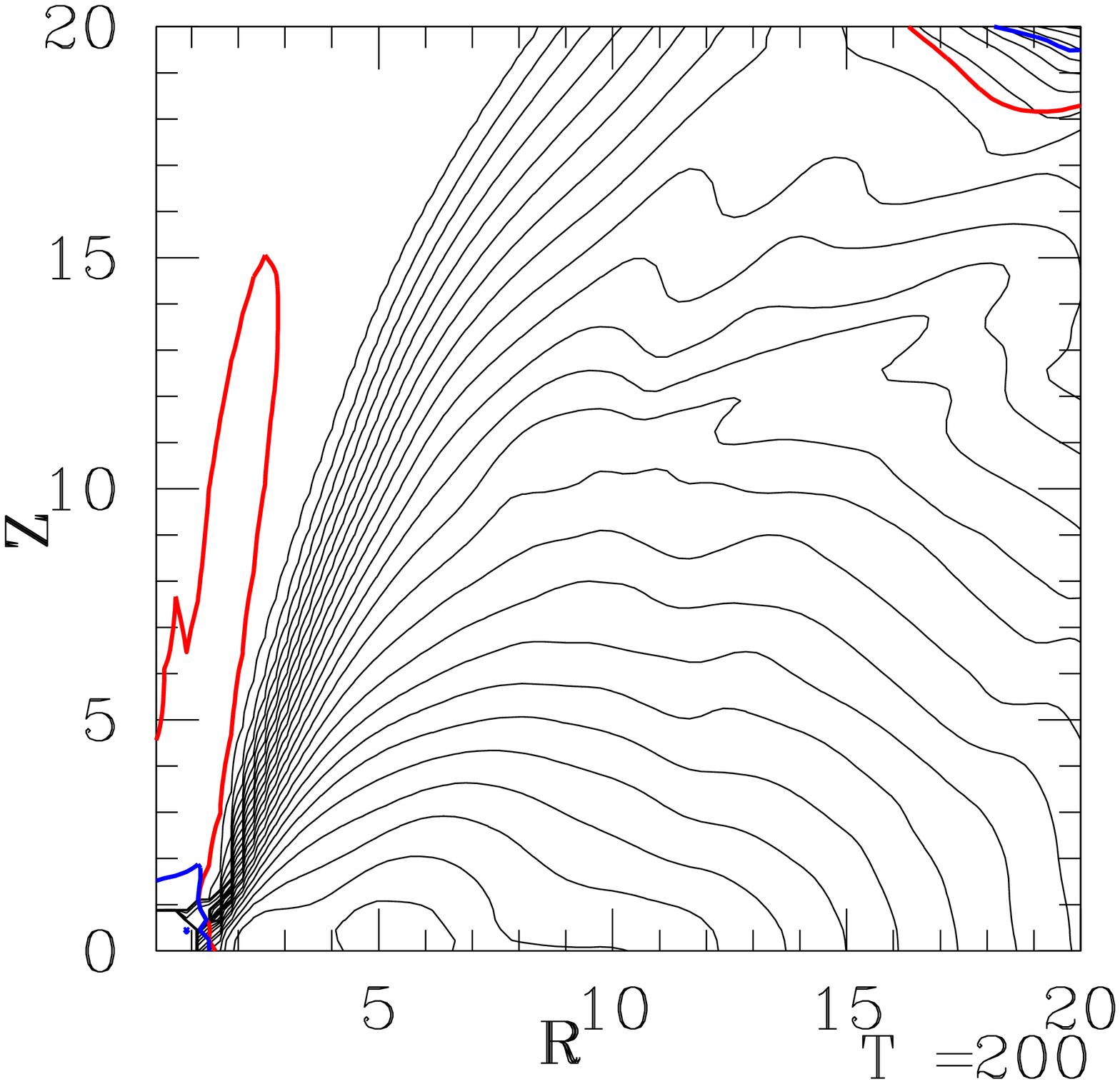}
\includegraphics[width=4.2cm,height=4.cm]{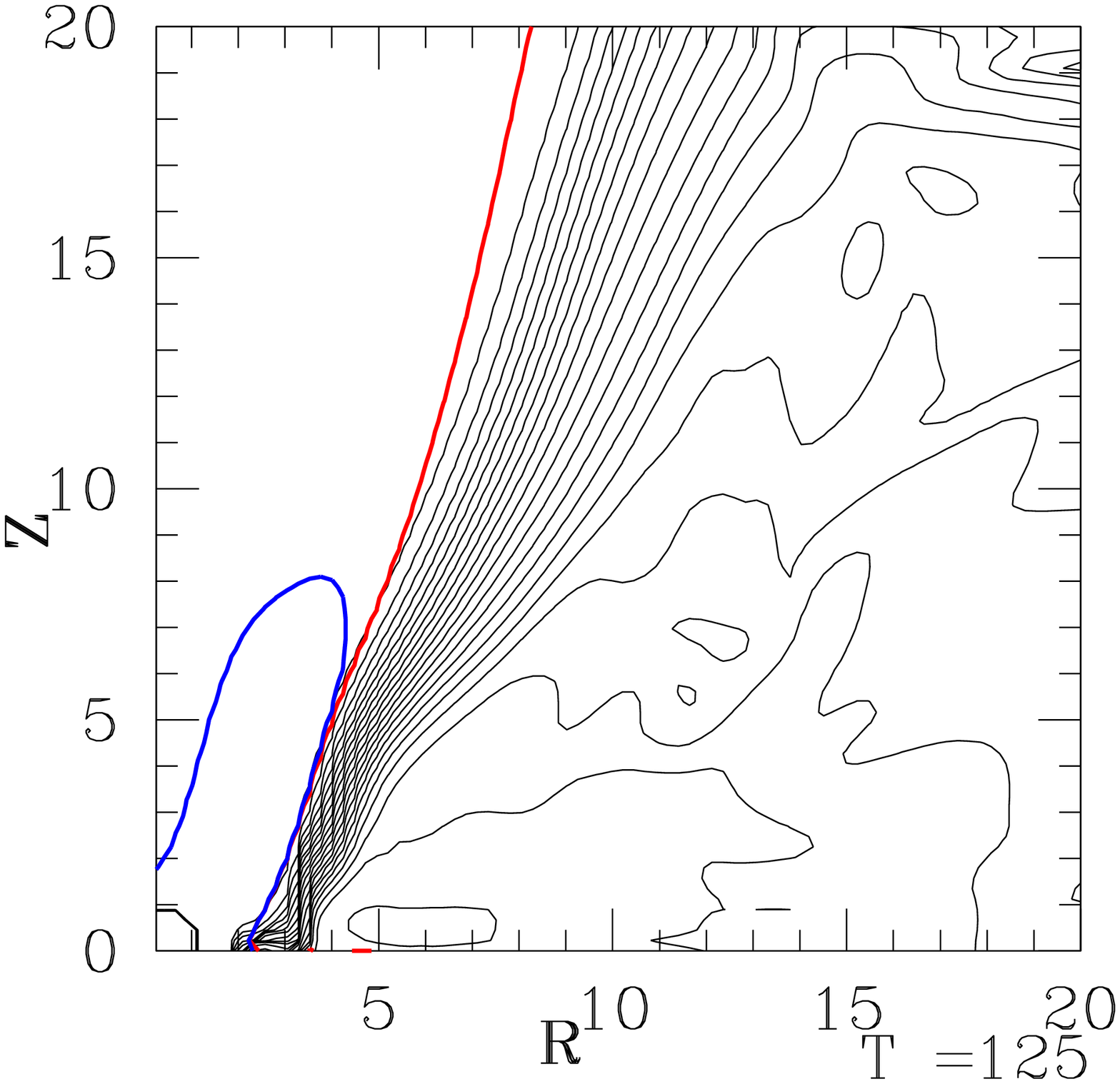}
\caption{Density isocontours in simulations S1a and S1b are shown in black
solid lines in the {\it Left } and {\it Right} panels, respectively. In the
left panel 25 isocontour lines are spanning logarithmically from $0.23\rho_0$ to
$8.3\times 10^{-7}\rho_0$, along a line parallel to the axis of symmetry at
R=10 from the disk equator to outer-Z boundary. In the right panel isocontour
lines span the same way from $0.02\rho_0$ to $1.2\times 10^{-6}\rho_0$. Lines
where $\beta$ and $\beta_1$ equal unity are shown in thick red and blue solid
lines, respectively.
}
\label{betaS1ab}
\end{figure}
We started with a disk in hydrostatic balance and, as a reference,
performed a hydrodynamic simulation without magnetic field. The
stellar rotation rate in the case shown in Figure \ref{rhovb0} was set
to $\Omega_\ast=0.15$, but we tried smaller and larger values, with
the similar outcomes. After relaxation, the disk remained
stable for hundreds of revolutions, in a quasi-stationary state, connected
to the stellar equator, with the matter from the disk very slowly falling onto
the star. The disk got puffed up, similar to the situation in
\citet{cf04}, where there was no central object in the simulation, only
the disk. The reason for increase in the disk height is an increase in the
pressure gradient force in the axial direction throughout the disk. It is an
outcome of a slight heating-up of the disk in our simulation with
$\gamma=5/3$. In simulation with smaller $\gamma<4/3$, disk puffs-up less.
However, for reasons of easier comparison with other works, we decided to
keep $\gamma=5/3$ in our simulations here.

\subsection{Simulations with very small and with small magnetic
field}\label{res31}
We seek to understand the effects of magnetic diffusion on the launching of
matter from the innermost vicinity of an YSO. In our code, numerical
resistivity and numerical viscosity are of the similar order of magnitude.
By including physical resistivity, but not physical viscosity, we probe the
portion of parameter space with $Pr_{\mathrm m}<1$. The stellar rotation
rate remains $\Omega_\ast=0.15$, the same as in the case without the
magnetic field.

At very small pure stellar dipole magnetic field, of the
order of 0.1 G, we notice that very fast flow occurs above the star close
to the axis, carrying very small mass flux. Similar solutions are obtained
until we reach the stellar dipole magnetic field of the order of 10 Gauss,
for an YSO disk accretion rate of $10^{-6}M_\odot$~yr$^{-1}$. At this
field strength, in addition to the flow above the star close to the axis,
a slower ejection of matter at larger angle from the axis, occurs. 

We check in more detail this solution with medium magnetic field, of
the order of 10 Gauss, which we name simulation S1a. The field is not
strong enough to completely terminate the disk, which is still falling
radially onto the star with a very small velocity of
0.1v$_{\mathrm K,0}\sim 16.4$~km~s$^{-1}$. It is
possible that our resolution is insufficient for establishing the inner
disk radius, as the line of balance of gas and magnetic pressure is very
close to the stellar surface. The final state in simulation S1a is shown in
Figure \ref{S1abplots}. In the flow very close to the axis of symmetry,
velocity is reaching the order of few hundreds of v$_{\mathrm K,0}$, and
at around 1/4 of the R$_{\mathrm max}$ from the axis, velocity reaches the
order of few tens of $v_{\mathrm K,0}$, (see Figure \ref{velS1ab}), which
gives $1.6\times 10^3 km~s^{-1}$ in the YSO case). Other characteristic
velocities, like sound speed and Alfv\'{e}n velocity, are all much smaller
than the Z-component of velocity, which is the main contributor to total
velocity shown by arrows in Figure \ref{S1abplots}.

\subsection{Artificial axial flow}
Inspection of forces in the radial direction, shown in Figure
\ref{forceS1abinR}, shows that axial ejection is of numerical origin,
produced by an unrealistically large magnetic force. This force is
produced by the large gradient of magnetic pressure along the axis, which
decreases inversely proportional to the radial distance $R$. That the
axial ejection is artificial, follows also from the observation that it
appears in runs with very small magnetic field. It is not the case with
the ejection at larger angle: it appears only when magnetic field is
above some critical value, which is of the order of 10 Gauss. Further
from the axis, magnetic force contributes to the launching, but is
always smaller than the pressure gradient force.

We try to prevent formation of this artificial axial flow choosing
different density floors, but they hamper evolution of the
system through the violent relaxation, and simulations stop or become
unphysical -- with density governed by the density floor. Another method
would be to add outflowing gas from the stellar pole, but it is overcome
by the strong artificial magnetic field, and axial flow anyway appears.
Since this flow is easily discernible from other flows as being
generated by numerically obtained magnetic force, we decided to leave
it, as a numerical effect in the simulations.

\subsection{Simulations with larger magnetic field}\label{res32}
\begin{figure}
\includegraphics[width=4.2cm,height=4.cm]{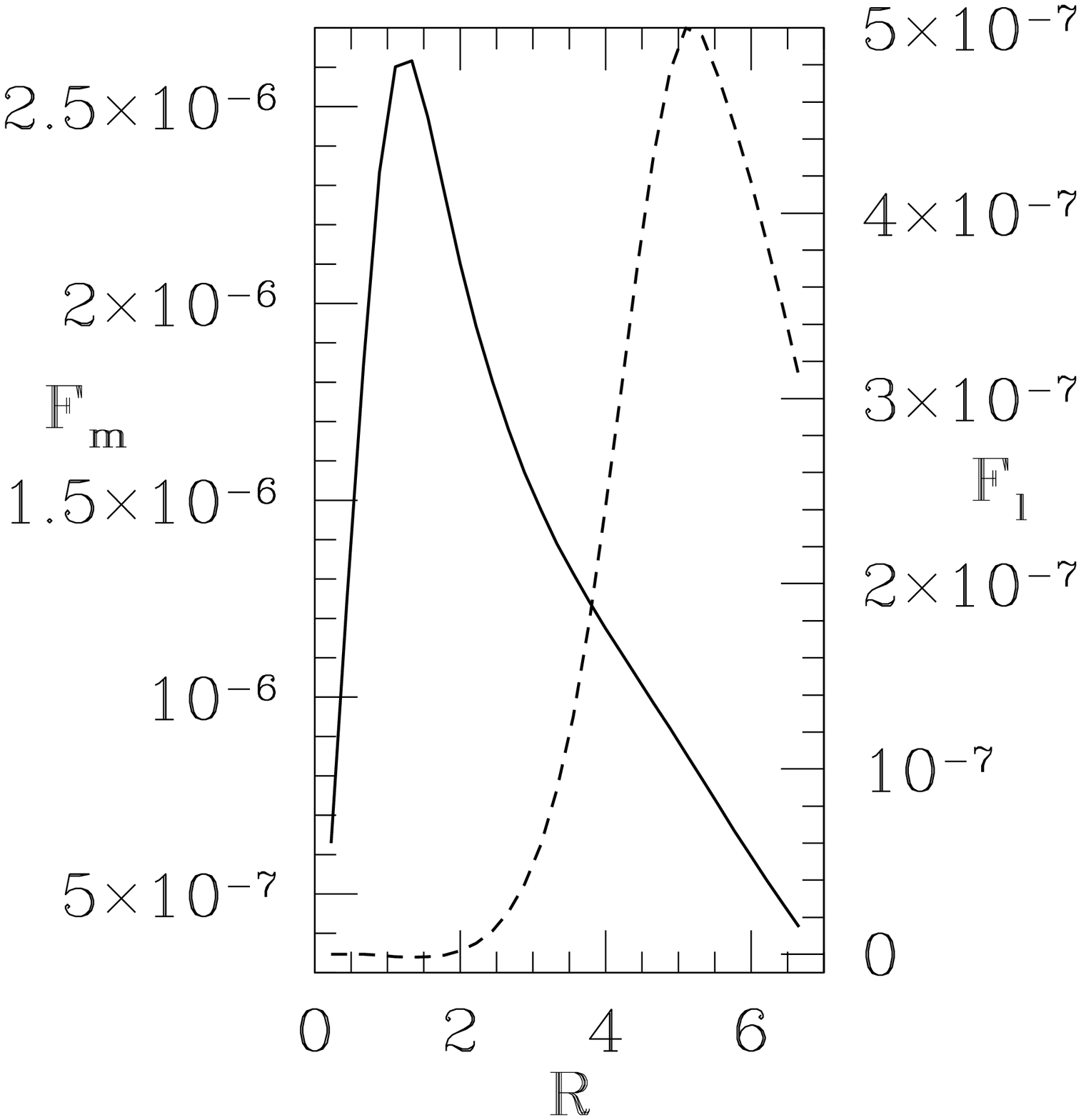}
\includegraphics[width=4.2cm,height=4.cm]{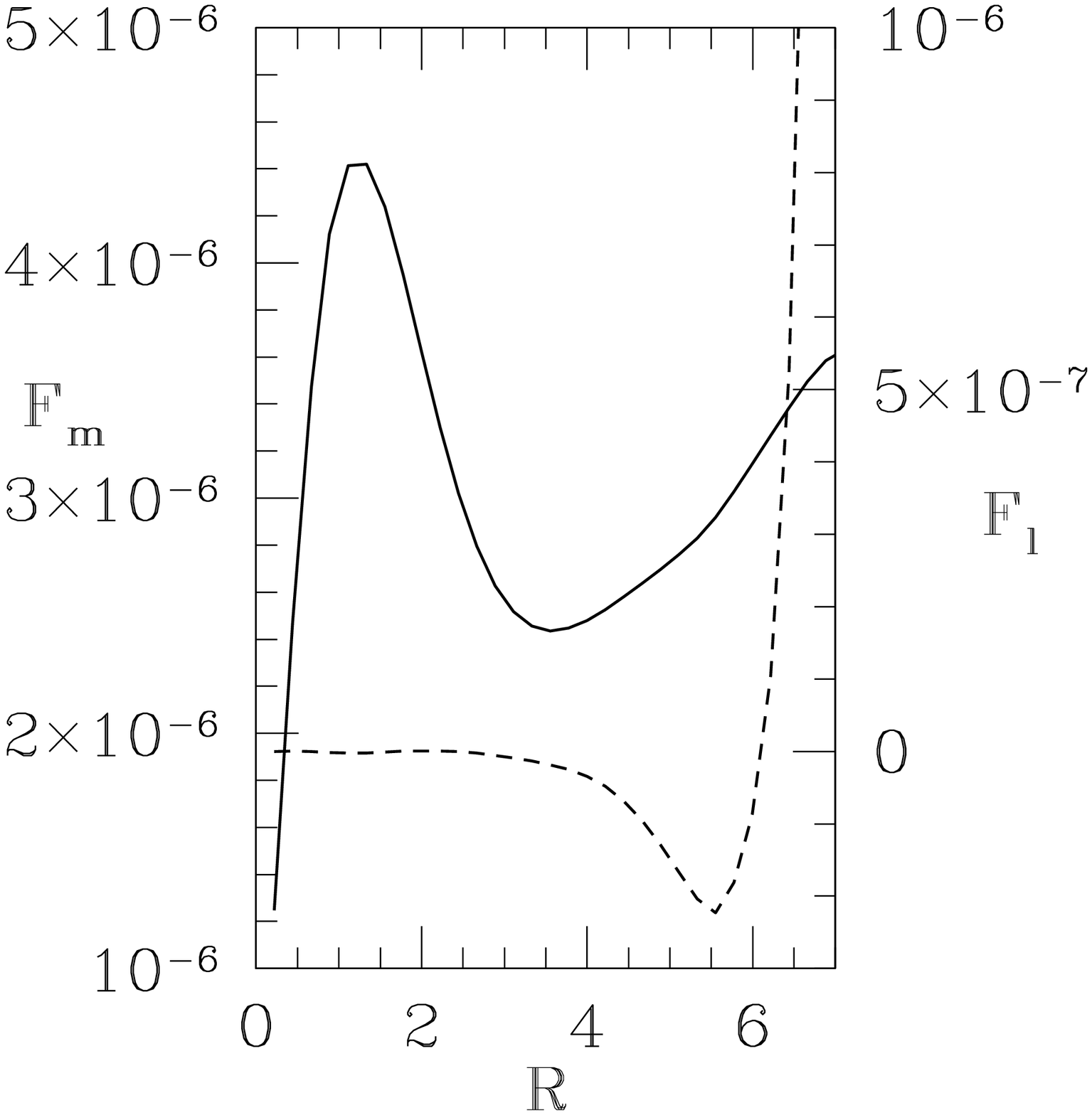}
\caption{In solid line are shown mass fluxes F$_{\mathrm m}$, and in
dashed line the angular momentum fluxes F$_{\mathrm l}$ in the simulation
S1a ({\it Left panel}) and S1b ({\it Right panel}), at T=200 and T=125,
respectively. Slices are taken along the outer Z-boundary. Fluxes are
calculated for only the 1/3 of R$_{\mathrm max}$ part of the box, where
ejection occurs. Note the different scales at the left and right side of
plots.
}
\label{mlflS1ab}
\end{figure}
\begin{figure}
\includegraphics[width=4.2cm,height=4.cm]{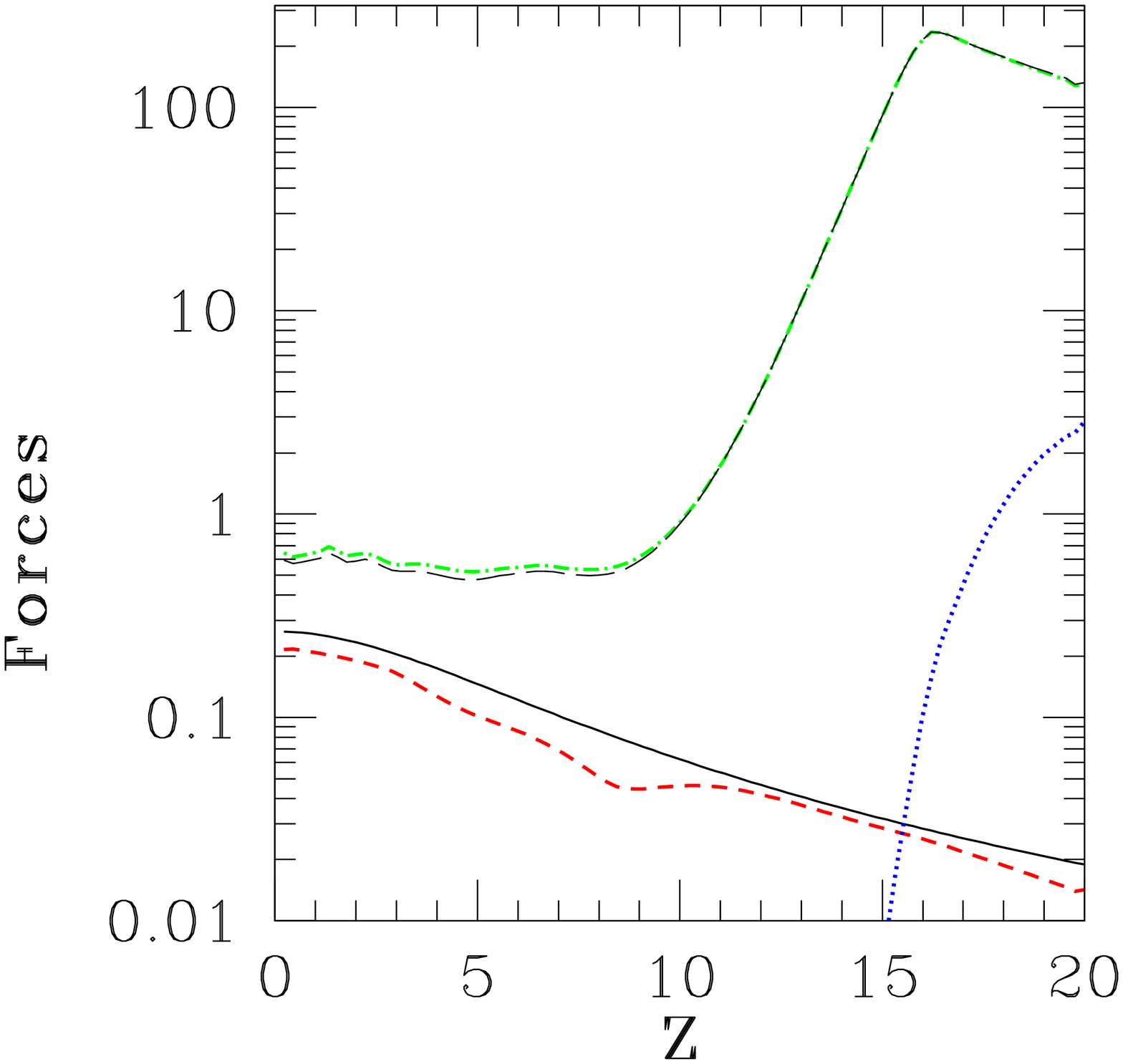}
\includegraphics[width=4.2cm,height=4.cm]{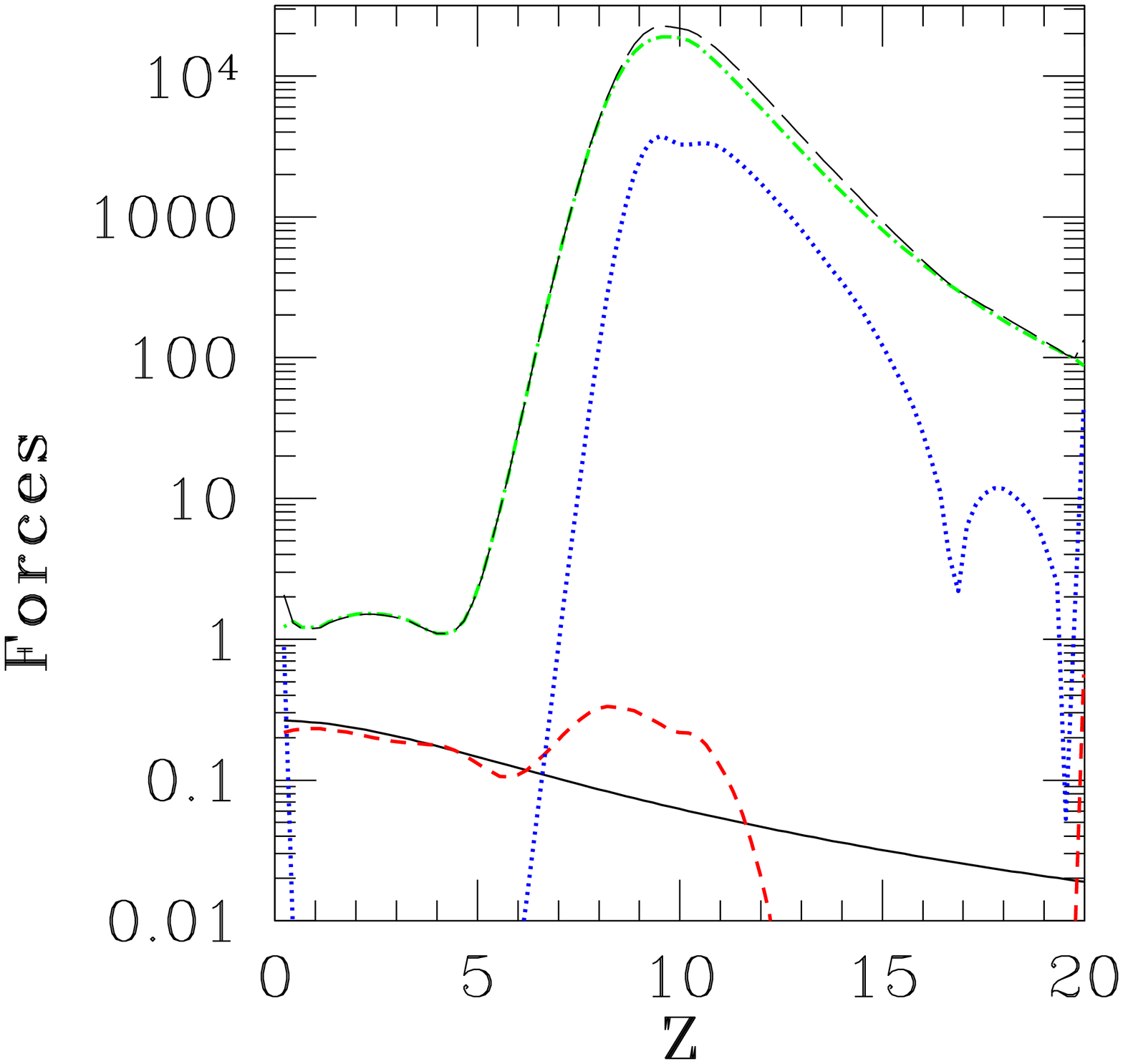}
\caption{Forces in simulation S1a at T=200 ({\it Left panel}) and S1b at
T=125 ({\it Right panel}), in a logarithmic scale. In black solid, green
dot-dashed, red short-dashed, blue dotted and black long-dashed lines
are shown absolute value of the gravitational force, pressure gradient,
centrifugal, magnetic and total forces, respectively. Forces are taken
along slices parallel to the axis of symmetry, at R=5. 
}
\label{forsS1ab}
\end{figure}

With increase in magnetic field for an order of magnitude in comparison
to the field in simulation S1a, the disk terminates at a line where the
magnetic and ram pressure are equal (Figure \ref{betaS1ab}). At this line
the plasma $\beta$ parameter, which is the ratio of the gas and magnetic
pressure $\beta=8\pi p/B^2$, equals unity. In the same Figure we also
draw the line where the parameter $\beta_1=8\pi(p^2+\rho v^2)/B^2$
equals unity. The flow is magnetically dominated when one of those
parameters becomes less than unity. Now there
is no radial infall onto the star any more. We name this simulation S1b.
In Figure \ref{velS1ab} we show the velocities in the launched matter.
The part of the outflow nearby the axis is a numerical effect, as in
simulation S1a, produced solely by the magnetic force, shown in Figure
\ref{forceS1abinR}. The slower flow, launched by the combination of pressure
gradient and magnetic forces, is now launched at a larger angle than in
simulation S1a.

We calculate the mass flux F$_{\mathrm m}$ and the angular momentum flux
F$_{\ell}$ in each half-plane, above or below the disk equator. They are
defined as:
\beqa
F_{\mathrm m}=\int_0^R 2\pi\rho {\mathrm v}_Z R dR\ ,\nonumber\\
\quad F_\ell=\int_0^R 2\pi \left(\rho{\mathrm v}_Z
{\mathrm v}_\phi R-\frac{B_Z B_\phi R}{4\pi}\right)dR\ .
\eeqa

In Figure \ref{mlflS1ab} is shown radial distribution of flux in
simulations S1a and S1b in the first 1/3 of $R_{\mathrm max}$. Both
fluxes are very small, with the mass flux of the order of
$10^{-6}\dot{M}_0$, for a few orders of magnitude smaller than the
estimate in the optical jets from observations, usually inferred to be
10\% of the disk accretion rate. Mass and angular momentum fluxes from the
artificial flow closest to
the axis (for $R<0.5$) contribute very little to the total fluxes, as
they are even an order of magnitude smaller. From Figure \ref{mlflS1ab}
we read that maximum of ejection in simulation S1a occurs under an angle
of $\Theta=3^\circ$ from the axis, and drops steeply to half of the peak
value until $\Theta=9^\circ$ from the axis. In the simulation S1b,
matter is launched into a wider angle, from $\Theta=3^\circ$ to
$\Theta=20^\circ$.

Which force is launching the matter above the disk gap? In Figure
\ref{forsS1ab} we show forces along the ejecta. The force exerted
because of the pressure gradient is contributing significantly
to the total force. Hence, fast, but very light ejection of
matter in our simulation is driven by a combination of pressure
gradient and magnetic forces. Ejection is
directed outwards from the star-disk system, in presence of ongoing
magnetic reconnection along the boundary between the stellar and disk
components of the magnetic field. Matter from the corona is launched
along this boundary layer, so that reconnection is
pushing the ejecta along the current sheet which is established at the
boundary. Our disk is puffed-up; it is probable that
for thinner disk in simulations the angle of ejecta would be larger,
as then magnetic field higher above the gap would be less pressed
towards the axis by the disk. The ejected matter, which originates from
the corona, and not directly from the disk, is very fast and light,
as shown in Figures \ref{velS1ab} and \ref{mlflS1ab}. This is the
main difference of our result with results from the Table \ref{tabla1}.

Because of role of reconnection in launching, our results are similar
to situation in solar context, when micro-flares are launched into
a solar corona, only that in our case matter is launched from the   
magnetosphere between the star and the disk. To distinguish our ejections
from other cases of more massive outflows in the star-disk system,
we name them {\em micro-ejections}, as they are tiny in mass flux,
and localized nearby the boundary between the stellar and the disk magnetic
field. Position of our micro-ejections seems to suggest that magnetic
field producing them could have similar geometry as a helmet streamer
in X-wind model \citep{shuetal94}.

Varying the accretion rate and strength of the magnetic field does not
change the nature of our result: micro-ejections remain light and fast.

What is the reason for launching under a wider angle? Is it because of
larger magnetic field, or conditions near the disk gap? Forces along
the flow in Figure \ref{forsS1ab} show that the reason for launching is
the same, only with much larger pressure gradient force in simulation
S1b. Mass and angular momentum fluxes remain very small, but above
the value which could be set by the density floor, set in our
simulations to $5\times 10^{-8}$ in code units. The axial region
remains almost evacuated because of the fast ejections, which have
a tiny mass and angular momentum flux.

In a setup as described, simulations with larger magnetic field, of the
order of kG, tend to stop during, or not long after, the relaxation,
because of numerical problems. To study the quasi-stationary state, in
which small fluxes we obtain could evolve further (and eventually increase
or disappear), simulations lasting for hundreds of rotations are needed,
with a stable disk gap and realistic magnetic field strength.

\subsection{Long-lasting simulations with fixed disk gap}
\begin{figure}
\includegraphics[width=4.1cm,height=3.9cm]{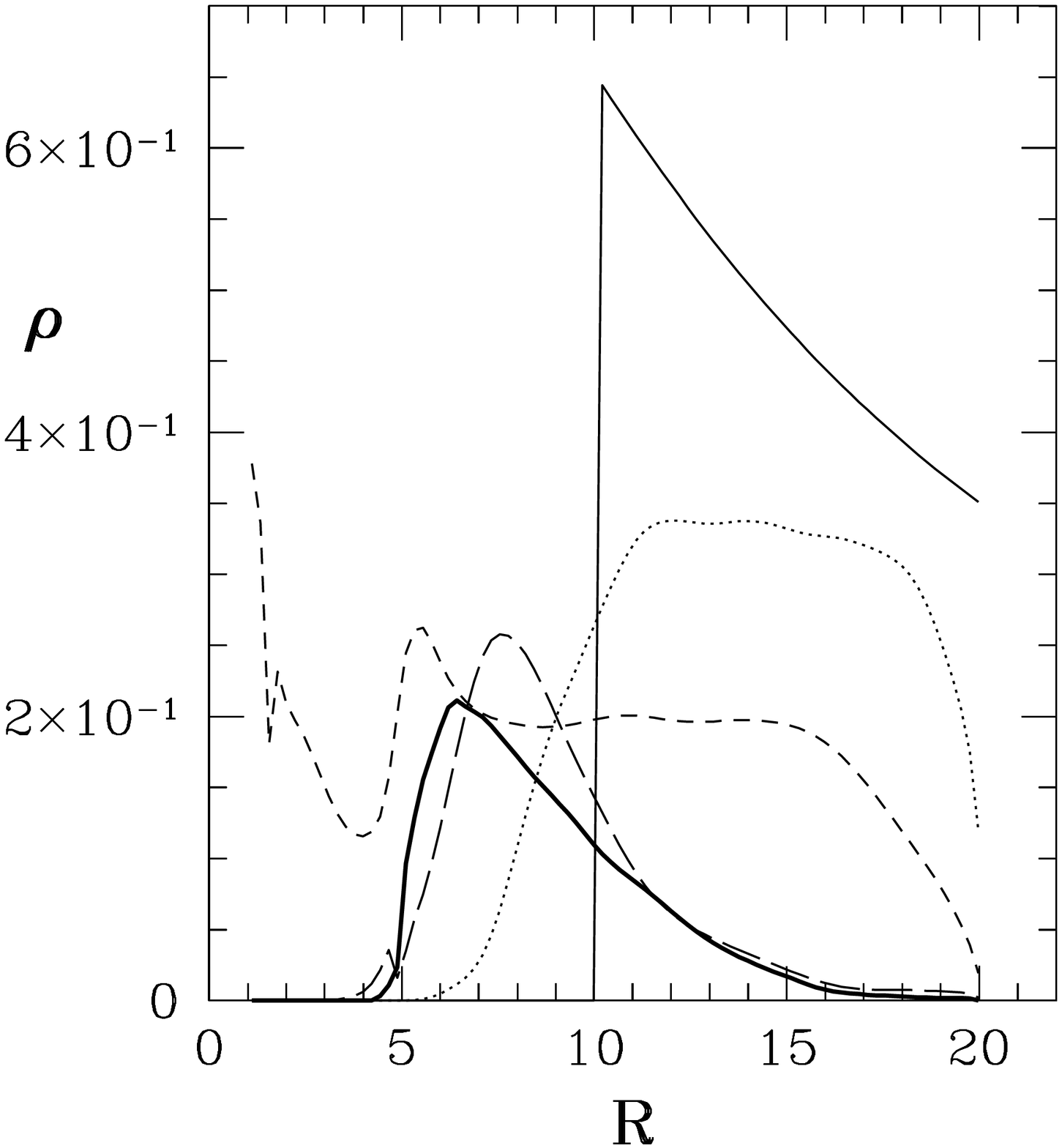}
\includegraphics[width=4.3cm,height=4.1cm]{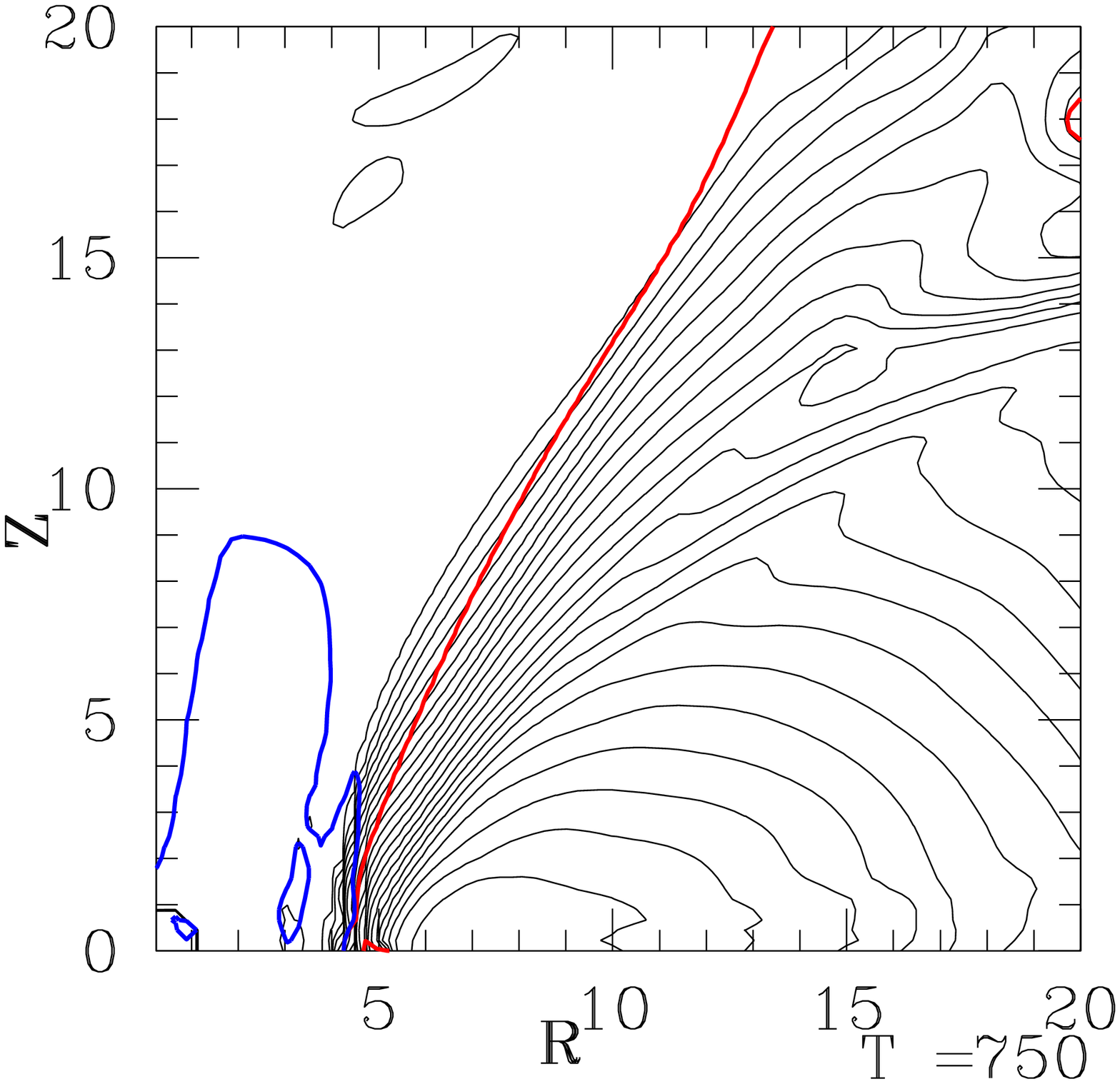}

\caption{In the {\it Left panel} is shown change in time of the disk
density profile in simulation S2 along the equatorial plane. We show
densities at T=0,5,10,100,850 in thin solid, dotted, dashed, long
dashed and thick solid
line, respectively. The disk density is substantially modified only during
the relaxation, afterwards it does not change much. In the {\it Right panel}
are shown density isocontours at T=750 in black solid lines, in 25
logarithmically spanning lines along the line parallel to the axis of
symmetry, ranging from $0.11\rho_0$ at R=10 in the equatorial plane of the
disk, towards $5\times 10^{-8}\rho_0$ at the outer-Z boundary. Positions
where $\beta$ and $\beta_1$ are unity shown in thick red and blue solid
lines, respectively.
}
\label{S2rhoint}
\end{figure}
\begin{figure}
\includegraphics[width=4.2cm,height=4.cm]{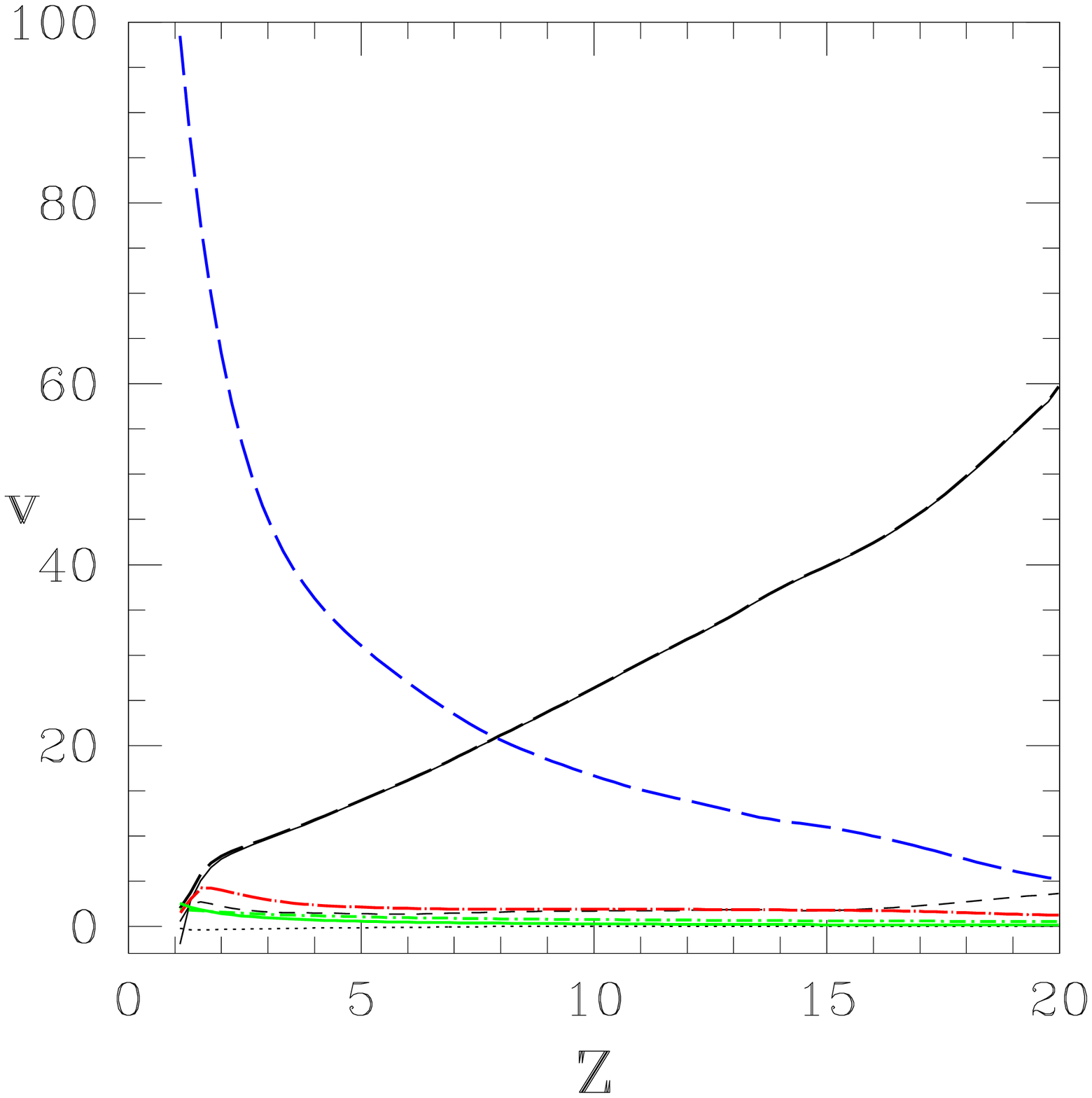}
\includegraphics[width=4.2cm,height=4.cm]{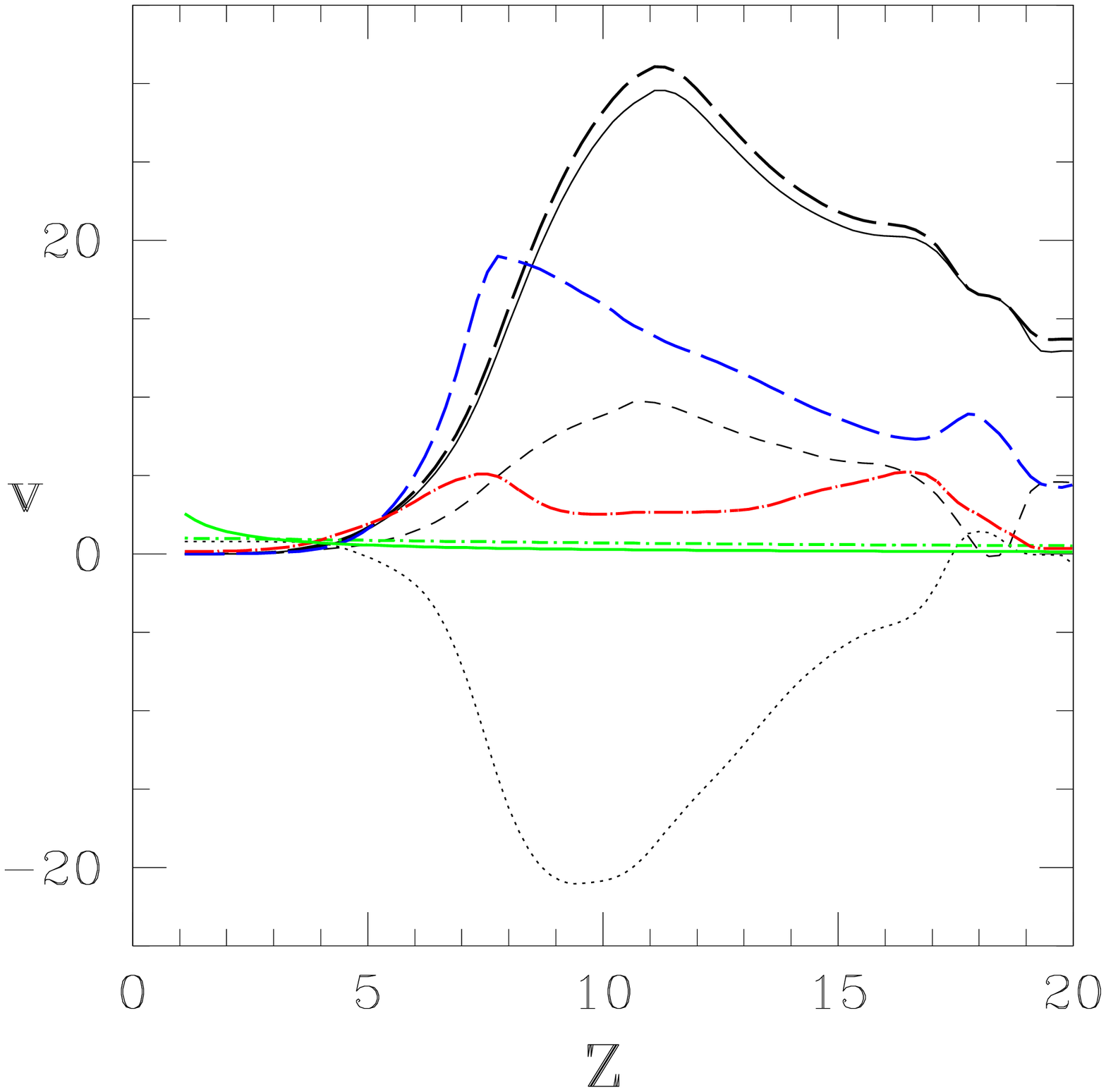}
\caption{Velocity profiles in simulation S2, in units of v$_{\mathrm
K,0}=164~$km~s$^-1$, for results from the bottom
panel in Figure~\ref{rhovS2} at T=750. Profiles along the propagation
direction of micro-ejection in the Z-direction, parallel to the
symmetry axis, at
$R=R_\ast$ we show in the {\it Left panel}, and at a larger distance
$R=5 R_\ast$ in the {\it Right panel}. Components of the velocity in Z,
R and toroidal direction are shown in thin solid, dashed and dotted black
lines, respectively. Poloidal, Alfv\'en, escape velocity and sound speed
velocities are plotted in the long-dashed (black), long-dashed (blue),
dot-short-dashed (green) and dot-long-dashed (red) line, respectively.
}
\label{velS2}
\end{figure}
\begin{figure}
\includegraphics[width=8cm,height=4.5cm]{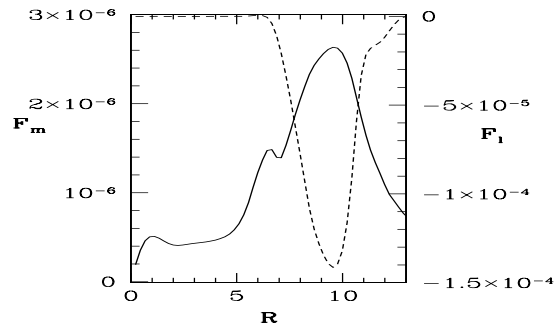}
\caption{
In solid line is shown mass flux F$_{\mathrm m}$ in units of $\dot{M}_0$,
and in dashed line the angular momentum flux F$_{\mathrm l}$ in the
corresponding units in the simulation S2 at T=750. Scales from the left and
right side of the plot are for
F$_{\mathrm m}$ and F$_{\mathrm l}$, respectively. Slices are taken along
the outer Z-boundary. We show the fluxes only for part of the box where
the micro-ejection is located.
}    
\label{S2mfluxinr}
\end{figure}
\begin{figure}
\includegraphics[width=8cm]{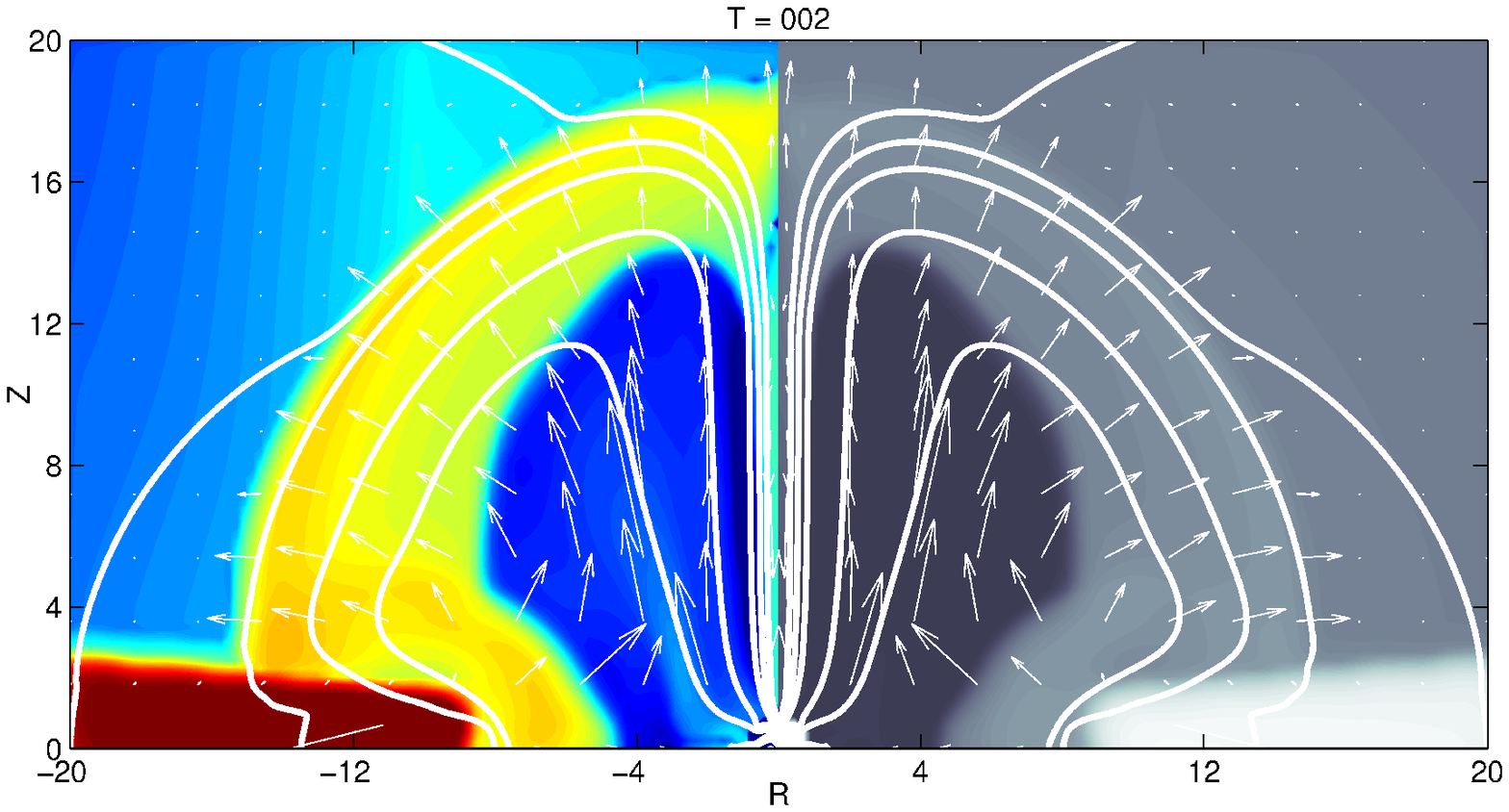} 
\includegraphics[width=8cm]{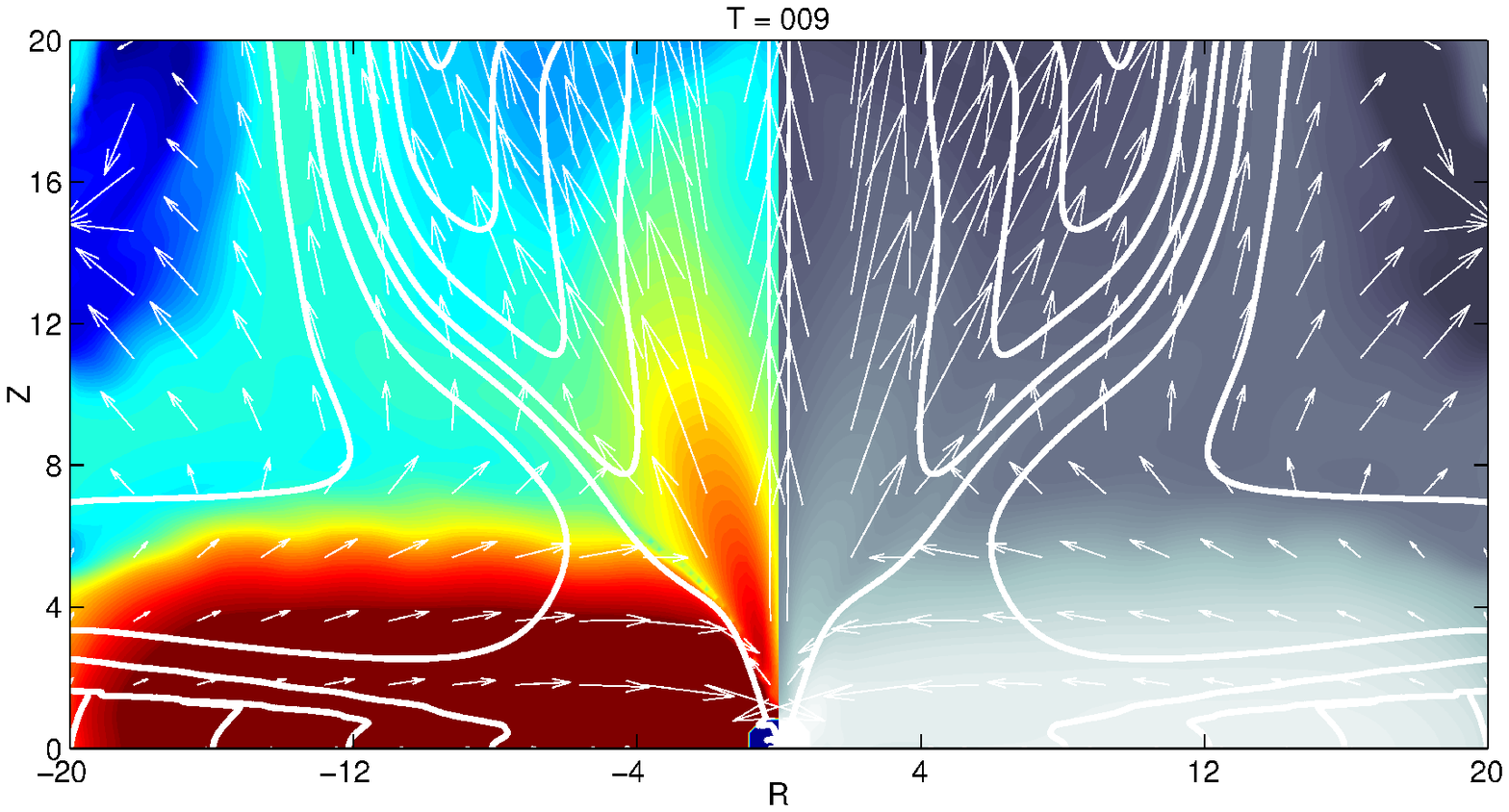} 
\includegraphics[width=8cm]{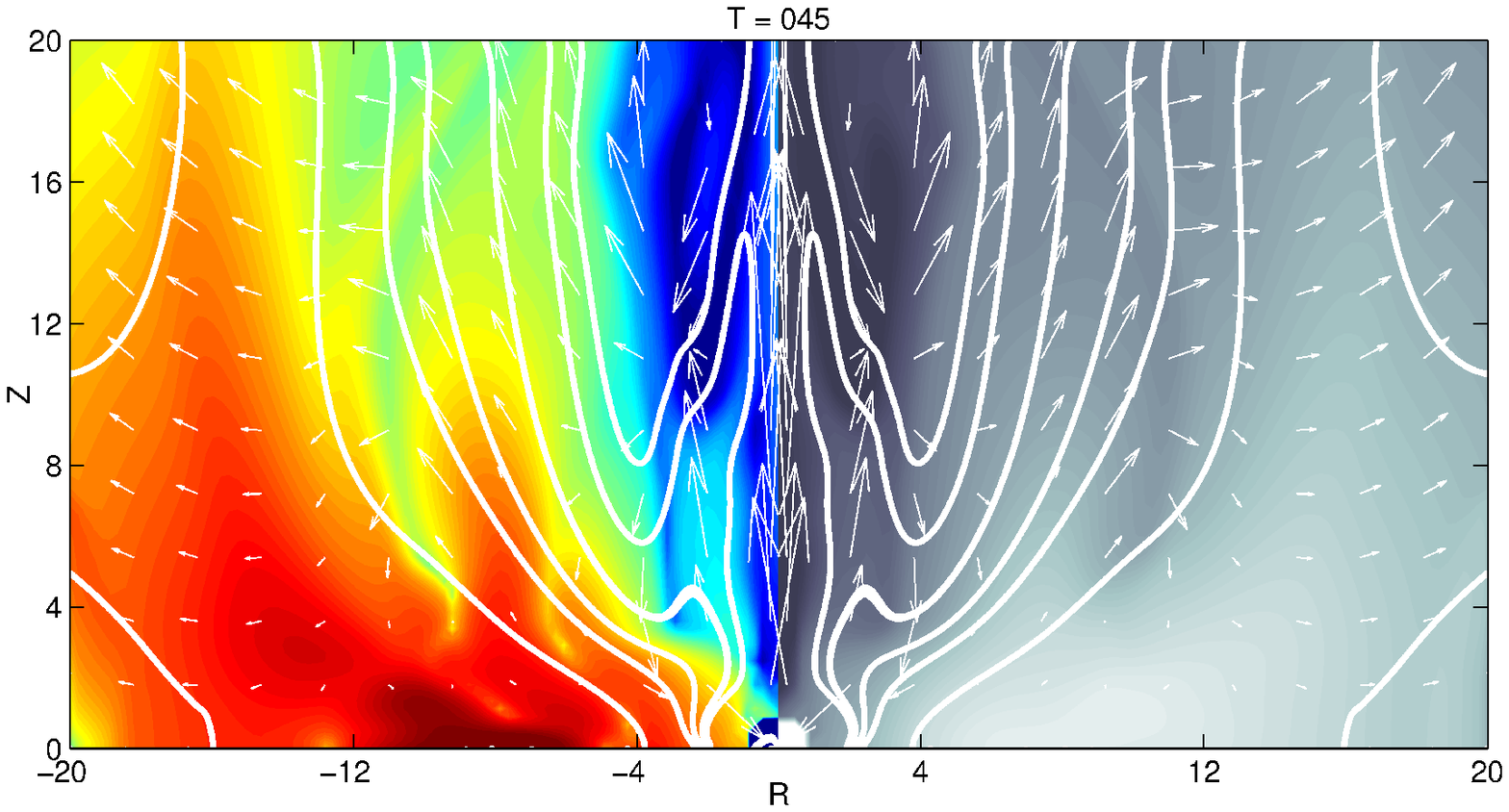}
\includegraphics[width=8cm]{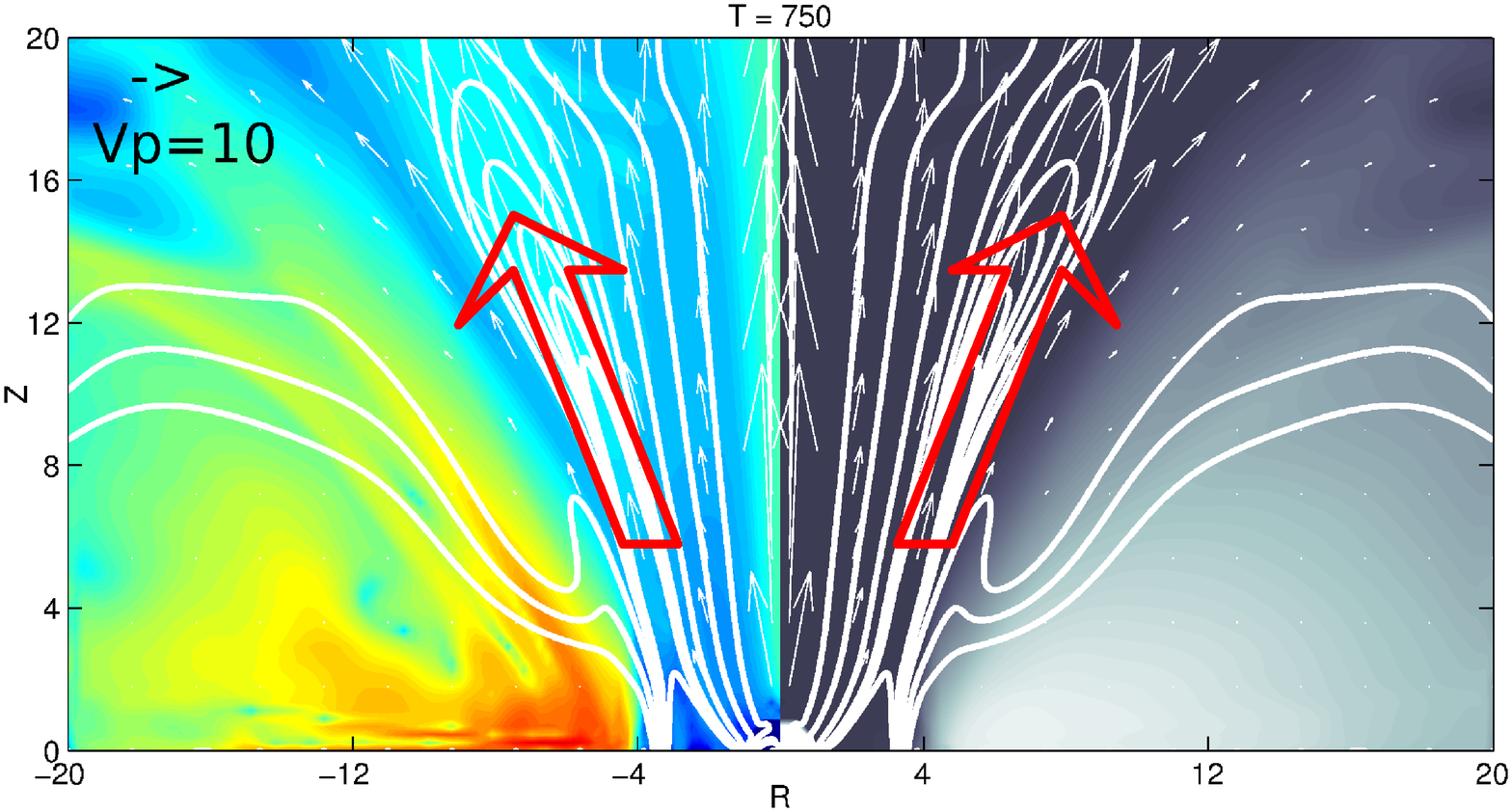}
\includegraphics[width=8.3cm]{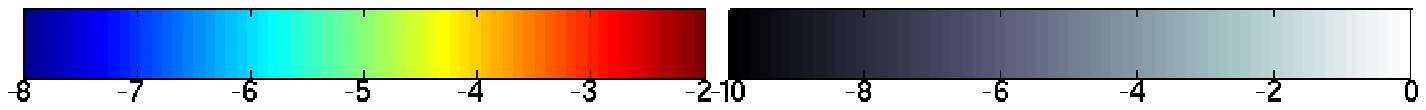}
\caption{Snapshots in our simulation S2. The left half
of each panel shows the poloidal mass flux $\rho {\rm v_p}$, and the right
half shows the density. Lines and arrows have the same meaning as in Figure
\ref{S1abplots}. Both plots are in logarithmic color scales, shown at the
bottom of the panels, with mass flux in units of $\dot{M}_0$ and density in
units of $\rho_0$. The initial magnetic field is a pure stellar magnetic
dipole, with value $B_\ast=30$\,G if assumed disk accretion rate is
$10^{-6}M_\odot/$yr$^{-1}$. {\it Top} to {\it bottom}
are shown characteristic stages discussed in this work: I) initial
relaxation when the magnetic field
is swept in and pinched near the disk mid-plane toward the star, II)
inflation and reconnection that end up opening the field, and strong infall
of matter onto the star from the disk, III) retraction of the disk matter
towards the corotation radius, with a transient inflow of matter onto the
star, and the light bullets of fast matter expelled along the axis-which is
a numerical, artificial flow, IV) final, quasi-stationary stage, with a
light, fast micro-ejection launched along the boundary layer between the
stellar and the disk field. For clarity, we additionally marked
micro-ejection with large red arrows.
}
\label{rhovS2}
\end{figure}
\begin{figure*}
\includegraphics[width=4.4cm,height=4.4cm]{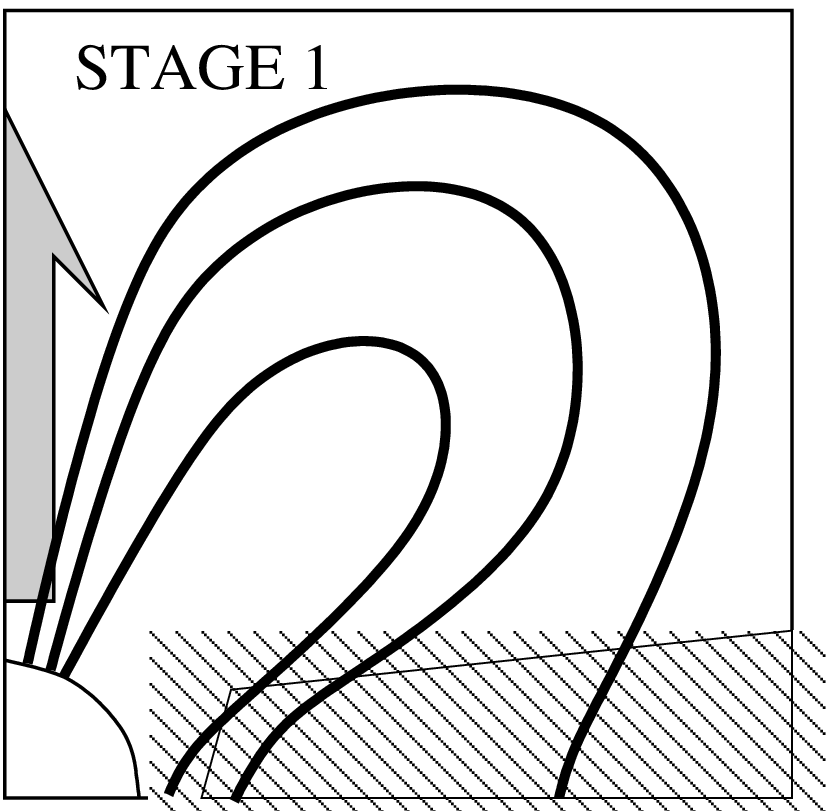}  
\includegraphics[width=4.4cm,height=4.4cm]{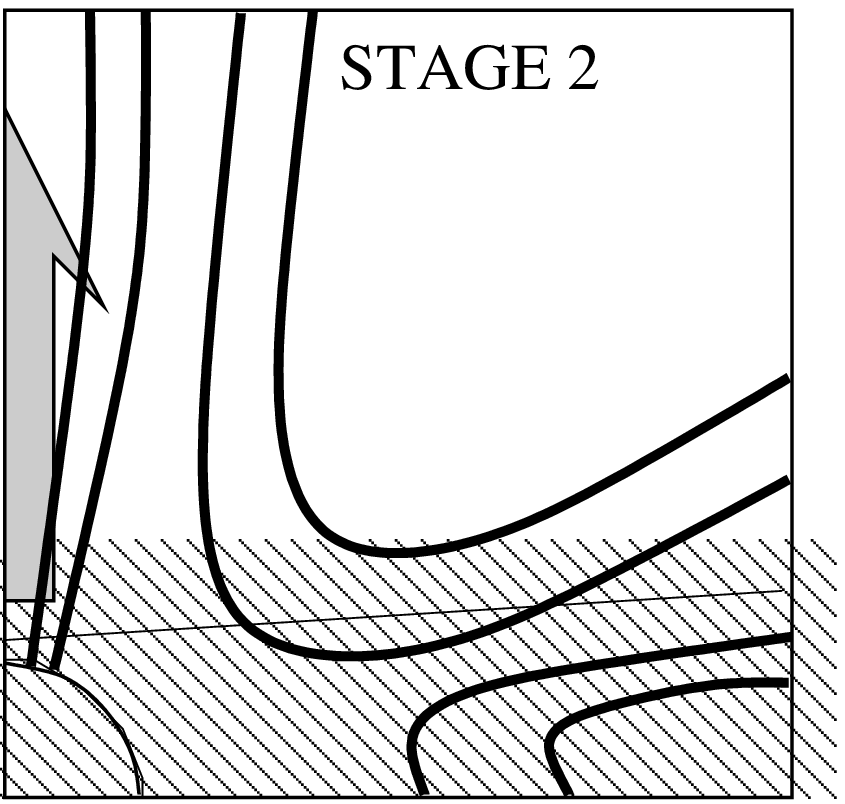}
\includegraphics[width=4.4cm,height=4.4cm]{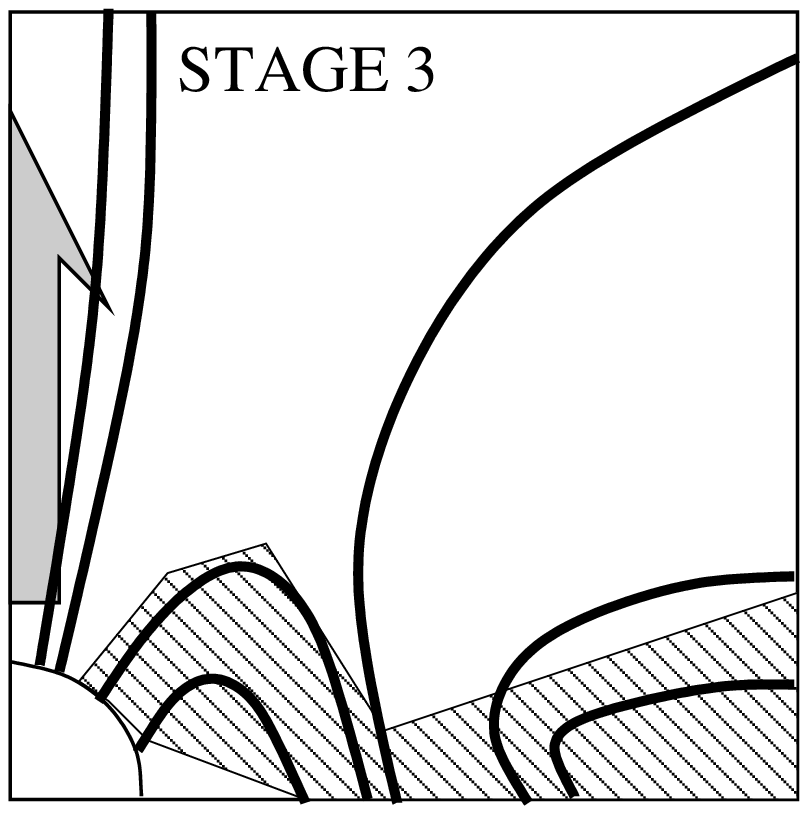}
\includegraphics[width=4.4cm,height=4.4cm]{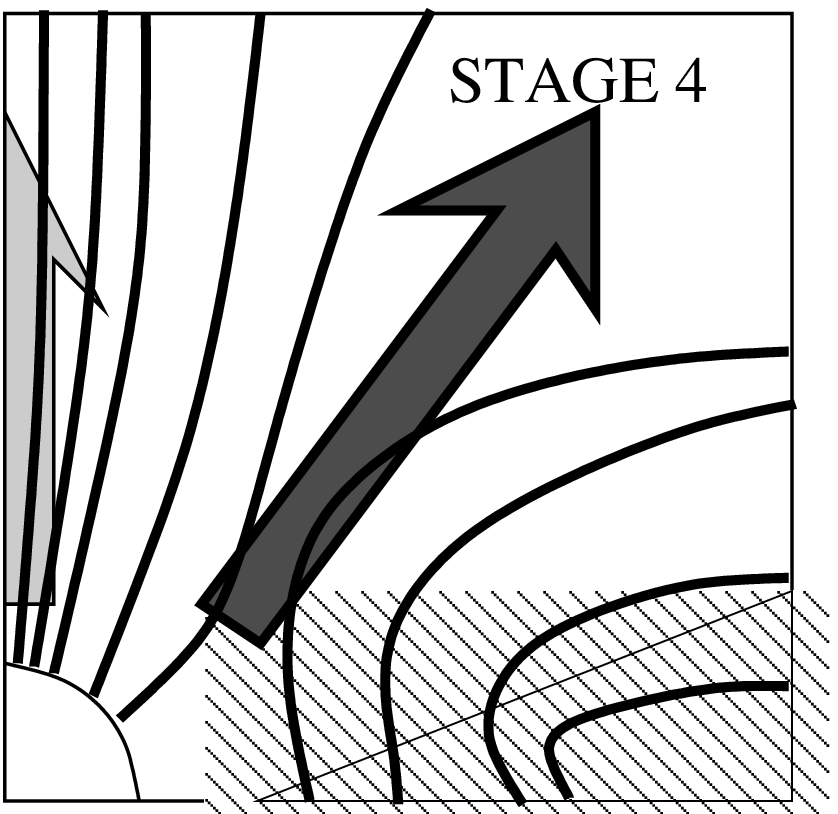}
\caption{A schematic sketch of the evolution in star-disk interaction in our
simulations. Stage one: the initial stellar dipole gets pinched during the
relaxation, when matter flushes in toward the central star; Stage two: the
magnetic field lines are open after reconnection, and disk matter can reach the
surface of the star. Our simulations here differ from most of the works
in Table 1 in the violent initial conditions which bring the disk onto the star. 
For the setups with a stable disk, such situation occurs only in the regime
of very weak stellar magnetic field. In the case of strong magnetic
field, configurations will be as sketched in \citet{gods99a}, where
the disk terminates in the magnetosphere, away from the star;
Stage three: the disk matter retracts and a funnel flow forms from the
disk inner radius and accretes matter onto the star; Stage four: the
system reaches a quasi-steady state, with magnetic field re-organized
into a configuration consisting of an open stellar field and field
footed in the disk. The arrows indicate the directions of the matter flow.
In the first three stages, artificial flow of a numerical origin forms (marked
with gray shadow) along the axis. It is of negligible mass flux. In the fourth
stage, along the boundary between the stellar and the disk magnetic
field, launched is a micro-ejection, which reaches a quasi-stationary state
(marked with black shadowed arrow) together with magnetic field.
}
\label{geommag}
\end{figure*}

To obtain longer lasting simulation, we devise a simulation with the
disk gap numerically imposed, as described in \S\ref{boundc}. Such
simulation, dubbed S2 here, has been performed with a part of the
disk mid-plane inside the disk gap defined as an open boundary. It means
that the disk truncation radius is not determined
self-consistently\footnote{For a large enough magnetic field, of the order
of 100~G as in simulation S1b, such imposed disk gap is largely ignored by the
disk, as matter is lifted above the disk equatorial plane.}. For a changed
boundary condition, we check the change of density in time along the disk
equatorial plane in Figure \ref{S2rhoint}. Inflow of matter into the disk
from the outer boundary, which mimics the accretion of matter from the
outer part of the accretion disk, has to be well chosen. In the case of
a too large inflow of matter into the disk, it would pile up and the disk
would become unstable. On the other side, if the disk would become drained
of matter, it would change the conditions we want to investigate. In Figure
\ref{S2rhoint}, we see that after the relaxation, our reservoir of matter
in the simulation is not changing for too much. In the other panel of the
same Figure we show lines along which parameters $\beta$ and $\beta_1$
equal unity, marking positions where the flow becomes magnetically
dominated. For our purpose of obtaining simulation similar to
S1b, it is important that those lines are similarly positioned in
both simulations.

Snapshots at different times in simulation S2 are shown in Figure
\ref{rhovS2}. We show the density and poloidal mass flux
$\rho {\rm v_p}$ for the same time step in the right and left side
of the same panel, to stress that in the density plots micro-ejection
will typically not be visible even in logarithmic color grading.
Other features, as ejected plasmoid or accretion
flow onto the star are well seen in both the density and mass flux plots,
and have been described in the literature mentioned in \S\ref{intro}.
Micro-ejections which we report here are lighter, so that they are not easy
to contrast even in logarithmic plots with mass fluxes.

We show the velocity along the micro-ejection in the quasi-stationary state
in Figure \ref{velS2}. The flow is supersonic and sub-Alfv\'enic close to
the stellar surface, but at the middle of the computational box it
becomes super-Alfv\'enic. Both the poloidal and total velocity in
micro-ejection are larger than the escape velocity
$v_{\mathrm esc}=(2GM_*/R)^{1/2}$. The total velocity is for two
orders of magnitude larger than the escape velocity. In Figure
\ref{S2mfluxinr} we show the radial dependence of fluxes $F_{\mathrm m}$
and $F_\ell$ across the outer Z-boundary in the quasi-stationary state.

Now we obtained a long lasting simulation, which resembles S1b, but 
the magnetic field required for the launching of micro-ejection is
smaller for almost an order of magnitude. We
can identify four evolutionary stages in progression in a system of an
interacting magnetosphere with its surrounding disk. We show snapshots
taken at different times, in Figure \ref{rhovS2}. The time-evolution
proceeds in the similar way in simulations S1a and S1b, and the results are
{\em robust} in that they occur under a wide range of explored parameters,
although each with different details. An initially pure dipole magnetosphere
has already bulged out and brought some gas along with it at as few rotations
as $T=2$. Near the axis, some gas also flows out at high velocity due to
magnetic pressure that is gradually building up--because of unrealistic
large magnetic force there, we conclude that this flow is artificial, of
numerical origin. The matter in the disk has flown in at a magnetic
stagnation point around $8R_\ast$, where the magnetic field dragged in
with the gas is pinched. Around $T=9$, matter went through a magnetic
reconnection and ejected plasmoids. The reconnected and opened field
enabled the disk gas to flush into the stellar surface and,
at the same time, more violent gas flows are directed
outwards both from the axial region and from the disk.
At a later time, $T=45$, after a few occurrences of the magnetic
reconnection events, part of the field closes back to the stellar surface,
and part remains open, footed near the new truncation radius, which is in
this case pre-determined by the boundary conditions. Matter channels
through the field lines that are footed both in the disk and the stellar
surface. Matter along the axis is still expelled by a numerical effect,
in the form of bullets or, when more stabilized, in a more continuous
light, very fast stream. The bottom panel in Figure \ref{rhovS2} shows
the representative snapshot in this simulation at a much later time $T=750$.
The system settled into a configuration where the magnetic field has been
opened into space with either foot in the star or in the disk, and formed
loops that connect to both the star and the disk. The gas flows out from the
regions on top of the star in an artificial fast flow, and from the boundary
on top of the loops along the diverging field lines open to the space,
forming micro-ejection. The matter which is launched outwards originates
from the magnetosphere above the disk where one foot of the magnetic
field is rooted. The mass and angular momentum fluxes in the artificial,
axial flow are typically for at least one order of magnitude smaller than
in micro-ejection launched under larger angle. The disk material stays
slightly outside of the magnetic footpoint where the field is pinched,
around the truncation radius. The disk, after the relaxation, does not
differ much from the purely hydrodynamic case in Figure \ref{rhovb0}.

Similar steps have been observed in other simulations with violent
relaxation, as e.g. \citep{gods99b}. This process can be described by
four conceptual stages, shown in a schematic sketch in Figure
\ref{geommag}. Stage I is the initial relaxation from the highly
non-equilibrium state in our simulation, when the magnetic field is
swept in by matter infalling from the disk towards the star, and pinched
near the disk mid-plane. The magnetic loops are twisted,
inflating\footnote{The inflation of the magnetic field lines occurs
because differential rotation at the footpoints of the magnetic field
loops which thread the star and the disk, tends to open the field lines.
It has been described in e.g.\ \citet{gh60}, \citet{Aly80} and
\citet{lov95}.} and forming plasmoids. The gas that flows with the
field swirls in and is gradually accelerated in the axial region by
magnetic pressure being built up. Stage II takes the scene after the
system goes through a reconnection that ends up opening the
field, enabling strong infall of matter onto the star from the disk. The
artificial axial flow, steady or in a series of bullets, forms as a
result of magnetic pressure from the twisted field, built because of
numerical effect nearby the axis. Stage III follows when the disk
matter retracts towards the corotation radius, and a time-variable inflow of
matter funnels onto the star from the inner disk truncation radius. The
system may have several passages through the first three stages and finally
move onto the quasi-steady state when the magnetic field is pinched and
strong enough to balance the ram pressure of the disk gas, truncating
the disk near the magnetospheric radius. The magnetic field settles into a
geometry where the field is open into the space both axially and
conically, with some loops anchoring both in the star and the
disk. Matter flows out quasi-stationary along the open field lines,
forming the fast micro-ejection along the current sheet formed between
the disk and the stellar component of the magnetic field.

Evolution should proceed in
a similar way for simulations with slowly rotating star which, as ours,
start with non-equilibrium initial conditions and with not too
strong stellar magnetic field, with disk reaching the star
during the violent relaxation phase. If the disk retracts is
depending of the strength of magnetic field: for a small field, the
gap does not form. For a sufficient large field, the disk is truncated,
and a disk gap is formed.

\section{Properties of micro-ejections}\label{props}
\begin{figure}
\includegraphics[width=7.5cm,height=5.cm]{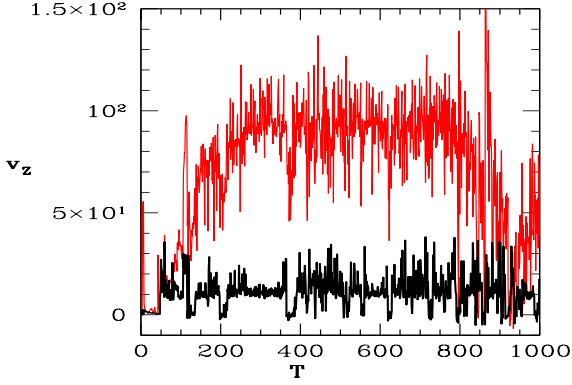}
\caption{Velocity in the axial direction in simulation S2, throughout the
simulation. We show the velocity in the micro-ejection at
$(R,Z)=(0.4R\ast,Z_{\mathrm max})$ in a thin (red) solid line, and
in at $(R,Z)=(7.1R_\ast,Z_{\mathrm max})$ in a thick
(black) solid line. In the former case average value in the quasi-stationary
state is of the order of 80v$_{\mathrm K,0}$, and in the latter it is of the
order of 10v$_{\mathrm K,0}$.
}
\label{S2v1fig}
\end{figure}
\begin{figure}
\includegraphics[width=4.2cm,height=4.5cm]{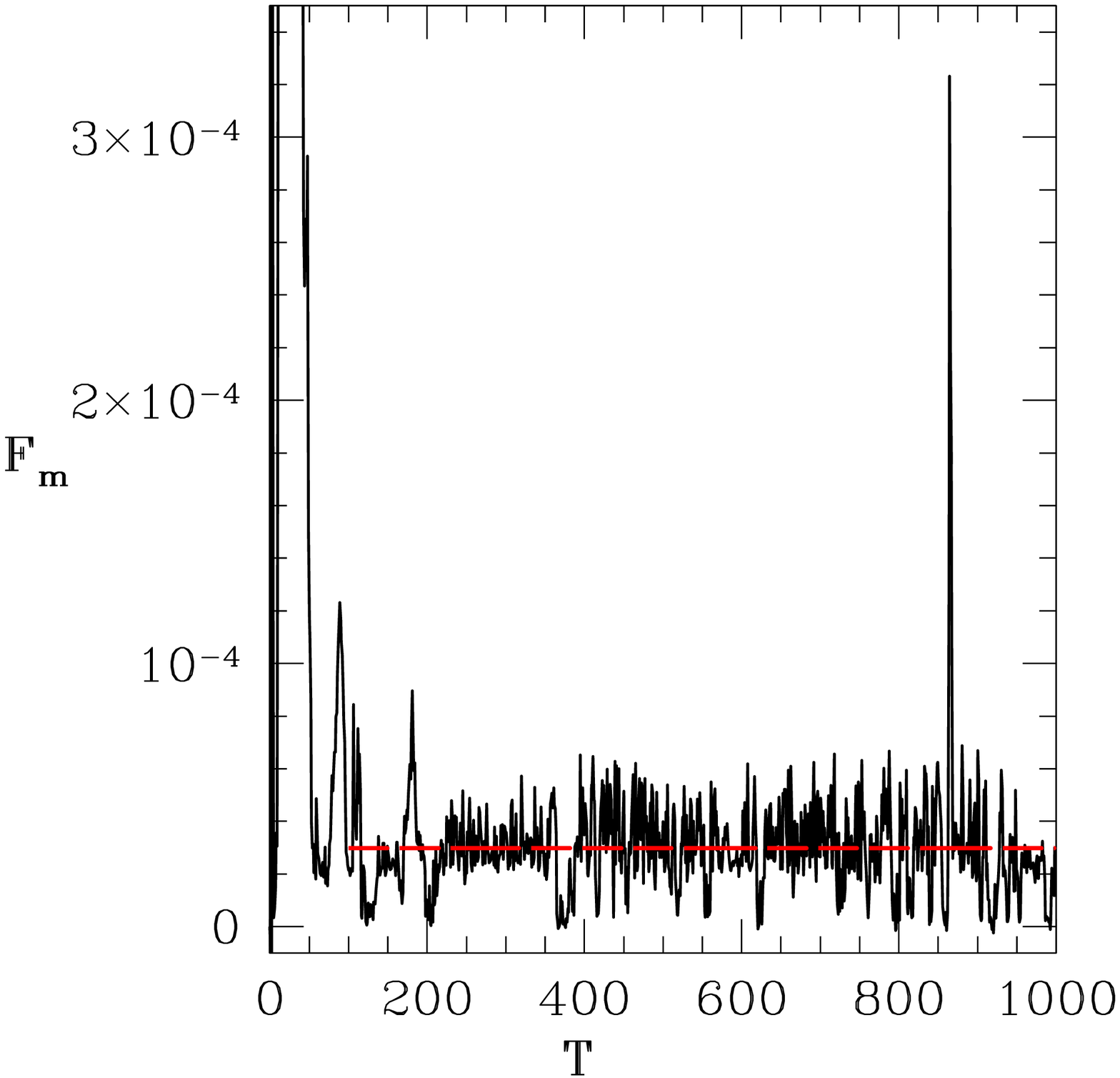}
\includegraphics[width=4.2cm,height=4.5cm]{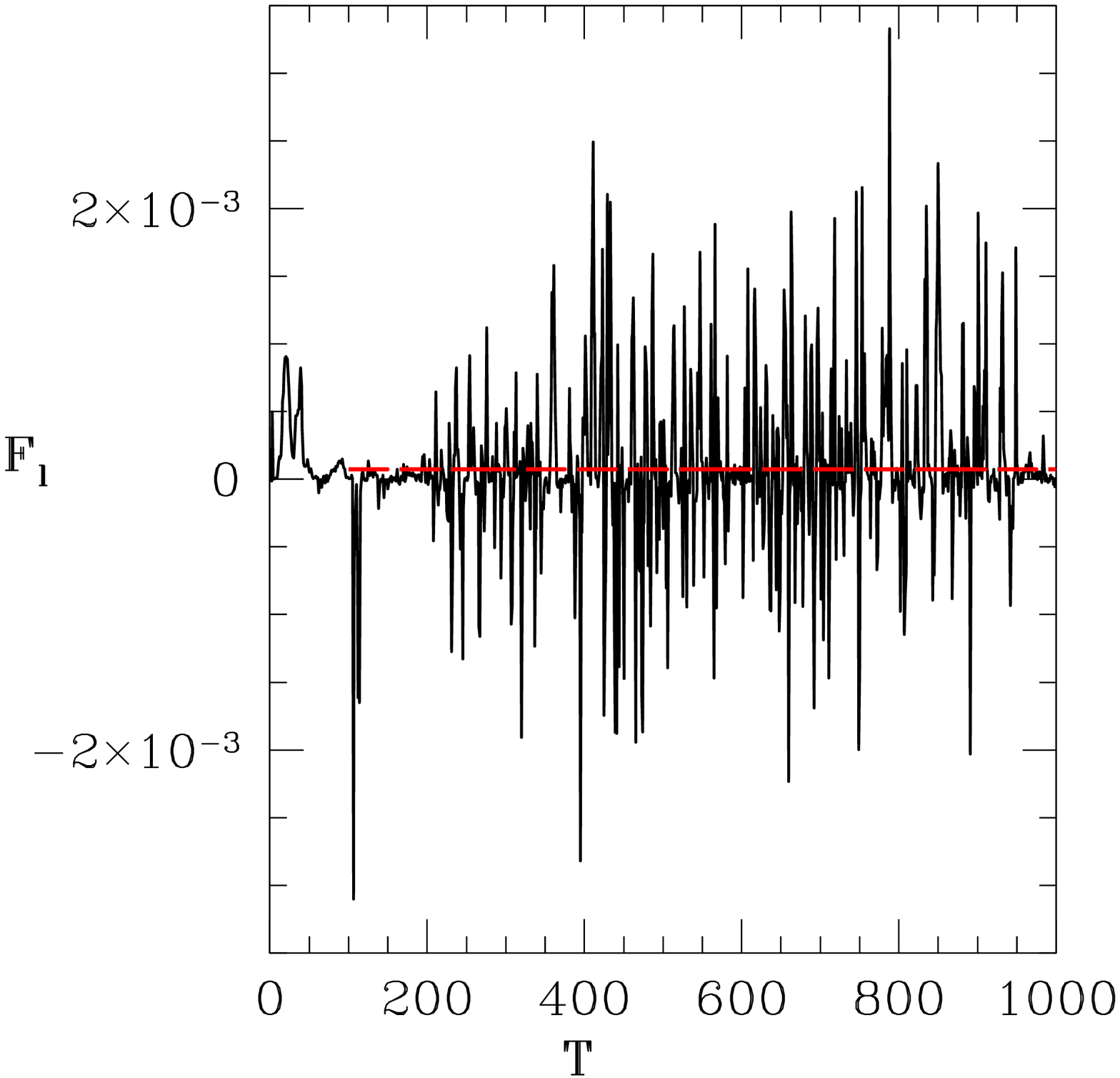}
\caption{
Time evolution of mass ({\it Left panel}) and angular momentum
({\it Right panel}) fluxes in simulation S2, parallel to the axis of
rotation, along the ${\mathrm Z_{max}}$ boundary for 1/3 $R_{\mathrm max}$
part of the box. We show the fluxes after
the relaxation in solid (black) line. In dashed (red) line we show
the average value, computed starting from T=100, when the flow becomes
quasi-stationary. The mass flux average value is $3.0\times 10^{-5}\dot{M}_0$,
and the angular momentum flux average value is $7.4\times 10^{-5}$.
}    
\label{S2fluxes}
\end{figure}
\begin{figure}
\includegraphics[width=7.5cm,height=5.cm]{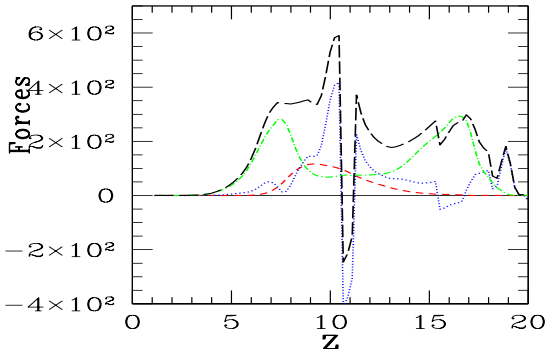}
\caption{Forces in simulation S2 at T=750 along slice parallel to the axis
of symmetry, at R=5. In black solid, green dot-dashed, red short-dashed,
blue dotted and black long-dashed lines are shown gravitational, pressure
gradient, centrifugal, magnetic and total forces, respectively. Micro-ejection
is driven by a combination of pressure gradient and magnetic forces.
}
\label{forsS2}
\end{figure}
\begin{figure}
\includegraphics[width=4.2cm,height=4.5cm]{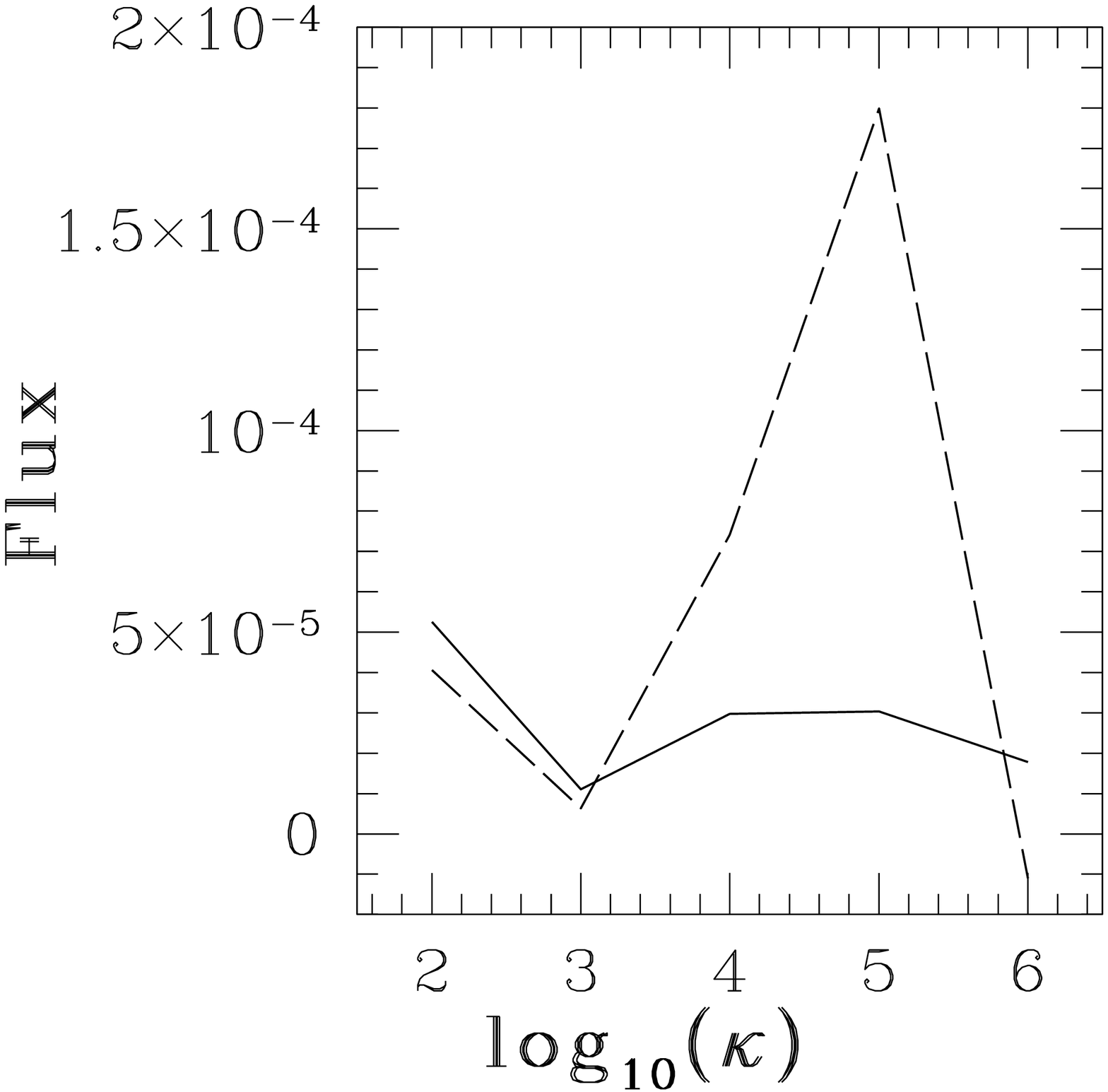}
\includegraphics[width=4.2cm,height=4.5cm]{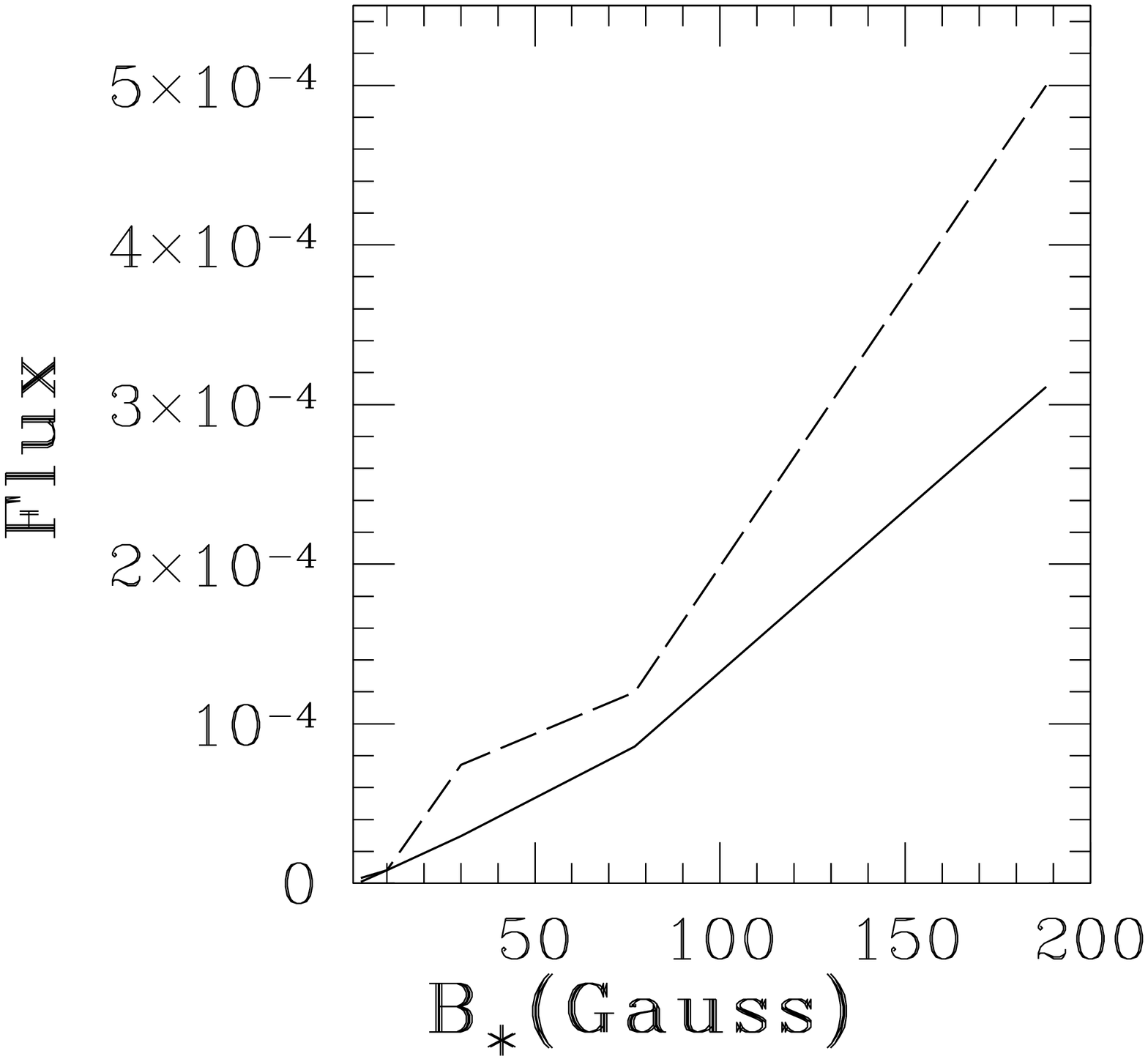}
\caption{Average mass and angular momentum fluxes
F$_{\rm m}$ and $F_{\ell}$ in our simulations with changing density contrast
$\kappa$ between the disk and the corona are shown in the {\em Left} panel,
and with changing stellar magnetic field strength in the {\em Right} panel.
In solutions with $\kappa=(10^2,10^3,10^4,10^5,10^6)$ we show the mass flux
in solid line, and the angular momentum flux in dashed line. Results for
the mass flux for $\kappa=10^2$ in our simulations are always larger for a
factor of 2 to 3, than for larger, more realistic values of $\kappa$. In
simulations with increasing magnetic field, we show the mass flux in units of
$\dot{M}_0$ in solid line, and the angular momentum flux in the
corresponding units in dashed line. With stellar magnetic field
$B_\ast=(3,10,30,77,188)G$, with disk accretion rate $10^{-6}M_\odot/$year,
both fluxes increase with increasing magnetic field. If we assume
$10^{-8}M_\odot/$year, magnetic field strenghts are 1/10 of those values.
}
\label{mfskappaiB}
\end{figure}
\begin{figure}
\includegraphics[width=3.8cm,height=4.5cm]{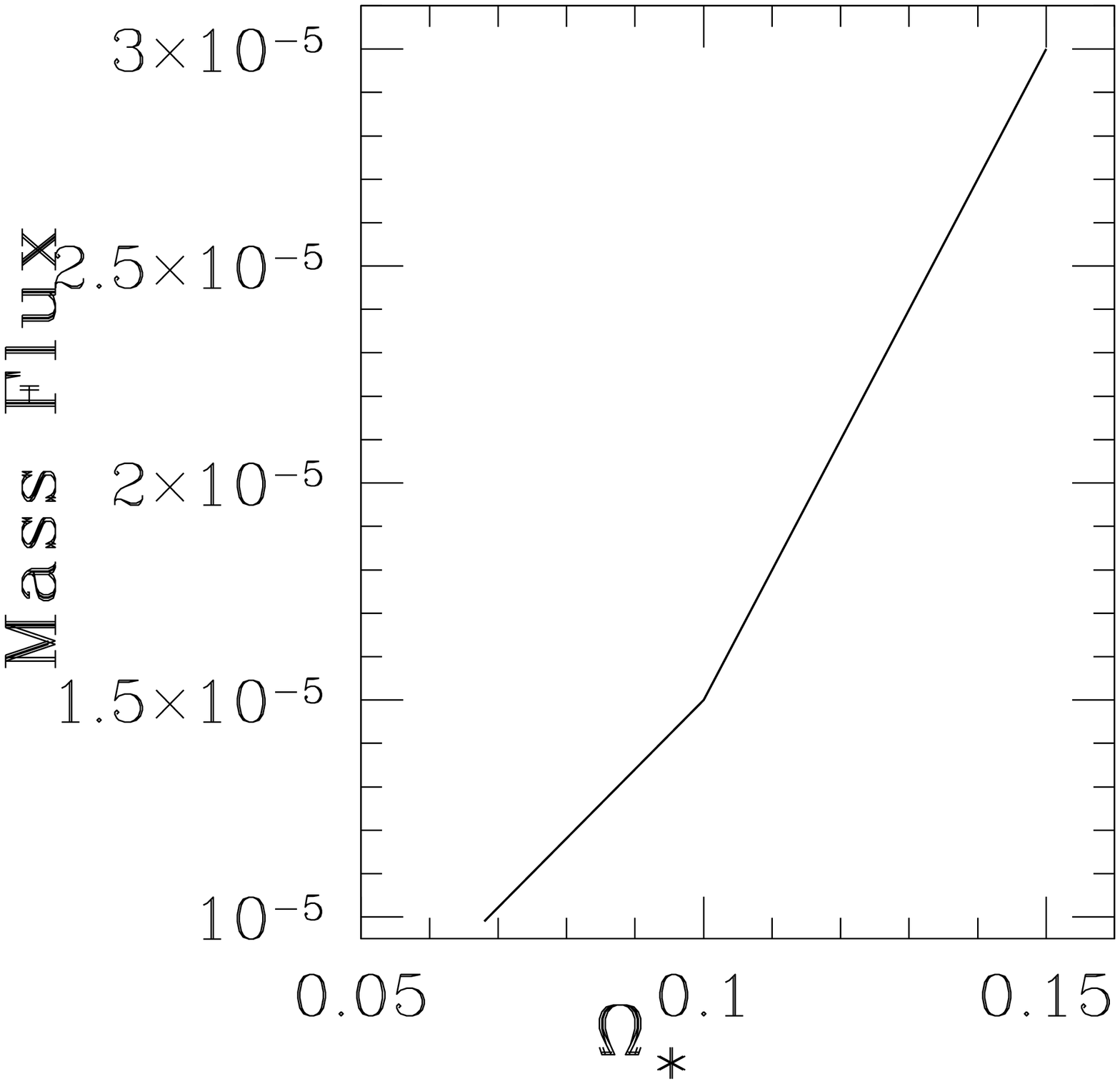}
\includegraphics[width=4.6cm,height=4.5cm]{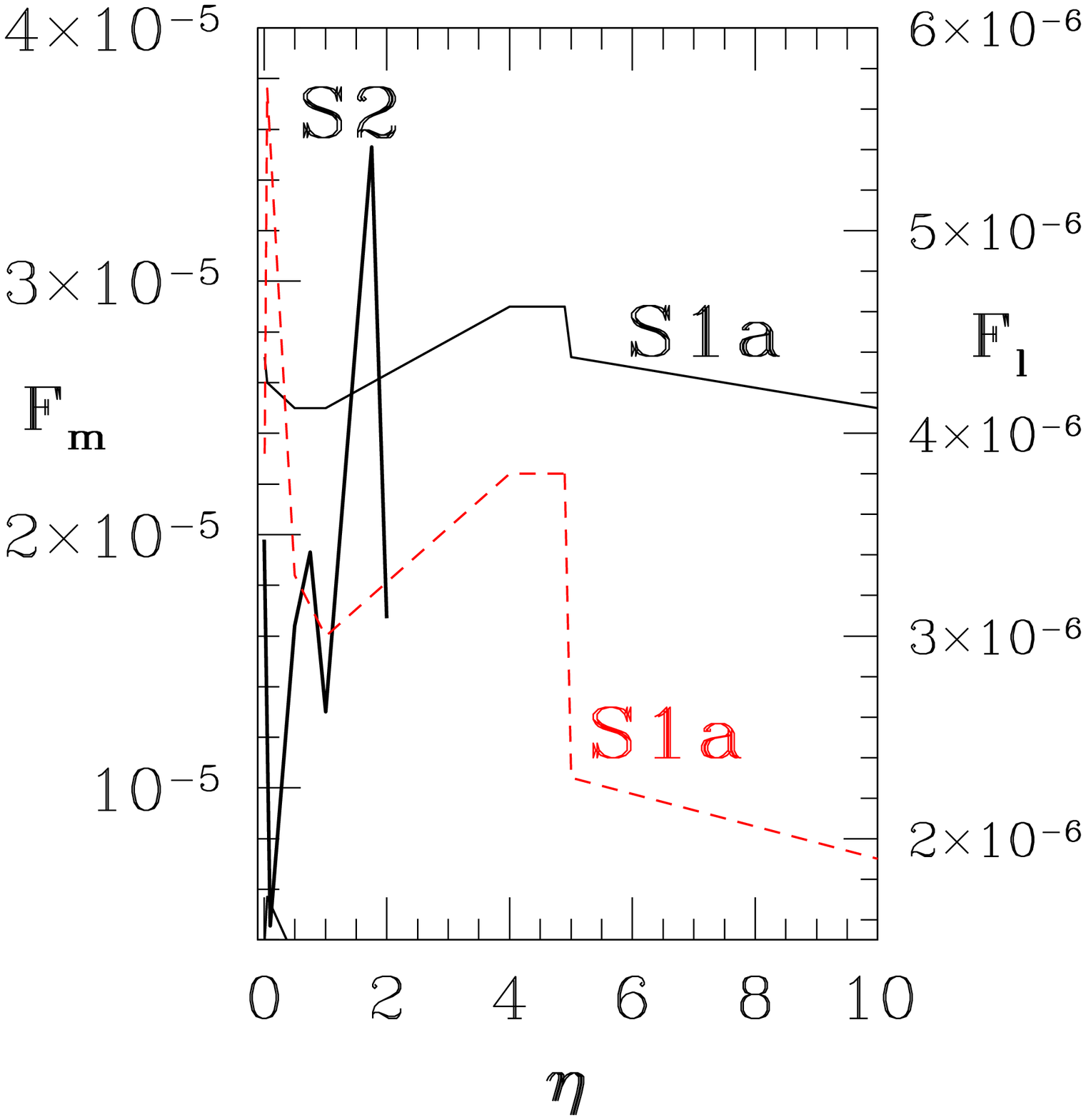}
\caption{In the {\em Left} panel we show change of mass flux with
variation of stellar rotation rate $\Omega_\ast$. Mass flux in
micro-ejection increases with increasing $\Omega_\ast$. In the
{\em Right} panel we show average mass fluxes F$_{\rm m}$ in
simulations S1a and S2 in dependence of maximum resistivity in
the computational box (the numerical resistivity is included).
Mass fluxes are shown in thick and thin (black) solid lines,
respectively. The average angular momentum flux
F$_{\rm l}$ in simulation S1a and its dependence on resistivity
is shown in the dashed (red) line. We do not show result for angular
momentum flux for simulation S2, as it varies randomly between
$-10^{-6}$ and $10^{-6}$. Values for both fluxes in simulation S1a for
$\eta=50$ are not shown in the figure, but are similar to values for
$\eta=10$. Note the different labels in the left and right sides of
the plot.
}
\label{fmeta}
\end{figure}
Here, we describe in more detail the properties of matter in
micro-ejection in the quasi-stationary solution in our simulation S2. In
Figure \ref{S2v1fig} we show the change of velocity in chosen points in
the computational domain. Oscillations show a sign of instability working at
a short time scale, so that the flow in micro-ejection varies even in
the quasi-stationary phase. The reason is that magnetic field is
permanently reshaped by reconnections along the flow --- we discuss it in
more detail in \S\ref{recon}. In Figure \ref{S2fluxes} we show the time
evolution of those fluxes. To avoid
influence of the disk flaring in presenting the results, we integrate fluxes
only to $R_{\mathrm max}/2$ at each time step. The fluxes reach
quasi-stationarity after relaxation and stabilization of the system.
Reconnection events along the boundary layer between the stellar and the
disk fields contribute to the oscillation at the timescale of one
rotation period. Following the evolution of ejections
longer into the quasi-stationary phase, we measure that the
contribution to the total fluxes from the artificial flow very close
to the axis is at least one order of magnitude smaller than from the
ejections launched under larger angle.

Forces along micro-ejection, in a slice parallel to the axis at R=5,
are shown in Figure \ref{forsS2}. The same as in simulations S1a and
S1b, the combination of pressure gradient and magnetic forces is
driving a micro-ejection.

\subsection{Dependence of results on the density contrast, magnetic
field, stellar rotation rate and resistivity}
Simulations in the literature, as we show in Table \ref{tabla1}, were
performed for various contrasts of the disk to corona density $\kappa$.
What is the influence of this parameter in our simulations?

We show results for variation of this parameter in simulations set-up like
simulation S2 in Figure \ref{mfskappaiB}. There are
no significant differences in the average fluxes in the range of $\kappa$
from $10^{3}$ to $10^{6}$. For $\kappa=10^{2}$, the mass flux is always larger, so
that it is double or, in some setups, triple the value from other cases.
We checked that this trend, which is illustrated here for a case of
sub-Keplerian rotation of the disk, is true also in the case of Keplerian
rotation profile. This means that results of simulations with
$\kappa=10^{2}$ could be unrealistic in the case of launching
of astrophysical outflows. This is not surprising, taking into account
that the realistic value of $\kappa$ is probably somewhere between
$10^{8}$ and $10^{5}$ for YSOs-this value will inevitably strongly
depend on local conditions.

We also check how the mass flux in simulation S2 depends on the
magnetic field strength. In Figure \ref{mfskappaiB} are shown the average
fluxes for magnetic fields up to 200 G, for YSOs with a disk
mass accretion rate of $10^{-6}M_\odot$~yr$^{-1}$. Both the mass and angular
momentum fluxes increase with increasing magnetic field strengths.
Our fluxes are very small, as the source of matter is magnetosphere
above the gap, and not directly the disk surface. Simulations like those
from the Table \ref{tabla1} with stellar fields of about 1 kG may
truncate the disks and launch stable, massive outflows, with the matter
from the disk inner radius.

Our simulations are performed for {\em slow} stellar rotation. We check how
fluxes change with variation of stellar rotation rate $\Omega_\ast$.
Results for mass fluxes are shown in Figure \ref{fmeta}, for
$\Omega_\ast=$0.068, 0.1 and 0.15, which correspond to corotation radii
6.0R$_0$, 4.6R$_0$ and 3.5R$_0$, respectively. Mass flux increases with
increasing stellar rotation rate, i.e. decreases with increasing corotation
radius. Angular momentum fluxes do not show clear dependence. 

Because of violent relaxation, a simulation can fail at early stages,
so that it is not easy to perform a parameter study of dependence on
resistivity. Each of the runs has its preferred level of resistivity,
to last for for hundreds of rotations into the quasi-stationary state.
We choose S2 as a representative case because it lasted more than
500 Keplerian rotations at $R_0$ with unit maximum resistivity.
In Figure \ref{fmeta} we show that average mass and angular momentum fluxes
do not depend on resistivity. For comparison, and to show that the same is
true also in the case of self-consistent disk gap boundary, we show along
also the mass flux in simulation S1a -- in both cases variation remains
small, within the same order of magnitude. With larger magnetic
field in simulation S1b, it is difficult to obtain quasi-stationary
results for a range of values of resistivity, so that we do not provide
values for simulations in S1b. Resulting fluxes could depend on the model
of resistivity, as it will enable or prevent reconnection to occur. In
simulations with higher resolution than in our simulations here, this
dependence could be more visible, as then the numerical resistivity
triggers with the lower value, leaving larger portion of the parameter
space for the physical resistivity.

\section{Reconnection and opening of magnetic field lines}\label{recon}
\begin{figure}
\includegraphics[width=7.5cm,height=5.cm]{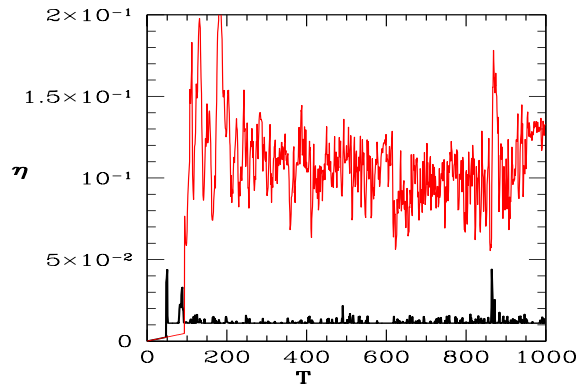}
\caption{Change of resistivity with time in simulation S2, in units of
$\eta_0=10^{19}$cm$^2$~s$^{-1}$. We show
$\eta$ in two positions (R,Z) in the computational box: at the exit
region of the micro-ejection in the position (7,$Z_{\mathrm max}$), and in the
middle of the computational box, in the position (10,10), in thick (black)
and thin (red) solid lines, respectively. As expected in our model with
resistivity in the disk corona dependent on the density, closer to the disk,
resistivity is larger.
}
\label{etaint}
\end{figure}
\begin{figure}
\includegraphics[width=4.25cm,height=4.5cm]{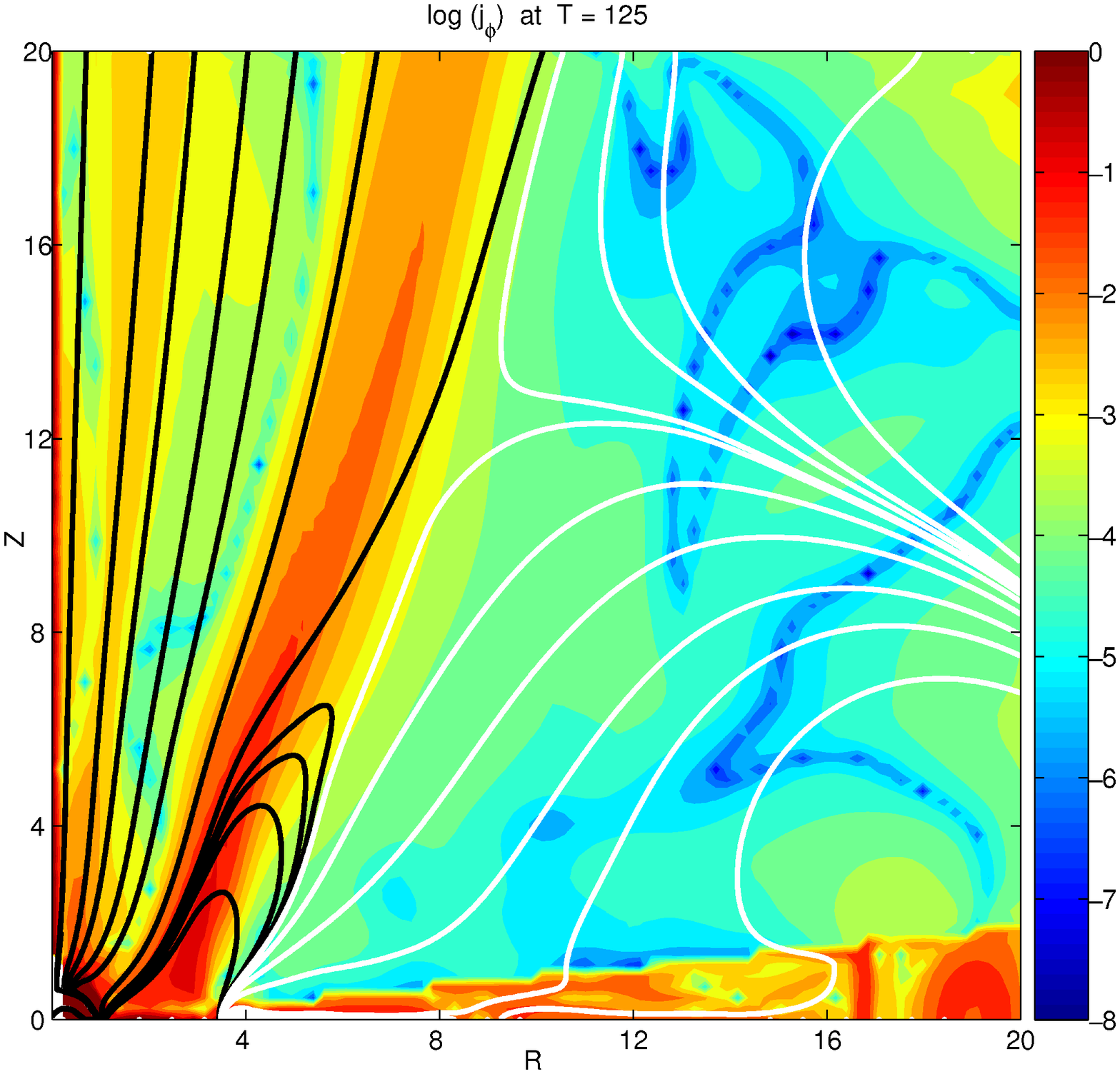}
\includegraphics[width=4.25cm,height=4.5cm]{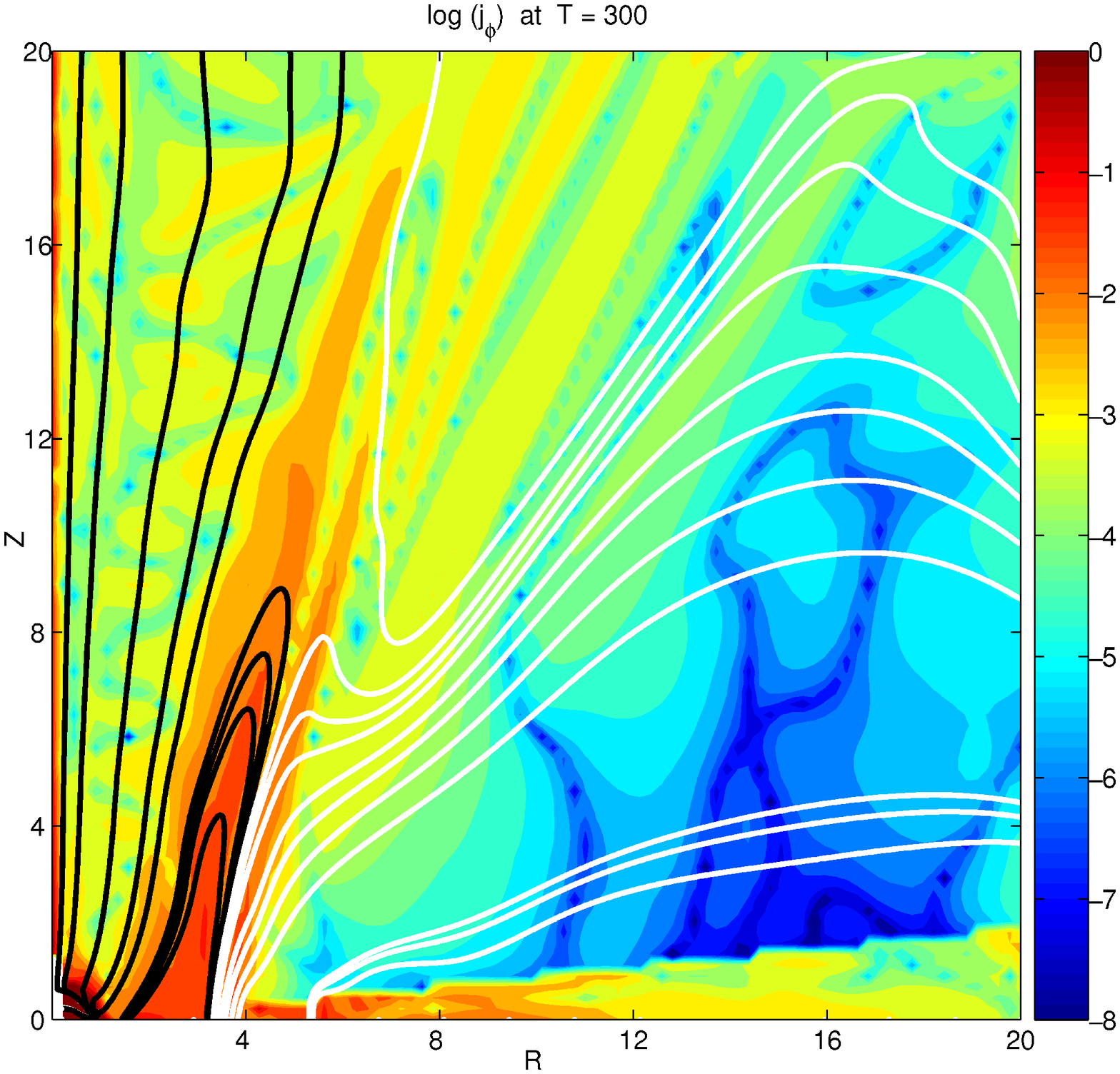}
\caption{
In the {\em Left} panel is shown absolute value of the toroidal current
$j_\phi$ at T=125 in simulation S1b, in logarithmic color grading, in
code units. In solid lines are drawn poloidal magnetic field lines, with
lines along which magnetic field vector points towards the star, painted
in black color, and those where the field
points in the opposite direction, in white color. Similar result is
obtained in simulation S2 at T=300, as shown in the {\em Right} panel.
Reconnection is ongoing along the boundary sheet throughout the
quasi-stationary state, and pushes the matter out of the magnetosphere,
in addition to the force caused by the pressure gradient. In the
simulation S1a, such current sheet forms closer to the star and under
smaller angle, measured from the axis.
}
\label{currsh}
\end{figure}
\begin{figure}
\includegraphics[width=6.5cm,height=6.cm]{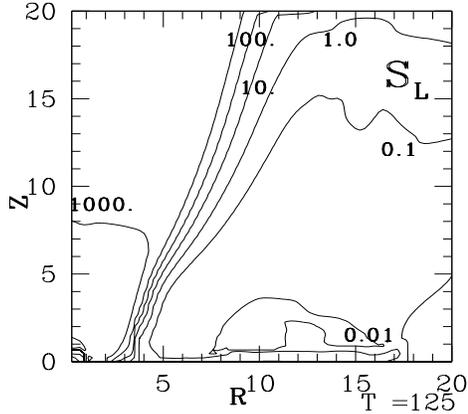}
\caption{
Isocontours of Lundquist number S$_L$ in S1b, at T=125.
Values of $S_L$ for some isocontours are labeled on the plot.
}
\label{freyn}
\end{figure}

Reconnection is essential for launching outflows from a star-disk
magnetosphere. If, for some reason, reconnection does not occur,
a magnetic wall forms and the evolution of the system will be different.
In the simulation with non-resistive corona, magnetic field does not relax
in the same way as in the simulation with a resistive corona. Without
resistivity, the corona relaxes from the initial condition with
a steeper gradient between the stellar and the disk component of the
magnetic field, because of slower reconnection process, as seen e.g. in
\citet{fen09}, where the disk was treated as a boundary condition,
with resistivity included in a whole domain.

Resistivity facilitates the reconnection, and shape
the geometry of the final magnetic field. Even if physical resistivity is
not included in the code, there is unavoidable numerical resistivity,
which can be estimated as $\eta_{\rm num}=v\Delta x$, where
$\Delta x$ is the grid spacing and the characteristic velocity is the
poloidal Alfv\'{e}n velocity. As numerical resistivity is different from
physical, results obtained with non-resistive code could be
unphysical. On the other side, chosen physical model of resistivity can
also affect the applicability of the results \citep{gods97}.

We included physical resistivity in the code, modeled in dependence of the
matter density, and allowed it to vary in time. In Figure \ref{etaint}
we show variation of resistivity with time in two positions in the
computational box. Reconnection is ongoing along the boundary between the
stellar and the disk magnetic field. The toroidal component of the
current $j_\phi$ in Figure \ref{currsh} shows the position of a current
sheet in the computational box in the quasi-stationary state in
the simulation S1b. The sheet forms close to the boundary
line where the star and the disk fields, which are in our case of opposite
direction, meet. Reconnection is ongoing along the boundary during the
time-evolution, into the quasi-stationary state. It looks similar in
simulation S2, and in simulation S1a a current sheet starts from the
magnetosphere closer to the star. Reconnection pushes the matter outwards
from the magnetosphere in the disk gap, in addition to the force caused
by the pressure gradient and magnetic force. To show the spatial distribution
of resistivity, we compute the Lundquist number S$_L$ in the quasi-stationary
state of simulation S1b, in the Figure \ref{freyn}, with labeled values for some
isocontours of the value of S$_L$.

\section{Elsasser criterion for launching}\label{elsase}
Launching by a magnetospheric accretion-ejection is
the result of interaction of the magnetosphere and the innermost
portion of the disk. In our setup, with the resistivity modeled by the
density, line where the Lundquist number S$_L=1$ shows to be useful
indicator of the launching region, but this might be dependent on the model
of resistivity. The best, and most general indicator is the line where the
magnetic and matter pressure are equal ($\beta=1$).

Another number which we find to indicate where from the launching is
possible, is the Elsasser number. It is defined by the ratio of the Z
component of the Alfv\'en velocity and
$\eta\Omega_{\rm K}$ (see \citet{sal07} and references therein):
\beq
\Lambda \equiv \frac{V_{\mathrm AZ}^2}{\eta \Omega_{\rm K}} > 1 \,.
\eeq
In some systems, when the flow is mainly in the Z-direction, $\Lambda$ can
be similar or equal to the Lundquist number S$_L$, or magnetic
Reynolds number Rm, but in general they are different, because velocity
of the outflow differs from the Alfv\'{e}n velocity. 

\begin{figure}
\includegraphics[width=4.25cm,height=4.5cm]{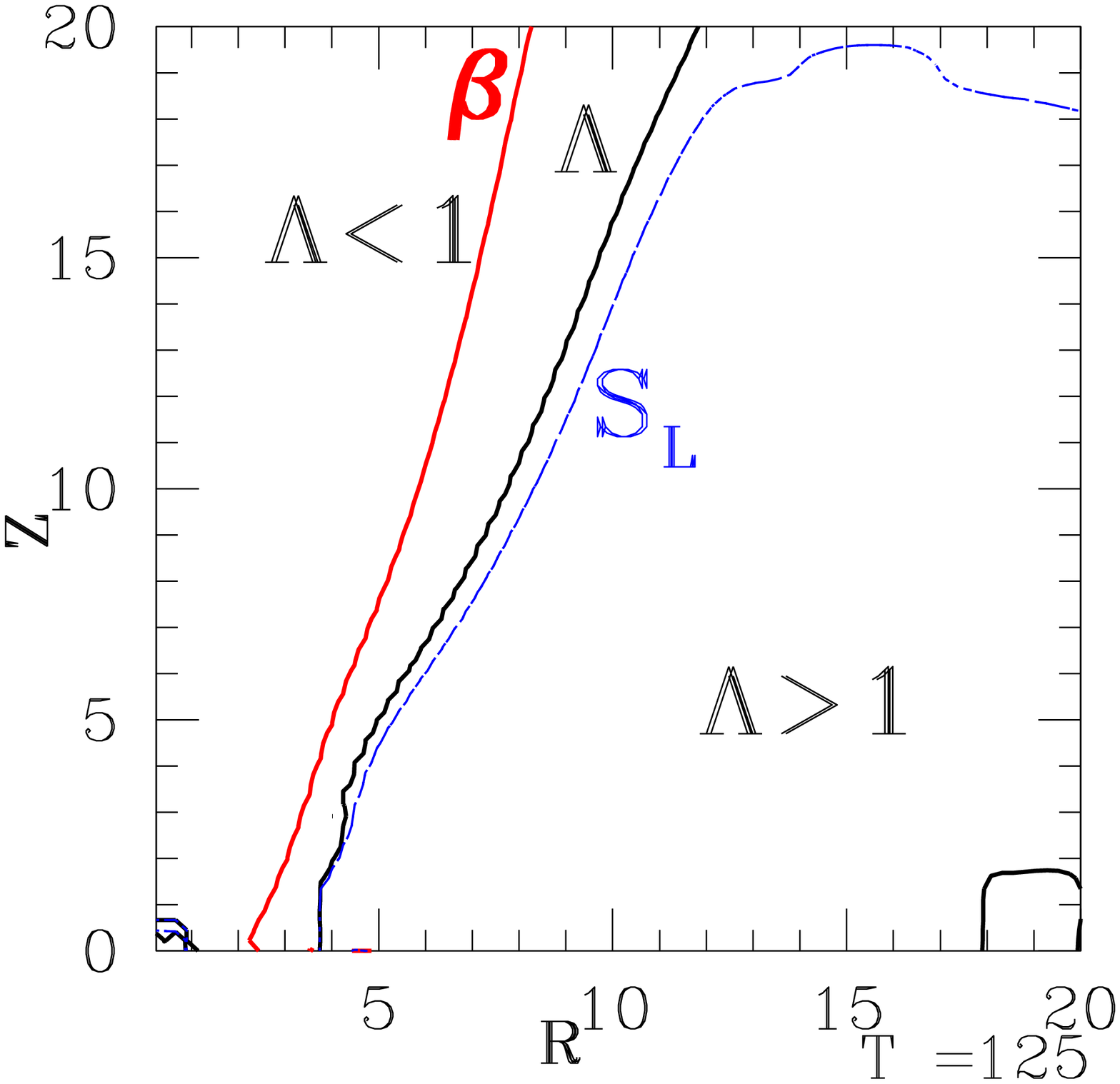}
\includegraphics[width=4.25cm,height=4.5cm]{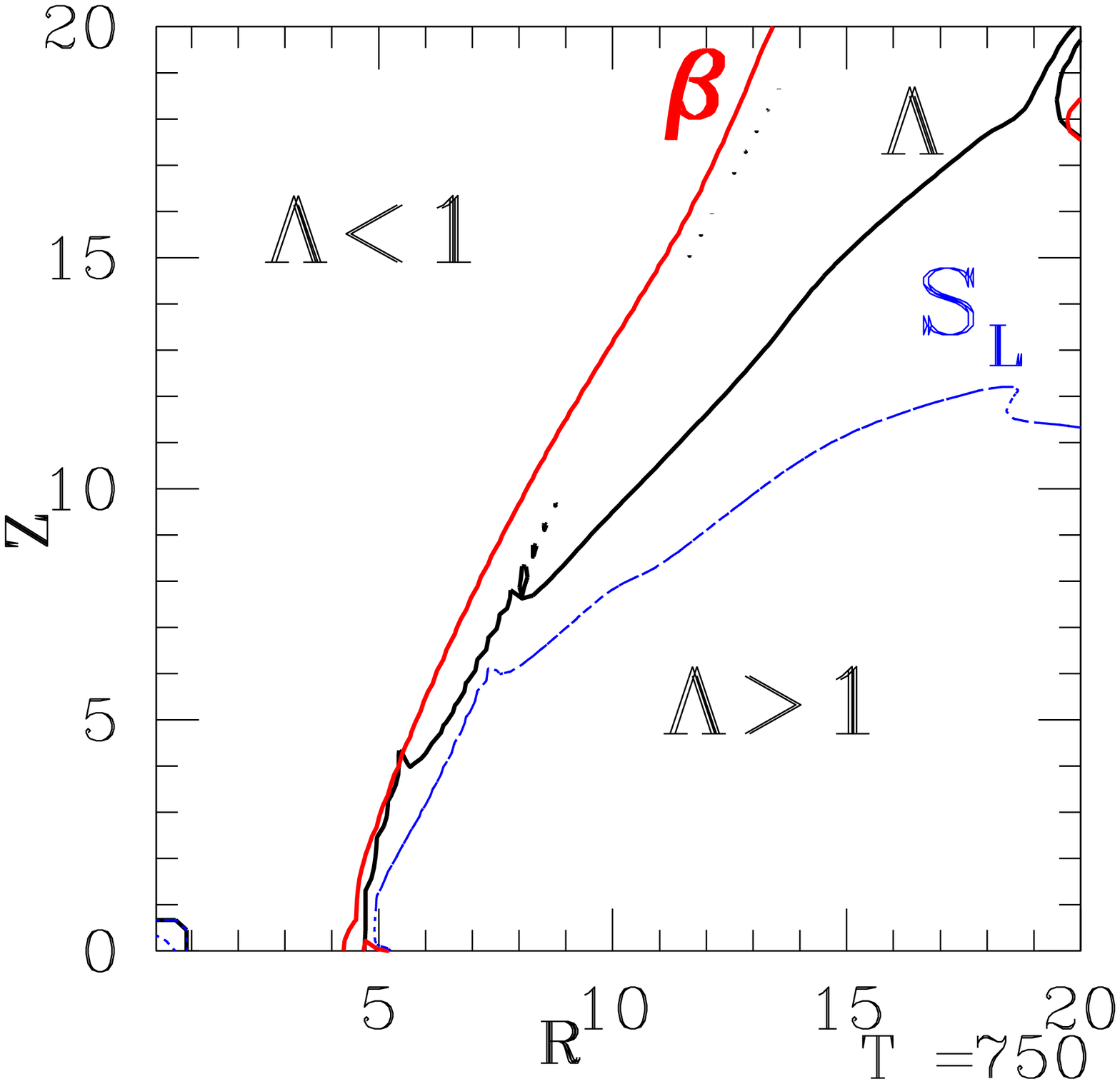}
\caption{
In thick (black) solid line is shown Elsasser number $\Lambda$ at T=125 and
T=750 in simulations S1b and S2, in the {\em Left} and {\em Right} panels,
respectively. Above the line in the computational box where magnetic
pressure equals matter pressure ($\beta=1$), and where the Elsasser
number $\Lambda$ equals unity (red solid line), launching of a flow
of matter is possible. Below those lines, the launching
of matter can not occur. For comparison, we show also the line where the
Lundquist number $S_L$ equals unity, in solid blue line.
}
\label{elsas}
\end{figure}

In Figure \ref{elsas}, we plot the Elsasser number $\Lambda>1$ in
simulations S1b and S2. We observe that $\Lambda\geq1$ is valid only in the
magnetosphere above the disk. For a small magnetic field of the order of
1\,G, there is no region in the box that could satisfy $\Lambda\geq1$, and
no launching is possible. For comparison, in the same figure we show the
lines where the magnetic and matter pressure are equal ($\beta=1$), and where
the Lundquist number equals unity. All three lines meet close to the disk
inner radius-the same is the case in simulations S1a-but then
separate higher above the disk. When compared with positions where ejection
is present, the line where $\Lambda =1$ and S$_L$=1 gives the
best separation between the locations where ejections do and do not occur,
in the cases of weaker magnetic field. The reason is possibly in
$\Omega_{\rm K}$ in the expression for $\Lambda$, which is still connecting
it with the disk, even in description of the (non-Keplerian) corona. We will
investigate this relation in further work, with more realistic disk. Line
with $\beta=1$ is always correctly following positions with the strongest
ejection.

\section{Disk truncation radius}\label{trunc}
\begin{figure}
\includegraphics[width=8.7cm]{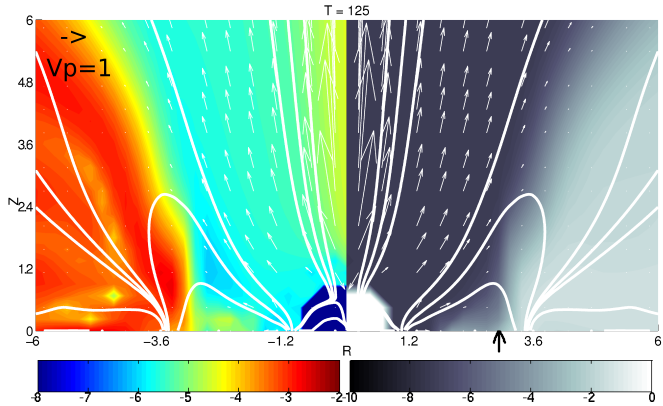}
\hspace*{.3cm}\includegraphics[width=8.3cm]{cbarsboth.eps}
\caption{
A zoom into the closest vicinity of the star from Figure \ref{S1abplots} in
simulation S1b, to show details in the disk gap, where the disk is
truncated. Meaning of colors and lines is the same as in Figure \ref{rhovS2}.
The stellar magnetic field is about 200 G, for the disk mass accretion rate
$10^{-6}M_\odot$~yr$^{-1}$. Black arrow marks the position of the truncation
radius of the disk, which is in this case $R_{\mathrm t}=3.0R_\ast$.
}
\label{S1bzoom}
\end{figure}
\begin{figure}
\includegraphics[width=4.2cm,height=4cm]{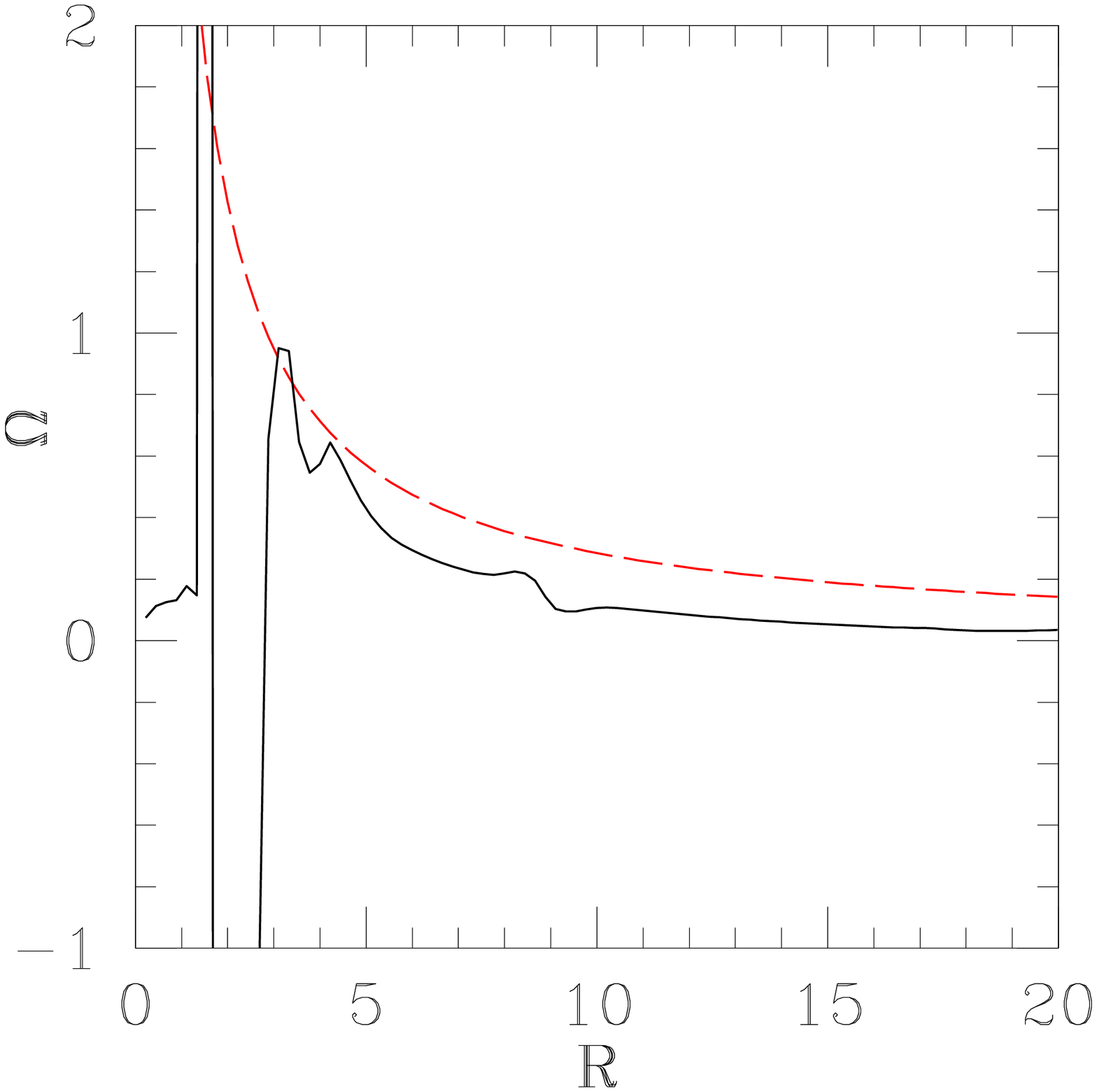}
\includegraphics[width=4.2cm,height=4cm]{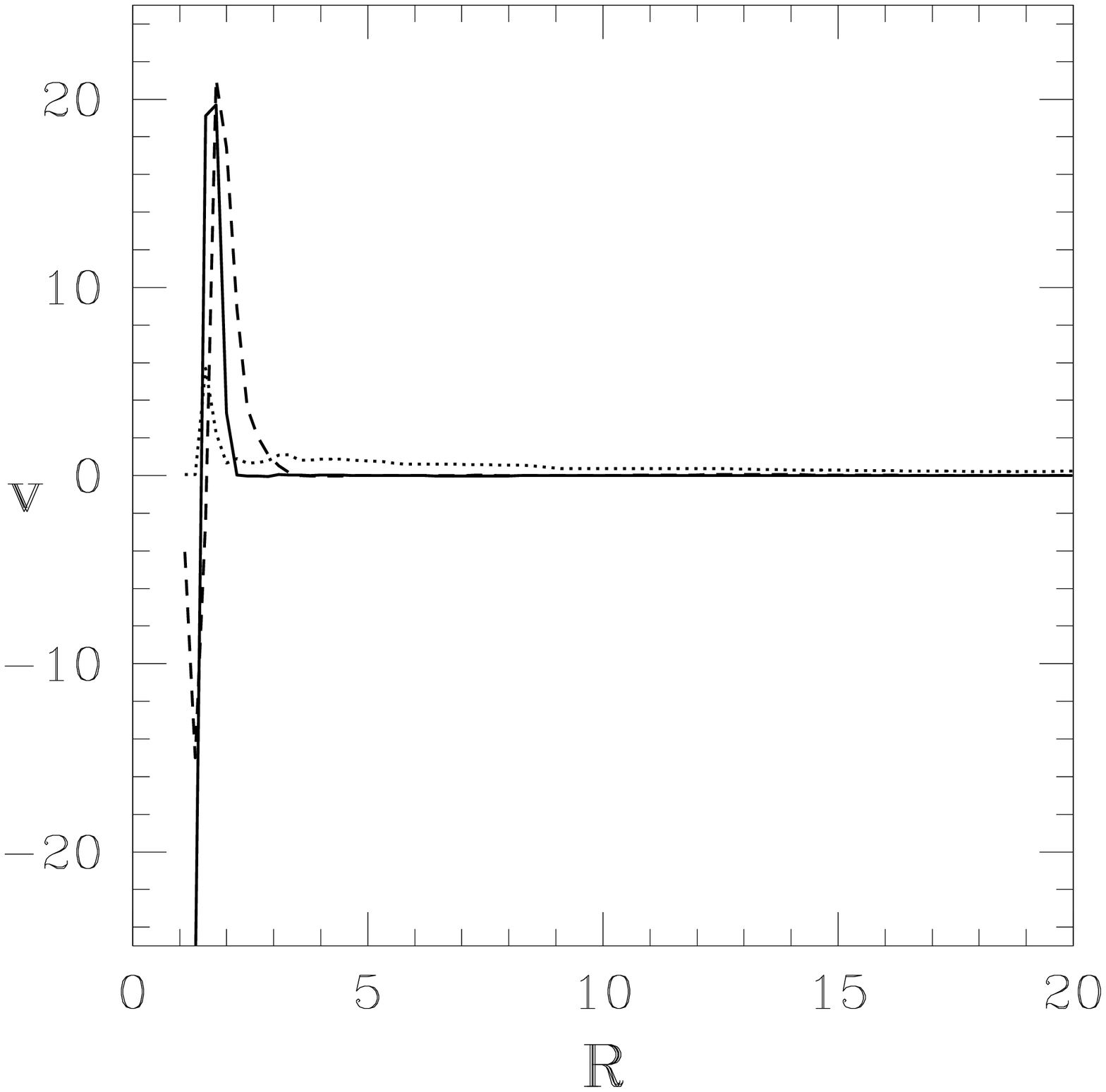}
\caption{
In {\it Left panel} in solid line is shown angular velocity profile along
the disk mid-plane at T=125 in simulation S1b. For comparison, in
dashed (red) line we show a Keplerian rotation profile. In the {\it Right
panel} we show, at the same time, components of velocity along the same
plane: v$_{\mathrm Z}$, v$_{\mathrm R}$ and v$_\phi$ in solid, dashed and
dotted lines, respectively.
}
\label{omegS1b}
\end{figure}
\begin{figure}
\includegraphics[width=4.2cm,height=4cm]{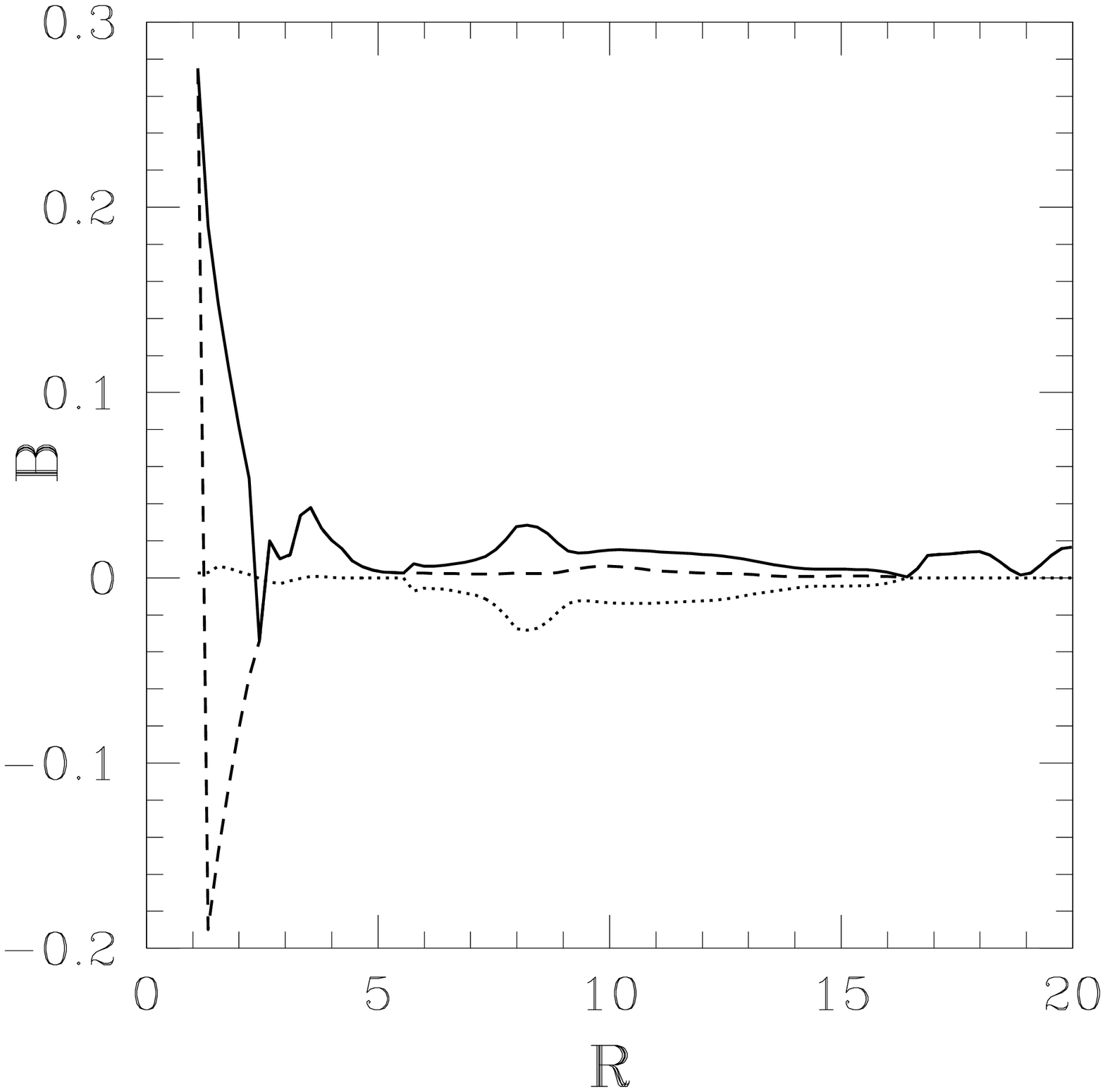}
\includegraphics[width=4.2cm,height=4cm]{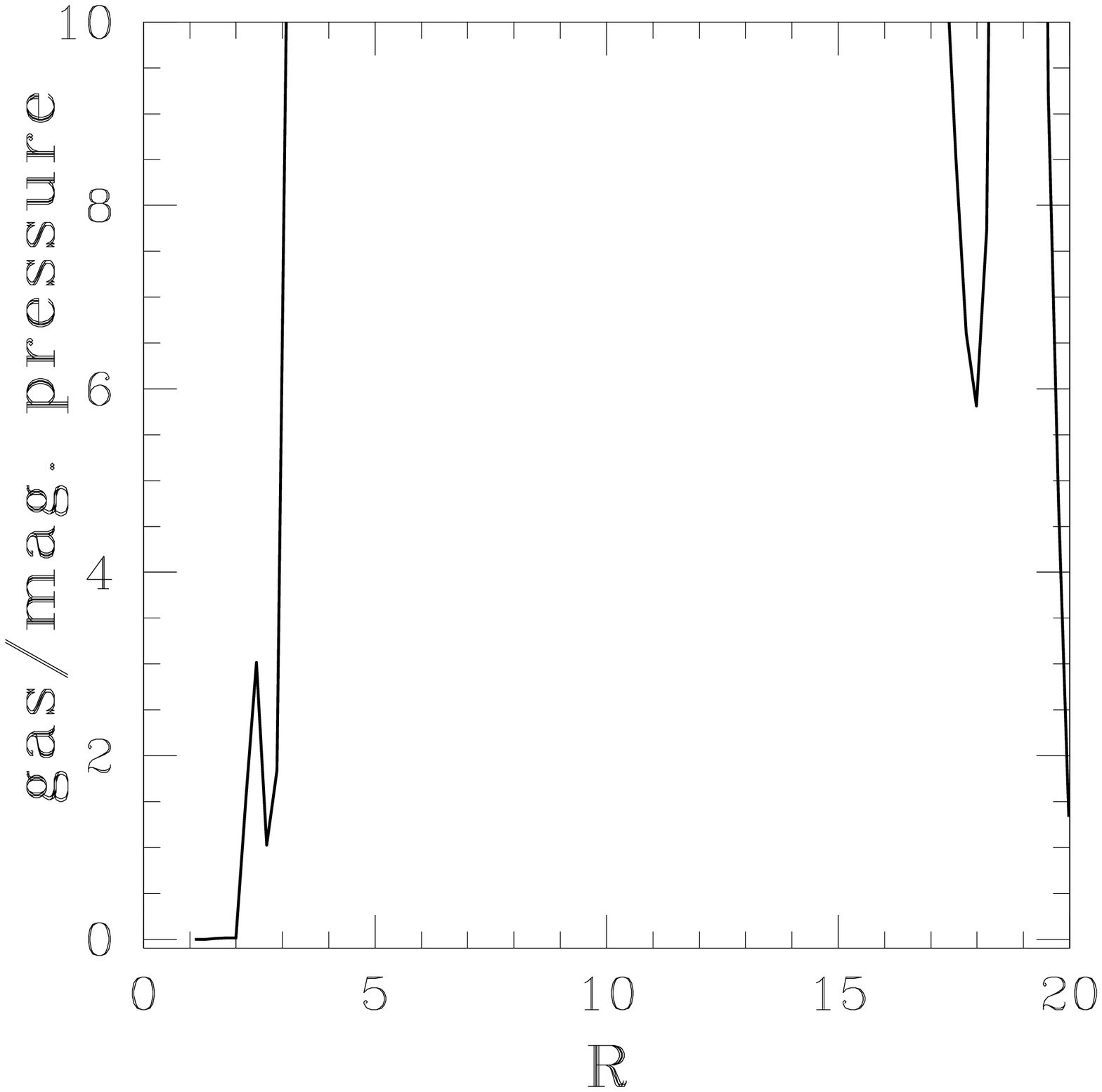}
\caption{
{\it Left panel} shows poloidal, toroidal and total
magnetic field along the disk mid-plane in simulation S1b at T=125 in
dashed, dotted and solid lines, respectively. Ratio of gas and magnetic
pressures along the same line is shown in the {\it Right panel}.
}
\label{bsprrat}
\end{figure}
In simulation S1b we maintained self-consistent boundary conditions at the
disk mid-plane, so we can try to estimate the disk truncation radius from
our results. The new truncation radius of the disk after the relaxation is
close to the line of balance of the disk ram pressure and the magnetic pressure,
$p+\rho v^2=B^2/8\pi$, located between a central object and the disk inner
radius. We show it in the \S\ref{res31}, where in the right panel of
Figure \ref{betaS1ab} we read that magnetic pressure is equal to the gas
pressure at the distance of about R=2R$_\ast$. To show details nearby the
disk gap, in figure Figure \ref{S1bzoom} we zoom close to the star
in simulation S1b. The disk terminates at $R_{\mathrm t}=3.0R_\ast$,
close to the position where the disk density drops steeply, and where the
magnetosonic Mach number equals unity at the equatorial plane, as matter is
being launched fast along the neutral line because of reconnection. The
angular velocity profiles along the disk mid-plane also show a dip at the disk
truncation radii, passing from the gap into the disk, as shown in Figure
\ref{omegS1b}. The magnetic field strength and the ratio between the gas and
magnetic pressure are shown in Figure \ref{bsprrat}. All show steep gradient
in quantities at crossing the disk truncation radius.

The disk truncation radius can be estimated as
$R_{\mathrm t}=\alpha_{\mathrm t}(B_\ast^4R_\ast^{12}/2GM_\ast\dot{M}_{\rm a}^2)^{1/7}$,
given by order of magnitude as the equilibrium of the ram
pressure of a spherical envelope in free fall and the magnetic pressure of
a stellar dipole \citep{el1977}. The non-dimensional factor
$\alpha_{\mathrm t}<1$ has been estimated to be 0.5 in
\cite{ghla78,ghla79a,ghla79b} and $\sim 1$ in \cite{osshu95}. In our
simulations the system departs significantly from the spherical infall case
and results in too large $\alpha_{\mathrm t}$, when compared to the
estimate above. Simulation S1b yields $\alpha_{\mathrm t}=4.17$. Simulations
in better resolution might improve matching with the theoretical prediction.

\cite{bes08} estimate $\alpha_{\mathrm t}$ to be equal
to $M_s^{2/7}$, where $M_s$ is the sonic Mach number measured at the disk
mid-plane, using the radial velocity of the matter in the disk for a comparison
with the sound speed. Their derivation was for the case with the accretion
column onto the star, but conditions should be similar also in the case of
ejection launched from the inner disk radius. For this estimate, inserting
the values from simulation S1b, we obtain $M_s^{2/7}=0.15$, which gives a
$R_{\mathrm t}=0.3R_\ast$. None of the two estimates matches our
result here. The reason is probably in disk puffing-up because of heating
produced in the disk, in our simulation with $\gamma=5/3$. An
investigation of location and stability of the disk truncation radius
should be done in a simulation with better disk computation. Further
improvement would also be to perform simulations in the complete R-Z
half-plane, to remove the constraint by the disk equatorial plane
boundary condition.

\section{Discussion}
In this work, we for the first time demonstrate launching fast,
light quasi-stationary flows of matter, which we call micro-ejections,
from the magnetosphere above the gap in the star-disk system. In our
simulations we included physical resistivity not only in the disk,
but in the whole computational box. We investigate effects of
anomalously large resistivity (when compared to microscopic
resistivity), modeled as a function of matter density, to the
launching of such micro-ejections in the case of slowly rotating star.

The physical resistivity, when included in previous simulations in the
literature, was limited to the disk, decreasing effectively to zero
out of it. The reconnection above the disk, which is necessarily
happening during such simulations, was at the mercy of numerical
resistivity, whose effects are different from physical resistivity
In order to better understand differences between a variety of
setups, we compare resistive simulations performed to date --- some
of them are shown in Table \ref{tabla1}. Resistivity can only
moderately modify the flow shape, on a slower time scale than the
formation of a magnetic wall \citep{lyb96,lyb03}, so that
if the resistivity is not large enough for reconnection to occur, the
geometry of the flow will be different. As sketched in Figure
\ref{geommag}, common phenomena have been identified in
our simulations, which occur, in those simulations of interacting
star-disk systems from Table \ref{tabla1} which have violent initial
relaxation and/or weak magnetic field. After the relaxation of the
system from unrealistic initial condition to a more evolved
configuration, magnetic reconnection adjusts the topology of compressed
field lines. Originally closed loops partially open. In this picture,
the resistivity plays an important role, as it enables the reconnection.

We model physical resistivity as an anomalous turbulent resistivity
dependent on the matter density $\rho$, following \cite{FC02}, with
$\eta=\eta_0 \rho^{1/3}$. Previously, resistivity has been included
in the whole computational domain in \citep{fen09}, but with the disk
only as a boundary condition. Those simulations followed the
propagation of outflow during few thousands of rotations of the
inner disk radius. With the disk included in the computation,
relaxation from initial conditions is much more complicated,
and it is not easy to follow the evolution of the system for hundreds
of rotations. We set up our resistive version of Zeus-3D code
without employing any special procedure to prevent the violent
relaxation from non-equilibrium initial conditions\footnote{Such
procedure could consist of the preparation of the particularly suitable
disk model, as in \citet{kuk03} or, as in the seminal paper of
\citet{R09}, slow introduction of the disk matter into the
computational box, with gradual increase of the stellar rotation rate.}.
Because of numerical difficulties with resistive MHD simulations in our
setup with violent relaxation, our simulations tend to cease during
relaxation, or last for too short after it for concluding about
the results with a stable disk gap at a longer timescale. To perform
the parameter study, we modified boundary conditions in the disk
equatorial plane, to enforce a stable disk gap.

For a sufficiently large stellar magnetic field, a
quasi-stationary micro-ejection forms. The magnetic field lines
are opened above the star, with the more or less episodic, very
light axial flow nearby the axis, which is artificial,
of numerical origin. The magnetosphere between the star and the disk,
along the boundary layer between the stellar and the disk field,
is the site of launching of slower and more massive micro-ejection
under a wider angle. The mechanism is similar to the launching of solar
micro-flares, only that in our case magnetosphere of the star-disk
system is the site of launching. Such events would leave trace in the
chemical and physical properties of the object, as indicated in
\citet{shuetal07}. The role of reconnection in accretion disks with
jets has been discussed also in \citet{EdP10}. We will address this
question in future work, together with the question of reconnection
with different models of resistivity.

In our parameter study we investigated influence of the density contrast
between the disk and corona, $\kappa$. There is no difference in results for
$\kappa$ from $10^{3}$ to $10^{6}$. For $\kappa=10^{2}$, the mass flux is
always larger for a factor of few, and we conclude that results of
simulations with $\kappa=10^{2}$ could be unrealistic. Another parameter we
checked was dependence of mass flux on stellar rotation rate. We found that
mass flux increases with stellar rotation rate. Both fluxes also increase
with increasing magnetic field--we vary the strength of the stellar dipole
in the range of (0.1--200)\,G for the disk accretion rate $10^{-6}M_\odot/$yr.
Both fluxes increase with increasing magnetic field. There is a limit to
increase of magnetic field in simulation. A very strong initial
field may cause strong shear motion at the beginning, and can largely
disturb the relaxation process, especially in the case of a non-resistive
corona, because pinching and reconnection of the magnetic field depend  
critically on the conditions in the magnetosphere \citep{lov95,gods97}.
Because resistivity enables the reconnection and reshaping of magnetic
field, we also investigated if fluxes in micro-ejection depend on
resistivity. We found that there is no such dependence. Without
resistivity, numerical or physical, there is no reconnection, but once it
occurs, the level of resistivity does not change the outcome of our
simulations. Further study of this dependence is needed, with different
resolutions. Our simulations do not scale well, and in the present
setup we could not perform such study. In ideal MHD simulations
performed with quasi-equilibrium initial conditions and slightly
higher resolution than in our simulations --- see e.g.\ \citet{rukl02} ---
numerical resistivity can mimic resistive effects, and such results could be
more realistic than those in the high resolution ideal-MHD simulations
\citep{Y86}. We note that in high resolution simulations, physical
resistivity should be included, or results could depend on numerical
resistivity, which is not related to physical quantities of the setup.
Any study of results is then necessarily flawed by purely numerical effects.
Different models for resistivity should be investigated, too. A model for
physical resistivity itself can affect the results, as was pointed out in
\cite{gods97}. 

To identify locations in which the launching is possible in our
computational box, we compare different indicators for a successful
launching. The best such indicator in general is the line where the   
magnetic and matter pressure are equal ($\beta=1$). We also find that
in cases with weak magnetic field, the Elsasser number
$\Lambda$, could serve as an indicator for successful launching.

There is no disk component of flow in our results, and this is why they
are very light. Matter in micro-ejections comes from the magnetosphere above
the disk gap, which is just a small part of the matter from the disk inner
radius, which slipped through the magnetic field lines.

\section{Summary}
We report results of our numerical simulations in the resistive MHD
regime of magnetospheric accretion and ejection in the closest vicinity
of a central object. Physical resistivity is non-negligible in the whole
computational box in our simulation, and varies in time. It enables
reconnection, which is essential for reshaping of the magnetic field
during simulations. For the first time, we show the launching of
long-lasting, fast micro-ejections in the case of purely resistive
magnetosphere of a slowly rotating star. Reconnection is part of the
launching mechanism of such ejections, similar to the solar
micro-flares, only that now launching occurs in the magnetosphere
above the disk gap.

We find that mass flux of micro-ejection increases with increasing
magnetic field strength and stellar rotation rate. For various density
contrasts between the disk and corona, or various resistivities,
difference in results is small and there is no clear dependence. As
for location where launching from the innermost magnetosphere is
possible, the line where the magnetic and matter pressure are balanced
is the best indicator where the strongest ejections occur. In the case of
weak magnetic field, the line where the Elsasser number $\Lambda$ equals
unity, might also serve as an indicator of launching region.

Micro-ejection found in our purely resistive simulations is launched by a
combination of pressure gradient and magnetic forces, in presence of ongoing
reconnection. It could be just a part of the overall outflow phenomena, or
transient, when the inner portion of the disk actually participates
dynamically and magnetically. Investigated solutions play a role in
re-distribution of the initial stellar dipole magnetic field by
reconnection into the stellar and disk fields, enabled by resistivity.

Our setup shares one caveat with most of the present works for a star-disk
problem: we do not include the stellar wind in simulations. It would
influence the solutions in the innermost magnetosphere. Together with
investigation of different models for resistivity, and stability in full
3D treatment, which could change because of different rate of reconnection,
we leave it for future work.

\acknowledgments
This work was supported by funding to Theoretical Institute for Advanced  
Research in Astrophysics (TIARA) in the Academia Sinica and National  
Tsing Hua University through the Excellence Program of the NSC, Taiwan.
M\v{C} performed part of work when resident in the European
Community's Marie Curie Actions - Human Resource and Mobility within
the JETSET network under contract MRTN-CT-2004005592 in Athens, Greece.
The authors thank Ruben Krasnopolsky, Oscar Morata, Jose Gracia and
Nektarios Vlahakis, for their very helpful discussions throughout the
project. We thank the LCA team and M. Norman for the possibility to use
the ZEUS-3D code. We acknowledge constructive criticism and useful
suggestions of a referee on a first version of this paper.

\end{document}